\documentclass[journal]{IEEEtran}

\usepackage{iycstyle}

\usepackage{algorithmic}
\usepackage{array}
\usepackage[caption=false,font=normalsize,labelfont=sf,textfont=sf]{subfig}
\usepackage{textcomp}
\usepackage{stfloats}
\usepackage{url}
\usepackage{hyperref}
\usepackage{verbatim}
\usepackage{graphicx}
\usepackage{balance}

\usepackage{bm}
\usepackage{tikz}
\usetikzlibrary{spy}
\usepackage{wrapfig}
\usepackage{booktabs}
\usepackage{xcolor}
\usepackage{cite}

\usepackage{filecontents}
\usepackage[resetlabels]{multibib}
\newcites{supp}{References}
\usepackage{cuted}
\usepackage{bbm}

\newcommand\blfootnote[1]{
  \begingroup
  \renewcommand\thefootnote{}\footnote{#1}
  \addtocounter{footnote}{-1}
  \endgroup
}

\def\boxitfig#1#2{%
    \smash{\color{red}\fboxrule=0.5pt\relax\fboxsep=0pt\relax
    \llap{\rlap{\hspace{0.7em}\raisebox{-0.7em}{\fbox{\phantom{\rule{#1}{#2}}}}}}}\ignorespaces
}

\def\boxittabb#1#2{%
    \smash{\color{red}\fboxrule=0.5pt\relax\fboxsep=0pt\relax%
    \llap{\rlap{\hspace{-1.8em}\raisebox{-0.25em}{\fbox{\phantom{\rule{#1}{#2}}}}}}}\ignorespaces
}

\def\boxittabbb#1#2{%
    \smash{\color{red}\fboxrule=0.5pt\relax\fboxsep=0pt\relax%
    \llap{\rlap{\hspace{-1.6em}\raisebox{-0.25em}{\fbox{\phantom{\rule{#1}{#2}}}}}}}\ignorespaces
}

  {\list{}{\leftmargin=#1\rightmargin=#1}\item[]}%
  {\endlist}

\usepackage{stackengine}
\def\deleq{\mathrel{\ensurestackMath{\stackon[1pt]{=}{\scriptstyle\Delta}}}}

\usepackage{xr}
\makeatletter
\newcommand*{\addFileDependency}[1]{
  \typeout{(#1)}
  \@addtofilelist{#1}
  \IfFileExists{#1}{}{\typeout{No file #1.}}
}
\makeatother

\begin{document}
\title{Self-supervised regression learning using domain knowledge: Applications to improving self-supervised  denoising in imaging}
\author{Il Yong Chun$^{\dagger}$,~\IEEEmembership{Member,~IEEE,} Dongwon Park$^\dagger$,~\IEEEmembership{Student Member,~IEEE,} Xuehang Zheng$^\dagger$, \\ Se Young Chun,~\IEEEmembership{Member,~IEEE,} and Yong Long,~\IEEEmembership{Member,~IEEE}
\thanks{
I.~Y.~Chun is supported in part by the Institute of Information \& communications Technology Planning \& Evaluation (IITP) grant funded by the Ministry of Science and ICT, South Korea (2019-0-00421, AI Graduate School Support Program (Sungkyunkwan University)), the Ingeborg v.F.~McKee Fund of the Hawai'i Community Foundation (20ADVC-106808), and
seed grants from Sungkyunkwan University.
D.~Park and S.~Y.~Chun are supported by Basic Science Research Program through the National Research Foundation of Korea (NRF) funded by the Ministry of Education (NRF-2017R1D1A1B05035810).
\textit{(Corresponding authors: Il Yong Chun, Se Young Chun, and Yong Long.)}

\textit{$\dagger$These authors equally contributed to this work.}

I.~Y.~Chun is with the School of Electronic and Electrical Engineering and Department of Artificial Intelligence, Sungkyunkwan University, Suwon, Gyeonggi 16419, Republic of Korea (e-mail: iychun@skku.edu).

D.~Park is with the Department of Electrical Engineering, Ulsan National Institute of Science and Technology, Ulsan 44919, Republic of Korea (e-mail: dong1@unist.ac.kr).

X.~Zheng and Y.~Long are with the University of Michigan - Shanghai Jiao Tong University Joint Institute, Shanghai Jiao Tong University, Shanghai 200240, China (email: zhxhang@sjtu.edu.cn; yong.long@sjtu.edu.cn). 

S.~Y.~Chun is with the Department of Electrical and Computer Engineering, INMC, Seoul National University, Seoul 08826, Republic of Korea (e-mail: sychun@snu.ac.kr).

This paper has supplementary material.
The prefix \dquotes{S} indicates the numbers in sections, figures, and tables in the supplement.
The prefix \dquotes{A} indicates the numbers in sections in the appendix.

}
}

\maketitle

\begin{abstract}
Regression that predicts continuous quantity is a central part of applications using computational imaging and computer vision technologies.
Yet, studying and understanding self-supervised learning for regression tasks -- except for a particular regression task, image denoising -- have lagged behind.
This paper proposes a general self-supervised regression learning (SSRL) framework that enables learning regression neural networks with only input data (but without ground-truth target data), by using a designable pseudo-predictor that encapsulates domain knowledge of a specific application. 
The paper underlines the importance of using domain knowledge by showing that under different settings, the better pseudo-predictor can lead properties of SSRL closer to those of ordinary supervised learning.
Numerical experiments for low-dose computational tomography denoising and camera image denoising demonstrate that proposed SSRL significantly improves the denoising quality over several existing self-supervised denoising methods. 

\end{abstract}


\section{Introduction}
\label{sec:intro}

Deep regression neural network (NN)-based methods that can accurately predict real- or complex-valued output have been rapidly gaining popularity in a wide range of computational imaging and computer vision applications including 
image denoising \cite{Vincent&etal:10JMLR, Xie&Xu&Chen:12NIPS, Zhang&etal:17TIP}, 
image deblurring \cite{Xu&etal:14NIPS}, 
image super-resolution \cite{Dong&etal:16TPAMI,Kim&Lee&Lee:16CVPR},
light-field reconstruction \cite{Chun&etal:20PAMI, Huang&etal:20ICASSP}, object localization \cite{Szegedy&Toshev&Erhan:13NIPS},
end-to-end autonomous driving \cite{Bojarski&etal:16NIPS-DLS}.
Yet, they lack a general self-supervised learning framework.

In training a regression NN $f : \bbR^{N} \rightarrow \bbR^{M}$,
the most prevalent supervised learning approach minimizes the mean square error (MSE) between what $f$ predicts from an input $x \in \bbR^N$ and a ground-truth target $y \in \bbR^M$:
\be{
\label{sys:sup}
\min_{f} \bbE_{x,y} \nm{ f(x) - y }_2^2.
}
Learning a denoising or refining NN uses \R{sys:sup} 
with $M=N$ -- dubbed Noise2True -- where $x$ is a corrupted image and $y$ is a clean (i.e., ground-truth) image.
However, it is challenging or even impossible to collect many clean images $y$ in many practical applications, motivating research on self-supervised learning for image denoising
\cite{Ulyanov&etal:18CVPR, 
Soltanayev&Chun:18NIPS, 
Krull&etal:19CVPR, 
Batson&Royer:19ICML, 
Laine&etal:19NIPS, 
Moran&etal:20CVPR,
Quan&etal:20CVPR,
Xu&etal:20TIP,
Hendriksen&etal:20TCI, 
Xie&Wang&Ji:20NIPS,
Krull&etal:20FCS,
Prakash&etal:20ISBI,
Prakash&Krull&Jug:21ICLR,
Khademi&etal:21WACV,
Huang&eta:21CVPR,Kim&Ye:21NIPS}
-- called self-supervised image denoising.
To learn a denoiser $f$ with single noisy images,
a popular self-supervised image denoising method, Noise2Self
\cite{Batson&Royer:19ICML} (see also the concurrent work \cite{Krull&etal:19CVPR}), and its sophisticated relaxation, Noise2Same \cite{Xie&Wang&Ji:20NIPS}, study the following MSE minimization problem:
\be{
\label{sys:self-sup}
\min_{f} \bbE_{x} \nm{ f(x) - x }_2^2.
}
These methods use some partitioning schemes in \R{sys:self-sup} to avoid that its optimal solution is just the identity mapping $\cI$. 
Partitioning schemes were originally proposed to train denoising NN only with one noisy observation per scene.
Noise2Noise \cite{Lehtinen&etal:18ICML} that learns a denoiser with pairs of two independent noisy images,
is a pioneer work for self-supervised image denoising.
Motivated by Noise2Noise, several self-supervised image denoising methods such as Noise2Inverse \cite{Hendriksen&etal:20TCI}, Neighbor2Neighbor \cite{Huang&eta:21CVPR}, and \cite{Soltanayev&Chun:18NIPS, Moran&etal:20CVPR, Xu&etal:20TIP}
emulate pairs of two independent noisy images,
by applying partitioning or adding simulated noise to single noisy measurements. Noise2Inverse~\cite{Hendriksen&etal:20TCI} emulates two independent noisy images by dividing measurement data, e.g., sinogram in computational tomography (CT), into two and reconstructing them.
Neighbor2Neighbor \cite{Huang&eta:21CVPR} emulates two independent noisy images from a single noisy image using a sub-sampling method.
In addition, there exist single-image self-supervised denoising methods that fit a denoising NN to each noisy image, such as deep image prior~\cite{Ulyanov&etal:18CVPR}, Noisy-As-Clean~\cite{Xu&etal:20TIP} and Self2Self~\cite{Quan&etal:20CVPR}.
All the aforementioned methods are \textit{blind} in a sense that one does not need to estimate noise parameters,
whereas some works estimate statistical parameters of noise, such as noise histogram \cite{Krull&etal:20FCS} and parameters of Gaussian mixture noise model \cite{Prakash&Krull&Jug:21ICLR}.
All the aforementioned methods except for \cite{Ulyanov&etal:18CVPR} have been developed based on some noise assumptions 
including pixel-wise independent noise
\cite{Krull&etal:19CVPR, Xie&Wang&Ji:20NIPS, Huang&eta:21CVPR}, zero-mean noise \cite{Lehtinen&etal:18ICML, Batson&Royer:19ICML, Xie&Wang&Ji:20NIPS}, and \dquotes{weak} noise \cite{Xu&etal:20TIP}.
Yet, the existing self-supervised denoising methods lack design flexibility that might relax noise assumptions and further improve the denoising performance of NNs.

This paper presents new insights on this topic.
The paper proposes a general self-supervised learning framework for regression problems, which we refer to as 
self-supervised regression learning (SSRL).
Proposed SSRL enables learning regression NNs with only input samples, by using a designable pseudo-predictor that can encapsulate domain knowledge of a specific application.
Our main results show that under some mild condition(s) (e.g., in image denoising, statistical noise properties in $x$), the \dquotes{better} pseudo-predictor is designed, the prediction error of the optimal solution of SSRL is expected to become closer to that of ordinary supervised learning or
the proposed SSRL loss becomes closer to the supervision counterpart. 
(We will define a \dquotes{good} pseudo-predictor when presenting the main results for different setups.)
In addition, a pseudo-predictor designed with good domain knowledge can substantially relax noise assumptions of existing self-supervised denoising methods. 
Numerical experiments for image denoising in low-dose CT and camera imaging with both simulated and intrinsically noisy (i.e.,~real-world) datasets -- corrupted by only single noise realization -- demonstrate that the proposed SSRL framework significantly improves denoising quality compared to several existing self-supervised denoising methods.
Put together, our findings provide new insights into how using good domain knowledge can improve self-supervised denoising, underscoring the benefits of understanding application-specific knowledge in SSRL.

The remainder of this paper is organized as follows.
Section~\ref{sec:theory} proposes the SSRL framework in different setups, 
and analyzes prediction errors of its optimal solution and SSRL loss function depending on a designable pseudo-predictor.
In addition, the section describes its relations to existing self-supervised denoising works, and explains that SSRL can relax noise assumptions of existing self-supervised denoising methods. 
Section~\ref{sec:know} explains how to use domain knowledge for designing a pseudo-predictor in low-dose CT and camera imaging applications with practical noise models, and proposes an empirical pseudo-predictor selection approach for image denoising if domain knowledge of specific applications is unavailable. 
Section~\ref{sec:result} demonstrates the outperforming performance of the proposed SSRL approach over the several existing self-supervised denoising methods, Noise2Self \cite{Batson&Royer:19ICML}, Noise2Inverse \cite{Hendriksen&etal:20TCI} or Neighbor2Neighbor \cite{Huang&eta:21CVPR}, Noise2Same \cite{Xie&Wang&Ji:20NIPS}, Noisy-As-Claen~\cite{Xu&etal:20TIP}, and Self2Self~\cite{Quan&etal:20CVPR}, with low-dose CT and camera image denoising experiments using both real and synthetic datasets.
In addition, Section~\ref{sec:result} provides some discussions based on experimental results therein.

\section{SSRL using domain knowledge}
\label{sec:theory}
The proposed SSRL loss is given by
\be{
\label{sys:SSRL}
\bbE_x \nm{ f(x) - g(x) }_2^2,
}
where $g : \bbR^N \rightarrow \bbR^M$ is a designable pseudo-predictor encapsulating domain knowledge of a specific application. We will incorporate some sophisticated setups in \R{sys:SSRL} such that $f$ obtained by minimizing \R{sys:SSRL} cannot merely be $g$. 
Although related theorems (see later) hold for any $M$, we mainly focus on practical image denoising applications.

\begin{figure}[t!]
\centering
\small\addtolength{\tabcolsep}{-9.5pt}

\begin{tabular}{cc}
 	
 		\raisebox{-.5\height}{
 			\begin{tikzpicture}
 			\begin{scope}[spy using outlines={rectangle,yellow,magnification=1.55,size=15mm, connect spies}]
 			\node {\includegraphics[width=34mm]{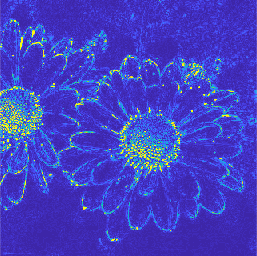}};
			\node [yellow] at (0.0,1.4) {\small $\text{PSNR} = 25.8$ dB};
 			\end{scope}
 			\end{tikzpicture}} &  		 
 		\raisebox{-.5\height}{
 			\begin{tikzpicture}
 			\begin{scope}[spy using outlines={rectangle,yellow,magnification=1.55,size=15mm, connect spies}]
 			\node {\includegraphics[width=34mm]{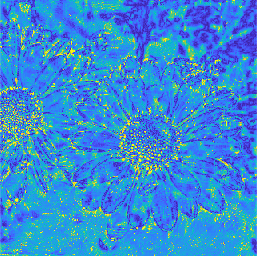}};
			\node [white] at (0.0,1.4) {\small $\text{PSNR} = 23.1$ dB};
 			\end{scope}
 			\end{tikzpicture}}
\end{tabular}
\vspace{-0.25pc}

\caption{Error map comparisons of denoised images from \R{sys:SSRL} using $g(x) = \mathrm{median} (x)$ {\bfseries (left)} and $g(x) = \mathsf{BM3D} (x)$ {\bfseries (right)} (blue and yellow denote $0$ and $50$ absolute errors, respectively). Peak signal-to-noise ratio (PSNR) values are averaged.}
\label{fig:motivate}
\vspace{-0.5pc}
\end{figure}

\subsection{Motivation} 
\label{sec:motivation}

This section empirically shows that understanding domain knowledge is important for designing $g$ in proposed SSRL loss \R{sys:SSRL}. The following camera image denoising examples demonstrate that \dquotes{well-designed} $g$ with good domain knowledge improves the denoising performance of learned $f$ via \R{sys:SSRL}.

Suppose that camera images are corrupted by salt-and-pepper noise. 
Consider two example setups for $g(\cdot)$,
median filtering and BM3D denoiser \cite{Mkinen&etal:20TIP}, denoted by $\mathrm{median}(\cdot)$ and $\mathsf{BM3D}(\cdot)$, respectively.
Fig.~\ref{fig:motivate} compares the denoising performance of minimum $f^\star$ with the two aforementioned $g$ setups:
$f^\star$ with median filtering significantly improved that with BM3D denoiser.
This is not surprising, as median filtering is widely known to be effective in reducing salt-and-pepper noise \cite[\S3.2]{Bovik:10book}.
This result emphasizes the importance of understanding domain knowledge of specific applications in proposed SSRL.

\subsection{Preliminaries}
\label{sec:prelim}

We first introduce the $\cJ$-complement between two functions $f$ and $g$:

\defn{
\label{def:Jind}
For a given partition $\cJ = \{ J_1,\ldots, J_B \}$ ($| J_1 | + \ldots + | J_B | = N$) of the dimensions of input $x \in \bbR^N$,
functions $f: \bbR^N \rightarrow \bbR^M$ and $g: \bbR^N \rightarrow \bbR^M$ are called $\cJ$-complementary,
if $f(x_{J^c})$ does not depend on $g(x_{J})$ for all $J \in \cJ$,
where $J^c$ denotes the complement of $J$, and $(\cdot)_{J^c}$ and $(\cdot)_J$ denote vectors restricted to $J^c$ and $J$, respectively. 
}

That is, $f$ and $g$ use information from outside and inside of $J$ to predict output and give pseudo-target, respectively.
In denoiser learning (where $M = N$), 
Definition~\ref{def:Jind} specializes to the
$\cJ$-invariance of $f$ \cite{Batson&Royer:19ICML}, by setting $g = \cI$.
Incorporating Definition~\ref{def:Jind} into the SSRL loss \R{sys:SSRL} 
is a straightforward approach to avoid that 
optimal $f$ is just $g$ in \R{sys:SSRL}.

Regardless of using the $\cJ$-complement in Definition~\ref{def:Jind},
the proposed SSRL framework uses the following assumption:
\bull{
\item[] \textit{Assumption~1)}
$f$ and $g$ are (Borel-)measurable.
}
Assumption~1 is satisfied if $f$ and $g$ are continuous. 
This condition is mild because many regression NNs $f$ are continuous -- where their modules, convolution, matrix-vector multiplication, rectified linear unit activation, max pooling, etc.~are continuous -- and one can design $g$ with measurable or continuous function.
If we do \textit{not} use  the $\cJ$-complement in SSRL, we additionally assume the following:

\bull{
\item[] \textit{Assumption~2)} $x_J$ and $x_{J^c}$ are conditionally independent given $y$, i.e., $p(x|y) = p(x_J | y) p(x_{J^c} | y)$.
\item[] \textit{Assumption~3)} $\bbE[g(x) | y] = y$.
}
In general, one may attempt to satisfy Assumption~2 by adding some randomized perturbations (independent of $y$) to either $J$ or $J^c$, similar to \cite{Moran&etal:20CVPR}.
In many imaging applications, Assumption~2 does not hold, e.g., noise has correlation between pixels, in denoiser learning, due to their complicated imaging physics.
It holds if
noise in each subset $J \in \cJ$ is conditionally independent from that in $J^c$, given $y$.
Assumption~3 gives an example for $g$ design.
Suppose that $x$ has non-zero-mean noise;
one then can design $g$ to make noise zero-mean using the domain knowledge.
(This will correspond to the direction for designing $g$ in SSRL in Sections~\ref{sec:SSRL:ind}--\ref{sec:SSRL}.)

The following two sections investigate two different SSRL approaches:
one uses the $\cJ$-complement in Definition~\ref{def:Jind}; the other does not.

\subsection{SSRL using domain knowledge with $\cJ$-complement}
\label{sec:SSRL:ind}

This section studies SSRL loss \R{sys:SSRL} minimization over $f$ that is $\cJ$-complementary of $g$.
Our first main result shows under Assumption~1 that 
the prediction error of the optimal solution $f^\star$ of SSRL \R{sys:SSRL} (using the $\cJ$-complement) is the sum of a deviation between its prediction and prediction of optimal $f^\ast$ from supervised learning \R{sys:sup}, and some irreducible error.

\thm{
\label{thm:soln}

Suppose that Assumption~1 holds.
Under the $\cJ$-complement setup, the optimal solution for the SSRL problem \R{sys:SSRL} is given by 
\be{
\label{eq:thm:soln}
f^\star (x) = \bbE [g(x_{J}) | x_{J^c}]
}
for each subset $J \in \cJ$.
The expected prediction error of $f^\star$ on an unseen input $x'$ is given by\footnote{
One can alternatively present the result \R{eq:thm:soln} by using the population model $y = f^\ast (x) + \zeta$, where $\zeta \in \bbR^M$ captures some inherent noise in $y$.
Under the classical assumption that $x \indep \zeta$ and $\bbE [\zeta] = 0$,
the expected prediction error of $f^\star$ is given by
$\bbE_{x,y} \| f^\star(x) - y \|_2^2 = \bbE_x \| f^\star (x) - f^\ast (x) \|_2^2 + \mathrm{Var}_\zeta ({\zeta})$.
}
\be{
\label{eq:thm:soln:err}
\bbE [ \| f^\star (x) - y \|_2^2 | x = x' ]
= \| f^\star (x') - f^\ast (x') \|_2^2 + \mathrm{Var}( y | x=x' ),
}
where $f^\ast$ is the optimal solution of supervised learning \R{sys:sup}, $f^\ast (x) = \bbE[y|x]$.

}
\prf{
See Section~\ref{sec:prf:thm:soln} in the appendix.
}

Result \R{eq:thm:soln} provides the benchmark under the $\cJ$-complement setup that 
SSRL \R{sys:SSRL} strives to approximate the conditional expectation in \R{eq:thm:soln}.
To better understand \R{eq:thm:soln}, consider two extreme cases.
First suppose that $x_J$ is a function of $x_{J^c}$. Then, $g(x_{J})$ is the optimal solution of SSRL in the sense of MSE, and one may not need to learn this as $g$ is already given.
The other extreme case is when $x_J$ and $x_{J^c}$ are independent.
In this case, $\bbE[ g(x_{J}) ]$ best approximates $y$ (in the MSE sense).

To reduce expected prediction error \R{eq:thm:soln:err},
our general aim in SSRL \R{sys:SSRL} is to design a \dquotes{good} pseudo-predictor $g$ that can lead the optimal solution of SSRL~\R{sys:SSRL} close to that of supervised learning~\R{sys:sup} particularly in the Euclidean norm sense.
If Assumption~1 is satisfied,
we suggest designing $g$ such that
\bulls{
    \item \textit{pseudo-prediction $g(x_J)$ becomes close to ground-truth $y$ in the expected sense given $x_{J^c}$;}}
we call such $g$ a \dquotes{good} or \dquotes{well-designed} pseudo-predictor.
Under the setup that $f$ in supervised learning \R{sys:sup} predicts output only from $x_{J^c}$ (so the optimal solution is $\bbE [ y | x_{J^c} ]$),
 using such well-designed $g$ will reduce the expected prediction error in \R{eq:thm:soln:err}.
For example, 
if $g$ is ideal such that $g(x_J) = y$ (given $x_{J^c}$),
the optimal solution of SSRL loss \R{sys:SSRL} becomes the supervision counterpart and the prediction error achieves the lowest bound $ \mathrm{Var}( y | x=x' )$ that is irreducible.\footnote{
It is well known that the prediction error of $f^\ast$ on $x'$ is $ \mathrm{Var}( y | x=x' )$ that is \textit{irreducible}.
Similar to this, \R{eq:thm:soln:err} has irreducible error captured by the second term.
}
To design good $g$, domain knowledge of specific application is crucial.
Domain knowledge includes noise properties in $x$, or can be captured by pre-trained NN via existing self-supervised denoising, such as Noise2Self \cite{Batson&Royer:19ICML} and Noise2Noise \cite{Lehtinen&etal:18ICML}.

Now, we investigate when the optimal solution of SSRL becomes equivalent to the supervision counterpart.
In particular, we additionally consider Assumptions~2--3.
If all Assumptions~1--3 are satisfied under the $\cJ$-complement setup, we can remove the third term in the following observation rewritten from SSRL loss \R{sys:SSRL}:
\be{
\begin{split}
\label{sys:SSRL,rewrite1}
&~ \bbE_x \| f(x) - g(x) \|_2^2 
\\
&= \bbE_{x,y} \| f(x) - y \|_2^2 + \| g(x) - y \|_2^2 - 2 \ip{ f(x) - y }{ g(x) - y }.
\end{split}
}
Our second main result under Assumptions~1--3 shows that the optimal solution of SSRL \R{sys:SSRL} with the $\cJ$ complement becomes the supervision counterpart (that is obtained by setting $f$ in \R{sys:sup} to predict output only from $x_{J^c}$).

\prop{
\label{thm:soln-all}

If all Assumptions 1--3 are satisfied, then under the $\cJ$-complement setup, the optimal solution for the SSRL problem \R{sys:SSRL} is given by 
\be{
\label{eq:thm:soln-all}
f^\star(x) = \bbE [y | x_{J^c}],
}
for each subset $J \in \cJ$.
}

\prf{
See Section~\ref{sec:prf:thm:soln-all} in the appendix.
}

Different from Theorem~\ref{thm:soln}, Proposition~\ref{thm:soln-all} that additionally uses Assumptions 2--3 does not show the effectiveness of using $g$ from the perspective of optimal solution.
Note, however, that using a desinable function $g$ can still substantially relax the noise assumption of several existing self-supervised denoising methods, $\bbE[x|y] = y$. (We will elaborate this in Section~\ref{sec:relax}.)

Proposition~\ref{thm:soln-all} reveals the conditions for optimal performance of SSRL.
We conjecture in SSRL that better satisfying the conditions leads to learn better $f$.
If Assumptions~1-2 are satisfied,
to better satisfy Assumption~3,
we suggest designing $g$ such that
\bulls{
\item \textit{$g(x_J)$ given $y$ is expected to becomes close to $y$;}
}
we call such $g$ a \dquotes{good} or \dquotes{well-designed} pseudo-predictor.
If one can design ideal $g$ that can perfectly satisfy $\bbE[g(x_J) | y] = y$ (Assumption~3 under the $\cJ$-complement setup) -- but not necessarily needing $g(x_J) = y$ -- then the optimal performance in Proposition~\ref{thm:soln-all} can be achieved.

\begin{figure*}[!t]
	\centering  	
	\hspace*{-0.5pc}\begin{tabular}{c}
		\includegraphics[width=150mm]{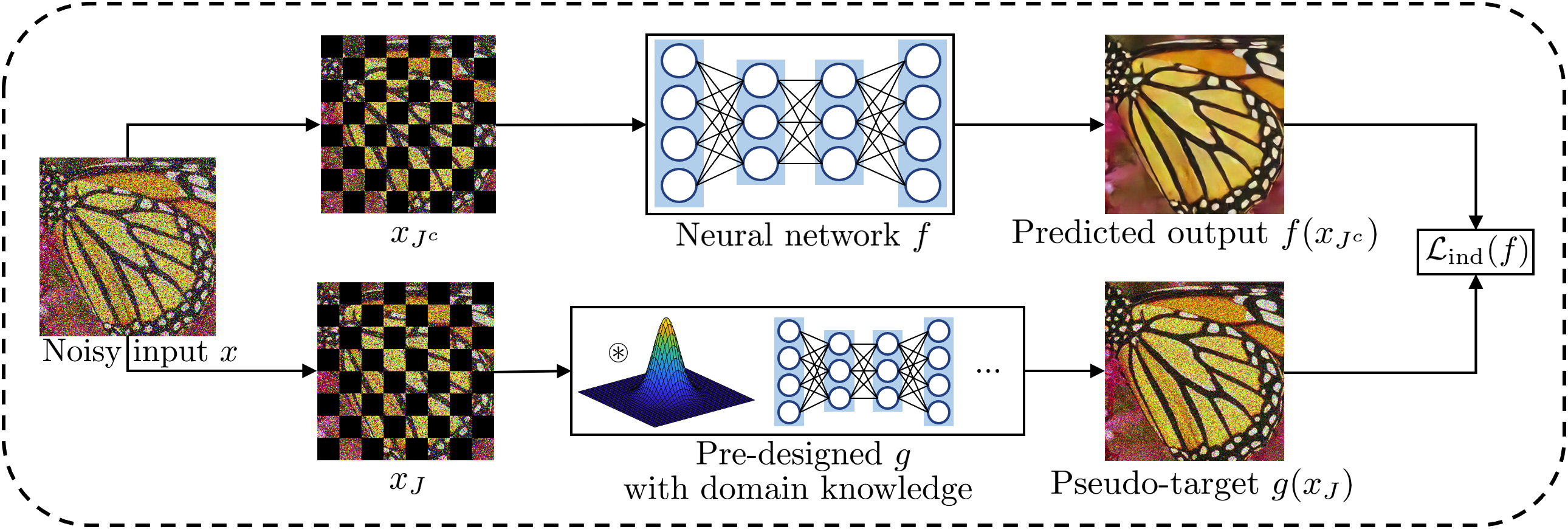}	
		\\

		\includegraphics[width=150mm]{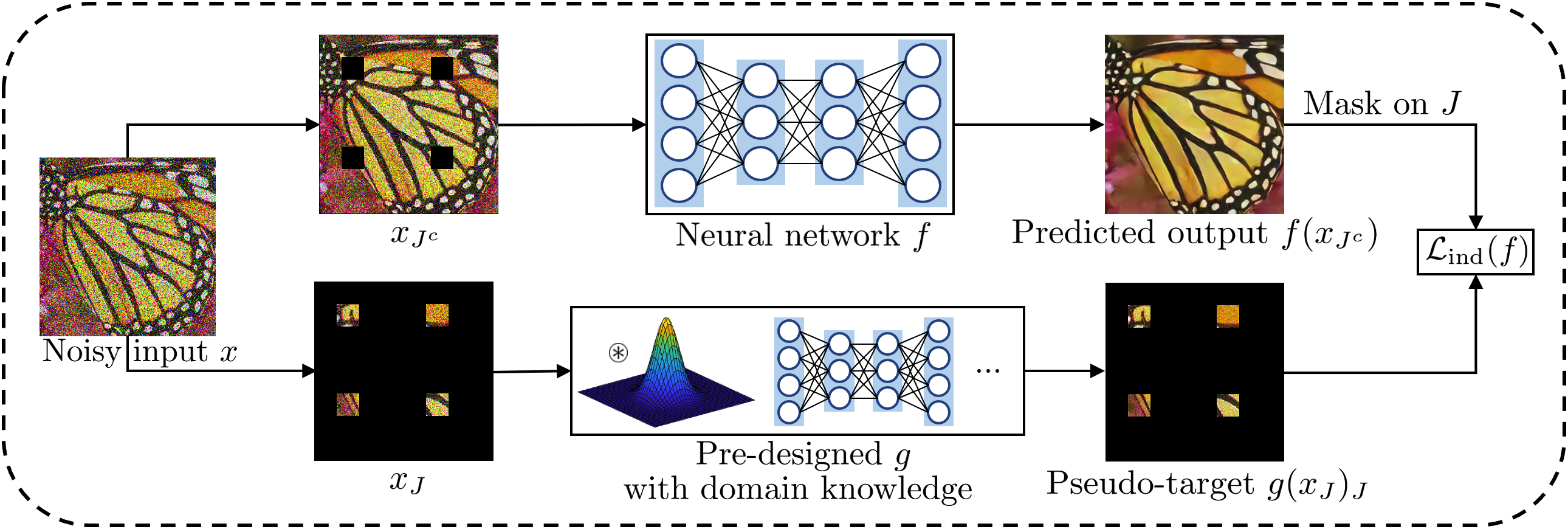}
	\end{tabular}
	
	\vspace{-0.25pc}
	\caption{Proposed SSRL models using the $\cJ$-complement in denoiser learning. 
	{\bfseries Top:} $f$ and $g$ use almost equal amount of information from input, i.e., $|J| \approx |J^c|$, where $J$ and $J^c$ are complementary checkerboard masks.
	{\bfseries Bottom:} $f$ and $g$ use unbalanced amount of information from input, specifically,  $|J^c| \gg |J|$.
	} 
	\label{fig:diagram} 
 	\vspace{-0.5pc}
\end{figure*}

Finally, the proposed SSRL loss using the $\cJ$-complement is given by
\be{
\label{sys:SSRL:Jind}
\cL_{\text{ind}}(f) \deleq \sum_{J \in \cJ} \bbE_x \nm{ f(x_{J^c}) - g(x_{J}) }_2^2. }
Fig.~\ref{fig:diagram} illustrates \R{sys:SSRL:Jind} with complementary checkerboard masks $J^c$ and $J$, where $f$ and $g$ use almost equal amount of information, 
and its variant, where $g$ uses much less information than $f$.
The variant computes MSE only on $J \in \cJ$; in this setup, it is challenging for $g$ to predict the entire image.

\subsection{SSRL using domain knowledge without $\cJ$-complement}
\label{sec:SSRL}

This section studies the SSRL loss \R{sys:SSRL} without using the $\cJ$-complement in Definition~\ref{def:Jind}.
Observe by \R{sys:SSRL,rewrite1} that 
$\bbE_{x,y} \| f(x) - y \|_2^2 + \| g(x_J) - y \|_2^2 
= \bbE_x \| f(x) - g(x_J) \|_2^2 + 2 \bbE_{x,y} \ip{ f(x) - y }{ g(x_J) - y }$.
Inspired by Noise2Same \cite{Xie&Wang&Ji:20NIPS},
the second proposed SSRL approach is to minimize an approximation of the right hand side in this equation that does \emph{not} assume that $f$ and $g$ are $\cJ$-complementary.
Our third main result finds an upper bound for the term $\bbE_{x,y} \ip{ f(x) - y }{ g(x_J) - y }$ without relying on the $\cJ$-complement (remind that this term vanishes if $f$ and $g$ are $\cJ$-complementary; see Proposition~\ref{thm:soln-all}).

\prop{
\label{thm:loss:b}
Assume that $\mathrm{Var} ( g (x_J)_m | y ) \leq \sigma^2$, $\forall m$.
Under Assumptions~1--3, the following bound holds:
\be{
\label{eq:thm:loss:b}
\bbE_{x,y} \ip{ f(x) - y }{ g(x_J) - y } \leq \sigma \sqrt{M} \cdot \big( \bbE_x \| f(x) - f(x_{J^c}) \|_2^2 \big)^{1/2}.
}
The following bound similarly holds for any $K \in \cK$: $\bbE_{x,y} \ip{ f(x)_K - y_K }{ g(x_J)_K - y_K } \leq \sigma \sqrt{|K|} \cdot \big( \bbE_x \| f(x)_K - f(x_{J^c})_K) \|_2^2 \big)^{1/2}$, 
where $\cK$ is a partition of $\{ 1,\ldots, M \}$, and $f(\cdot)_K$ and $g(\cdot)_K$  denote $f(\cdot)$ and $g(\cdot)$ restricted to $K$, respectively.
}
\prf{
See Section~\ref{sec:prf:thm:loss:b} in the appendix.
}

Using Proposition~\ref{thm:loss:b}, the proposed SSRL loss that does not rely on the $\cJ$-complement is given by
\be{
\label{sys:SSRL:b}
\begin{split}
 \cL(f) \deleq  &  \sum_{J \in \cJ} \bbE_x \| f(x) - g(x_J) \|_2^2 
 \\
&\hspace{1.5pc} + 2 \sigma \sqrt{M} \cdot \big( \bbE_x \| f(x) - f(x_{J^c}) \|_2^2 \big)^{1/2},
\end{split}
}
where $\sigma$ is given in Proposition~\ref{thm:loss:b}.
Here, a regression NN $f$ can use information from the entire input $x$,
whereas $\cL_{\text{ind}} (f)$ in \R{sys:SSRL:Jind} uses only partial input $x_{J^c}$ in $f$.
Proposition~\ref{thm:loss:b} suggests a direction for designing $g$ under its setup including that Assumptions~1--3 hold.
Under the setup in Proposition~\ref{thm:loss:b},
\dquotes{well-designed} $g$ such that $g(x_J)$ moves near $y$ given $y$ (i.e., $\sigma^2$ is small) will give a sharp bound in \R{sys:SSRL:b}, consequently leading \R{sys:SSRL:b} close to \R{sys:sup}.
Similar to the variant of $\cL_{\text{ind}} (f)$ (see Section~\ref{sec:SSRL:ind}),
if the amount of information between two partitions $J$ and $J^c$ is unbalanced in denoiser learning,
one can modify \R{sys:SSRL:b} to compute the MSE only on $J$ or $J^c$ in either both terms or the right term in \R{sys:SSRL:b}. 

We conjecture under the setup in Proposition~\ref{thm:loss:b} that the \dquotes{goodness} of $g$ is captured by $\sigma$ defined in Proposition~\ref{thm:loss:b}.
We support the conjecture with examples in Appendix~\ref{sec:SSRL:eg}.
Then, this goodness of $g$ balances the two terms in \R{sys:SSRL:b} via $\sigma$.
If $g$ is well-designed such that $g(x_J)$ moves near $y$, i.e., $\sigma^2$ is small, then the SSRL loss \R{sys:SSRL:b} relies more on the first term with good pseudo-target.
If $g$ is poorly-designed such that $\sigma^2$ is large, then \R{sys:SSRL:b} puts more weight more on the second term that can implicitly promote the $\cJ$-invariance of $f$ in \cite{Batson&Royer:19ICML}.

\subsection{Relation to previous self-supervised denoising works}

This section compares the proposed SSRL framework with previous self-supervised denoising works.

\subsubsection{SSRL using $\cJ$-complement}

In denoiser learning, the proposed SSRL loss \R{sys:SSRL} with the $\cJ$-complement of $f$ and $g = \cI$
specializes to \R{sys:self-sup} with the $\cJ$-invariance of $f$, i.e., Noise2Self \cite{Batson&Royer:19ICML}. 
Noise2Inverse \cite{Hendriksen&etal:20TCI} and
Neighbor2Neighbor \cite{Huang&eta:21CVPR} that emulate pairs of two independent noisy images by partitioning single measurements (e.g., CT ray measurements with independent noise and corrupted images with pixel-wise independent noise) can be viewed as Noise2Self.
Thus, SSRL loss \R{sys:SSRL} with the $\cJ$-complement of $f$ and $g = \cI$ specializes to the aforementioned Noise2Noise \cite{Lehtinen&etal:18ICML}-motivated self-supervised denoising methods.
(Noise2Noise also can be viewed by SSRL \R{sys:SSRL} with the setup above,
by constructing $x$ with stacking two independent noisy images, where an image is corrupted by two independent noise realizations.)

\subsubsection{SSRL without $\cJ$-complement}

The proposed SSRL loss \R{sys:SSRL:b} becomes the Noise2Same loss \cite[Thm.~2]{Xie&Wang&Ji:20NIPS},
by replacing $g(x_J)$ with $x$ and adding randomness to $K = J$ in the second term.
In practice,
one can tune $\sigma$ in \R{sys:SSRL:b} without knowing its exact value, similar to Noise2Same.
One might use the aforementioned conjectured behavior of \R{sys:SSRL:b} with expected performance of $g$.

\subsection{Relaxing noise assumptions of existing self-supervised denoising methods}
\label{sec:relax}

This section explains how proposed SSRL \R{sys:SSRL} can relax noise assumptions of existing self-supervised denoising methods. 
Noise assumptions of existing self-supervised denoising methods include additive white Gaussian noise (AWGN with known variance) \cite{Soltanayev&Chun:18NIPS}, 
pixel-wise independent noise \cite{Krull&etal:19CVPR, Xie&Wang&Ji:20NIPS, Huang&eta:21CVPR}, zero-mean noise \cite{Lehtinen&etal:18ICML, Quan&etal:20CVPR} or more generally $\bbE [x | y] = y$
 \cite{Batson&Royer:19ICML, Xie&Wang&Ji:20NIPS},
 and \dquotes{weak} noise \cite{Xu&etal:20TIP}.
Assumption~2 of proposed SSRL relaxes the AWGN and pixel-wise independent noise assumptions,
and is identical to the first assumption of Noise2Self \cite{Batson&Royer:19ICML}.
Assumption~3 of proposed SSRL can substantially relax the second assumption of Noise2Self, $\bbE [x|y] = y$, that is also (implicitly) used in Noise2Same \cite{Xie&Wang&Ji:20NIPS} and Noise2Noise \cite{Lehtinen&etal:18ICML}, by using a desinable pseudo-predictor $g$.
For example, 
$x$ is corrupted by additive non-zero-mean noise $e$ that is independent of $y$, i.e., $x = y + e$, then one can design $g$ as follows: $g(x) = x - \bbE[e]$, where $\bbE[e]$ can be estimated from calibration of imaging systems. 
Finally, SSRL is free from the \dquotes{weak} noise assumption in Noisy-As-Clean \cite{Xu&etal:20TIP}.
The next section will explain how one can design $g$ using domain knowledge and select $g$ if domain knowledge is unavailable in image denoising.

\section{Examples of how to design $g$ using domain knowledge, and empirical-loss approach for selecting $g$ in denoising in imaging}
\label{sec:know}

Understanding noise statistics or properties is the first step towards accurate image recovery in computational imaging.
First, this section describes how to use domain knowledge for designing a pseudo-predictor in two imaging applications with practical noise models:
\textit{1) low-dose CT denoising, and
\textit{2)} camera image denoising in mixed Poisson--Gaussian--Bernoulli noise.}
Both applications have complicated noise models or strong noise, where Assumptions~2--3 in Section~\ref{sec:prelim} do \textit{not} completely hold.
The noise in the first application is approximately zero-mean but it is likely non-independent and identically distributed (i.i.d.).
Noisy images in the second application are corrupted by i.i.d.~noise with non-zero mean.

In designing $g$ for each application, 
our particular aim is to design better $g$ than $\cI$ (the setup from existing self-supervised denoising methods),
following the suggested directions in Section~\ref{sec:SSRL:ind}.
To recap, well-designed $g$ -- in the sense of following the design directions in Section~\ref{sec:SSRL:ind}--\ref{sec:SSRL} -- will reduce the expected prediction error in Theorem~\ref{thm:soln}, better satisfy Assumption~3, or lead SSRL loss \R{sys:SSRL:b} close to the supervision counterpart, depending on different SSRL setups in Sections~\ref{sec:SSRL:ind}--\ref{sec:SSRL}.
We expect that such $g$ will lead to better performance of learned $f$ via SSRL, particularly compared to $\cI$. 
In designing $g$, we will investigate Assumptions 1--3.
Finally, we propose a $g$-selection approach for image denoising that calculates some empirical measure using only input training data.

\subsection{Designing $g$ for image denoising in low-dose CT}
\label{sec:know:ct}

\begin{figure}[t!]
 	\centering
 	\small\addtolength{\tabcolsep}{-9.5pt}
 	\begin{tabular}{cc}

 	    \raisebox{-.5\height}{
 			\begin{tikzpicture}
 			\begin{scope}[spy using outlines={rectangle,yellow,magnification=1.55,size=15mm, connect spies}]
 			\node {\includegraphics[width=34mm]{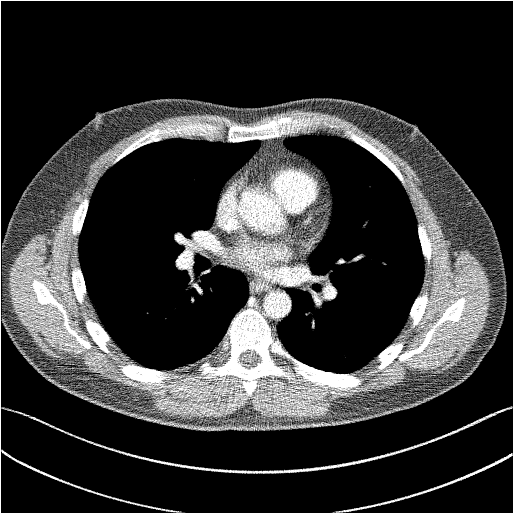}};
			\node [white] at (0.0,1.4) {\small $\text{RMSE} = 48.3$ HU};
 			\end{scope}
 			\end{tikzpicture}}
 			&
 			
 		\raisebox{-.5\height}{
 			\begin{tikzpicture}
 			\begin{scope}[spy using outlines={rectangle,yellow,magnification=1.55,size=15mm, connect spies}]
 			\node {\includegraphics[width=34mm]{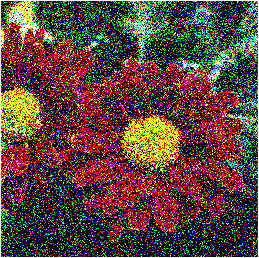}};
			\node [white] at (0.0,1.4) {\small $\text{PSNR} = 9.8$ dB};
 			\end{scope}
 			\end{tikzpicture}}
 			
    \vspace{-0.25pc}
 	\end{tabular}
\caption{An input noisy image in low-dose CT {\bfseries (left)} and camera image denoising in mixed noise {\bfseries (right)}. Root mean square error (RMSE) and PSNR values were averaged across all test samples. The noisy images were simulated using the practical noise models in Section~\ref{sec:know}.} %
    \label{fig:input}
    \vspace{-0.5pc}
\end{figure}

In X-ray CT (with a monoenergetic source), the pre-log measurement data is usually modeled by the Poisson model, i.e., $ \mathsf{Poisson}\{\rho_0 \exp(-[Ay]_l)\}$, $l = 1,\ldots,L$, where $\rho_0$ is the number of incident photons per ray, $A \in \bbR^{L \times N}$ is a CT projection system matrix, and $L$ is the number of measured rays. Using the quadratic approximation to the log-likelihood of a Poisson model, the post-log measurement $z \in \bbR^{L}$ given $y$ can be approximated as the following Gaussian model \cite{Sauer&Bouman:98TSP, Fessler:00BookCh}: $z | y \sim \cN( Ay, C )$, where $C \in \bbR^{L \times L}$ is a \emph{diagonal} covariance matrix and its diagonal elements become more nonuniform in lower-dose CT. 
This model suggests that post-log measurement may be modeled by $z = Ay + \varepsilon$, where $ \varepsilon \sim \cN(0, C)$.
The filtered back-projection (FBP) method \cite[\S3]{kak:book} performs computationally efficient CT reconstruction and has been widely used in commercial CT scanners \cite{Pan&Sidky&Vannier:09IP}. 
In low-dose CT, however, reconstructed image $x = Fz$ suffers from strong noise and streak artifacts, where $F \in \bbR^{N \times L}$ denotes a linear FBP operator, motivating research on learning denoising NNs.
Fig.~\ref{fig:input}(left) shows a noisy FBP image in low-dose CT.
Using the statistical results above,
we model that a reconstructed image by $F$ is corrupted by an arbitrary additive noise $e$: 
\be{
\label{eq:noise:ct}
x  = y + e, \quad e = (F A - I) y + F \varepsilon.
}
Low-dose CT uses all projection rays similar to standard-dose CT (but with substantially reduced dose)
where $F$ approximately inverts $A$,
i.e., $FA \approx I$,
so we conclude that under \R{eq:noise:ct}, $\bbE[e] \approx 0$ and $\bbE[x|y] \approx y$.
(See empirical results in Section~\ref{sec:add:empirical} that support $\bbE[e] \approx 0$.)

The above domain knowledge in low-dose CT indicates that Assumption~3 might \dquotes{approximately} hold with $g = \cI$, so 
handcrafting $g$ to have zero-mean $e$ to satisfy Assumption~3 may give a marginal impact to the performance of SSRL.
Assumption~2 unlikely holds because in FBP images, neighboring noise components are likely to be correlated, i.e., $\mathrm{Var} (e) \approx F C F^\top$ using $FA \approx I$ and noise model \R{eq:noise:ct}.
Considering these, we choose the design direction motivated by Theorem~\ref{thm:soln} (see Section~\ref{sec:SSRL:ind}).
We set $g$ as a pre-trained denoiser by the existing self-supervised denoising methods \cite{Batson&Royer:19ICML, Hendriksen&etal:20TCI, Xie&Wang&Ji:20NIPS}. Because such pre-trained $g$ will have some denoising capability,
we expect given $x_{J^c}$ that its prediction $g(x_J)$ becomes closer to $y$ than that using $g = \cI$, reducing the expected prediction error in Theorem~\ref{thm:soln}.
Our empirical results in Section~\ref{sec:add:empirical} support this.

Assumption~1 is satisfied as we use the conventional denoisiong NN, DnCNN \cite{Zhang&etal:17TIP} and (modified) U-Net \cite{Ronneberger&etal:15MICCAI}, that are a continuous function.

\subsection{Designing $g$ for image denoising in camera imaging}
\label{sec:know:nat}

The major noise sources in camera imaging (using charge coupled device) include
object-dependent photoelectrons in image sensors,
readout in camera electronics, and analog-to-digital converter and transmission errors that can be modeled by 
Poisson noise,
AWGN, and Bernoulli (i.e., salt-and-pepper) noise models \cite{Snyder:93JOSAA}, \cite[p.~90]{Bovik:10book}.
We use the following practical mixed Poisson--Gaussian--Bernoulli
noise model \cite{Snyder:93JOSAA, Batson&Royer:19ICML}:
\be{
	\label{eq:noise:nat}	
	\begin{split}
		x_n = \Pi_{[0,255]} (\mathsf{Bernoulli}_p(\mathsf{Poisson}(\lambda y_n)/\lambda + \epsilon_n) ), \\
		\quad \epsilon \sim \cN(0,\sigma^2_\epsilon I),
		\quad n = 1,\ldots,N,
	\end{split}
}
where $\Pi_{[0,255]}$ performs $8$-bit quantization and clips pixel values outside of $[0,255]$, 
$\mathsf{Bernoulli}$ substitutes a pixel value with either $0$ or $255$ with probability $p$ ($0$ and $255$ are coined with equal probability),
$\mathsf{Poisson}$ generates pixel intensity-dependent Poisson noise with gain parameter $\lambda$,
and $\epsilon$ is AWGN.
Fig.~\ref{fig:input}(right) shows a noisy image corrupted by the mixed noise model \R{eq:noise:nat}. 
If an image $y$ is corrupted only by the mixed Poisson--Gaussian noise,
$\bbE[x|y] = y$ in Assumption~3 with $g = \cI$ can be satisfied ($\bbE[ \mathsf{Poisson}(\lambda y_n)/\lambda + \epsilon | y] = y_n$, $\forall n$).
However, if Bernoulli noise is additionally considered as given in \R{eq:noise:nat}, 
the assumption $\bbE[x | y] = y$ will not hold
($\bbE[\mathsf{Bernoulli}_p(y_n) | y] = (1-p) y_n + 127.5 p$, $\forall n$).
The quantization-clipping operator $\Pi_{[0,255]}$ also makes it hard to satisfy $\bbE[x | y] = y$.

In camera image denoising with noise model~\R{eq:noise:nat}, 
Assumption~2 holds because the noise in \R{eq:noise:nat} is i.i.d.~and independent of $y$.
Considering this, we choose the design direction inspired by Proposition~\ref{thm:soln-all} (see Section~\ref{sec:SSRL:ind}). 
We handcraft $g$ with a simple operator, interpreting that aforementioned Bernoulli noise and clipping artifact as salt-and-pepper noise. 
Median filtering is a computational efficient method that is effective in reducing salt-and-pepper and impulse noises~\cite[\S3.2]{Bovik:10book}. 
We design $g$ by using weighted median filtering~\cite{brownrigg1984weighted:84ACM} to a pixel with intensity either $0$ or $255$ at each color channel.
We expect given $y$ that this $g$ design leads $g(x_J)$ closer to $y$ than $g = \cI$, by suppressing the salt-and-pepper noise effects cased by $\Pi_{[0,255]}$ and $\mathsf{Bernoulli}_p$.
Our empirical results in Section~\ref{sec:add:empirical} support this by showing that this particular $g$ \dquotes{approximately} satisfy Assumption~3 (under some mild assumptions).

In Assumption~1,
we conjecture that the above $g$ design is measurable (median operator is measurable under some conditions \cite{Rustad:04AAM}).

\subsection{Empirical-loss approach for selecting $g$ if domain knowledge unavailable}
\label{sec:add:guidelines}

If accurate domain knowledge of a specific application is unavailable, it would be challenging to explicitly design $g$.
In such cases in denoising, our general suggestion is to measure an existing self-supervised denoising loss for $g$, an upper bound of $\bbE \| g(x_J) - y \|_2^2$ or its variant, using only input training data.
The lower quantity implies that $g$ is better and implicitly encapsulates better domain knowledge. 
In low-dose CT denoising with the intrinsically noisy dataset \cite{Moen&etal:21MP},
empirical measures of the Noise2Self loss \cite{Batson&Royer:19ICML} -- $\bbE \| g(x_J)_{J^c} - x_{J^c} \|_2^2$ -- with setting $g$ as $\cI$ and pre-trained DnCNN by Noise2Self are $22044.5$ and $17062.0$ (in HU$^2$ where HU stands for modified Hounsfield unit), respectively.
In camera image denoising with the intrinsically noisy dataset \cite{Abdelhamed&etal:18CVPR},
empirical measures of the Neighbor2Neighbor loss \cite{Huang&eta:21CVPR} -- $\bbE \| g(x_{J}) - x_{J^c} \|_2^2$ -- with setting $g$ as $\cI$ and median filtering are $0.0052$ and $0.0048$, respectively.
We expect better SSRL performance with the selected $g$ designs over $\cI$.

\section{Experimental results and discussion}
\label{sec:result}

We evaluated the proposed SSRL framework in two practical imaging applications in Section~\ref{sec:know} with both simulated and intrinsically noisy datasets.
For these applications, we mainly focuses on comparisons with self-supervised denoising methods using single noisy input samples, particularly when statistical noise parameters are unavailable.
We compared the performances of the following self-supervised denoising methods using training datasets: 
Noise2Self \cite{Batson&Royer:19ICML}, Noise2Noise-motivated methods that emulate pairs of two independent noisy images --
Noise2Inverse \cite{Hendriksen&etal:20TCI} or Neighbor2Neighbor \cite{Huang&eta:21CVPR} -- 
Noise2Same \cite{Xie&Wang&Ji:20NIPS},
and corresponding SSRL to each aforementioned method.
In addition, we evaluated state-of-the-art single-image self-supervised denoising methods, Self2Self~\cite{Quan&etal:20CVPR} and Noisy-As-Clean~\cite{Xu&etal:20TIP}, and Noise2True \R{sys:sup} (as baseline).
Note that all the aforementioned methods are blind or in the blind setup, so one does \textit{not} estimate noise parameters.
For $f$ or $g$ in all the self-supervised denoising methods using training datasets, we used the conventional denoising NN architecture, DnCNN \cite{Zhang&etal:17TIP}, or modified U-Net used in Noise2Self. 
In testing the single-image self-supervised denoising methods, we used the default NN architectures in their implementation.
Sections~\ref{sec:setup} and \ref{sec:comp} include experimental setups, and results and discussion for/with synthetic datasets.
Sections~\ref{sec:setup:real} and \ref{sec:comp:real} include experimental setups, and results and discussion for/with intrinsically datasets.

\subsection{Experimental setup with simulated imaging datasets}
\label{sec:setup}

\subsubsection{Low-dose CT denoising}
\label{sec:setup:sim:ct}

We evaluated proposed SSRL with The 2016 Low Dose CT Grand Challenge data \cite{McC-Mayo}.
We selected $200$ regular-dose chest images of size $N = 512 \times 512$ and the $3$ mm slice thickness from four patients.
For training, we used $170$ ($85$\%) chest images from three patients; for tests, we used $30$ ($15$\%) chest images from the other patient.
We simulated low-dose sinograms using the Poisson model with the selected regular-dose chest datasets.
In particular, we simulated sinograms of size $L = 736 \times 1152$ (\quotes{detectors} $\times$ \quotes{projection views}), with fan-beam geometry corresponding to a no-scatter monoenergetic source with $\rho_0 = 5\times 10^4$.
We used FBP \cite[\S3]{kak:book} to reconstruct images with resolution $0.69~\text{mm} \times 0.69~\text{mm}$.
We evaluated the denoising quality by the most conventional error metric in CT application, RMSE in HU.
In Noise2Inverse setup, we emulated two independent noisy images by partitioning single sinograms with odd and even views and applying FBP to partitioned sinograms.

\subsubsection{Camera image denoising}
\label{sec:setup:sim:nat}

We evaluated SSRL with RGB camera image datasets, ImageNet ILSVRC 2012 Val \cite{russakovsky:15IJCV} and BSD 300 \cite{MartinFTM:01PCCV}.
For training, we used the ImageNet ILSVRC 2012 Val dataset with $20,\!000$ images; in testing trained NNs, we used the BSD 300 consisting of $300$ images. We simulated noisy images with the following imaging parameters in \R{eq:noise:nat}:
$\lambda = 30$, 
$\sigma_\epsilon = 60$, 
and $p = 0.2$ \cite{Batson&Royer:19ICML,Xie&Wang&Ji:20NIPS}.
We evaluated the denoising quality by the most conventional error metric in camera image denoising, PSNR and structural similarity index measure (SSIM).
In Neighbor2Neighbor setup, we emulated two independent noisy images from single noisy images by neighbor sub-sampling with $2 \!\times\! 2$-window \cite{Huang&eta:21CVPR}.
    
 \subsection{Experimental setup with intrinsically noisy imaging datasets (i.e.,~real-world imaging datasets)}
\label{sec:setup:real}

We also evaluated the proposed SSRL framework with intrinsically noisy low-dose CT and camera imaging datasets, where we do not have their complete noise properties/statistics.
We chose the publicly available Low Dose CT Image and Projection Data \cite{Moen&etal:21MP} and SIDD sRGB Data \cite{Abdelhamed&etal:18CVPR},
where both the datasets include standard-dose FBP or high-quality images so that one can run Noise2True experiments and obtain quantitative results.
We used the DnCNN architecture for all experiments. For each experiment, we used the same implementation setup (such as hyperparameters and masking scheme) as that of the corresponding simulated data experiment.

\subsubsection{Low-dose CT denoising}
\label{sec:setup:ct}

\begin{figure}[t!]
 	\centering
 	\small\addtolength{\tabcolsep}{-9.5pt}
 	\begin{tabular}{cc}
 	
 		\raisebox{-.5\height}{
 			\begin{tikzpicture}
 			\begin{scope}[spy using outlines={rectangle,yellow,magnification=1.55,size=15mm, connect spies}]
 			\node {\includegraphics[width=34mm]{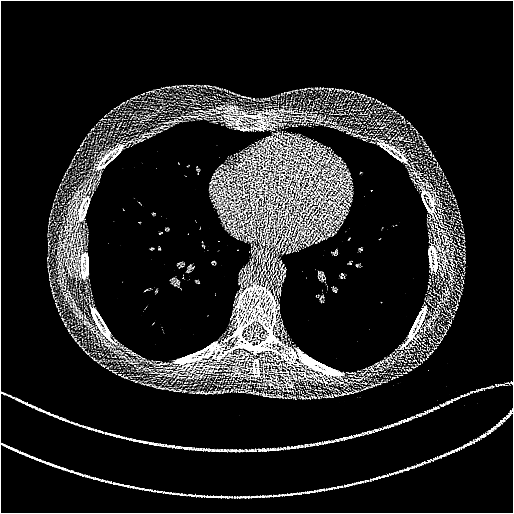}};
			\node [white] at (0.0,1.4) {\small $\text{RMSE} = 147.9$ HU};
 			\end{scope}
 			\end{tikzpicture}} &  		 
 		\raisebox{-.5\height}{
 			\begin{tikzpicture}

 			\begin{scope}[spy using outlines={rectangle,yellow,magnification=1.55,size=15mm, connect spies}]
 			\node {\includegraphics[width=34mm]{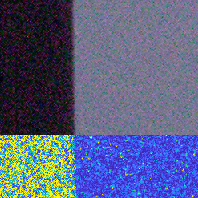}};
			\node [white] at (0.0,1.4) {\small $\text{PSNR} = 23.7$ dB};
 			\end{scope}
 			\end{tikzpicture}}
 			
    \vspace{-0.25pc}
 	\end{tabular}
	\caption{An input intrinsically noisy image in low-dose CT {\bfseries (left)} and camera imaging {\bfseries (right)}. RMSE \& PSNR values were averaged across all test samples.} 
	\label{fig:input:real}
	\vspace{-0.5pc}
\end{figure}

We evaluated SSRL with the Low Dose CT Image and Projection Data~\cite{Moen&etal:21MP}.
We followed the training-test data construction setup in Section~\ref{sec:setup} that was used in simulated data experiments; we remark that chest CT scans in the Low Dose CT Image and Projection Data use the $1$ mm slice thickness. 
Fig.~\ref{fig:input:real}(left) shows a intrinsically noisy FBP image.
We could not run Noise2Inverse experiments because the data \cite{Moen&etal:21MP} does not provide two independent half-view FBP images.
We thus ran experiments only with the following comparison setups: Noise2Self and Noise2Same.

\subsubsection{Camera image denoising}
\label{sec:setup:nat}

We evaluated SSRL with the SIDD sRGB Data \cite{Abdelhamed&etal:18CVPR}. 
For training, we used the SIDD sRGB training dataset with $400$ full images; in testing trained NNs, we used the SIDD sRGB validation dataset consisting of $2,\!140$ cropped images of size $240 \times 240$. 
Fig.~\ref{fig:input:real}(right) shows a intrinsically noisy camera image. 
We set $g$ as weighted median filtering, similar to simulated camera image denoising experiments in Section~\ref{sec:setup:sim:nat}.
We chose the representative comparison setup, Neighbor2Neighbor, from the experiments in Section~\ref{sec:setup:sim:nat}.


\begin{figure*}[!t]
 	\centering
 	\small\addtolength{\tabcolsep}{-9.5pt}
 	\renewcommand{\arraystretch}{0.8}
 	
 	\begin{tabular}{ccccc}
 		
 		 \small{Reference} & 
 		 \boxitfig{6.83cm}{0.65cm}
 		 \parbox{3.75cm}{\centering
 		 \small ~Noise2Self}  & 
 		 \specialcell[c]{\small {\bfseries Proposed SSRL} in \\ \small Noise2Self setup}&	
 		\boxitfig{6.83cm}{0.65cm}
 		\parbox{3.75cm}{\centering
 		\small ~Noise2Inverse} &
 		\specialcell[c]{\small {\bfseries Proposed SSRL} in \\ \small Noise2Inverse setup}
 		 
 		 \\ 
 		
 		\raisebox{-.5\height}{
 			\begin{tikzpicture}
 			\begin{scope}[spy using outlines={rectangle,yellow,magnification=1.7,size=8mm, connect spies}]
 			\node {\includegraphics[width=34mm]{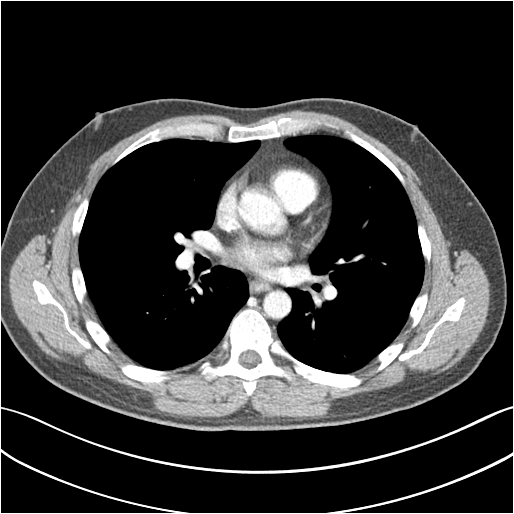}};
 			\spy on (-0.4, -0.9) in node [left] at (-0.9,-1.3);
 			\spy on (0.25, 0.25) in node [left] at (1.7,-1.3);
 			\end{scope}
 		    \draw[red, line width=0.2mm] (-1.1,-1.38) circle (1mm);
		    \draw[red, line width=0.2mm] (1.51,-1.40) circle (0.9mm);
 			\end{tikzpicture}} &

 		 \raisebox{-.5\height}{
 			\begin{tikzpicture}
 				\begin{scope}[spy using outlines={rectangle,yellow,magnification=1.7,size=8mm, connect spies}]
 					\node {\includegraphics[width=34mm]{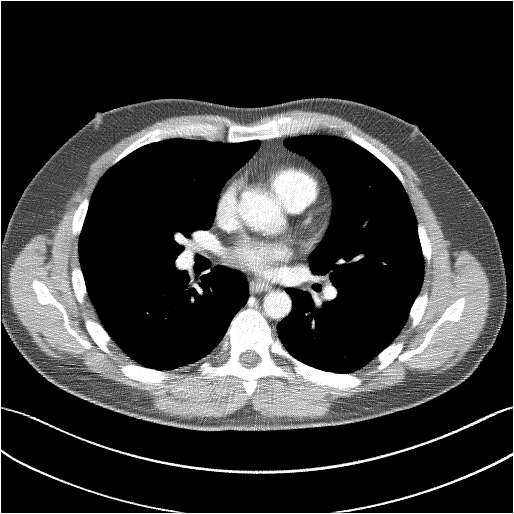}};
 					\spy on (-0.4, -0.9) in node [left] at (-0.9,-1.3);
 					\spy on (0.25, 0.25) in node [left] at (1.7,-1.3);
 					\node [white] at (0.0,1.4) {\small $\text{RMSE} = 47.8$ HU};
 				\end{scope}
 			    \draw[red, line width=0.2mm] (-1.1,-1.38) circle (1mm);
		        \draw[red, line width=0.2mm] (1.51,-1.40) circle (0.9mm);
 		\end{tikzpicture}} &
 		
 		\raisebox{-.5\height}{
 			\begin{tikzpicture}
 				\begin{scope}[spy using outlines={rectangle,yellow,magnification=1.7,size=8mm, connect spies}]
 					\node {\includegraphics[width=34mm]{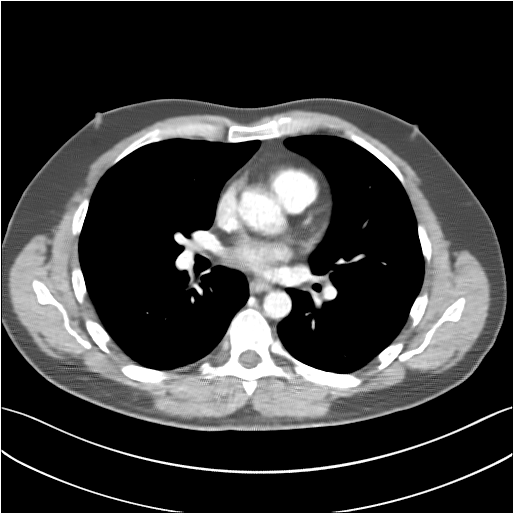}};
 					\spy on (-0.4, -0.9) in node [left] at (-0.9,-1.3);
 					\spy on (0.25, 0.25) in node [left] at (1.7,-1.3);
 					\node [white] at (0.0,1.4)  {\small  \color{yellow}{$\text{RMSE} = 25.0$ HU}};
 				\end{scope}
 				\draw[red, line width=0.2mm] (-1.1,-1.38) circle (1mm);
		        \draw[red, line width=0.2mm] (1.51,-1.40) circle (0.9mm);
 		\end{tikzpicture}}&
 			
 		  \raisebox{-.5\height}{
 		  	\begin{tikzpicture}
 		  		\begin{scope}[spy using outlines={rectangle,yellow,magnification=1.7,size=8mm, connect spies}]
 		  			\node {\includegraphics[width=34mm]{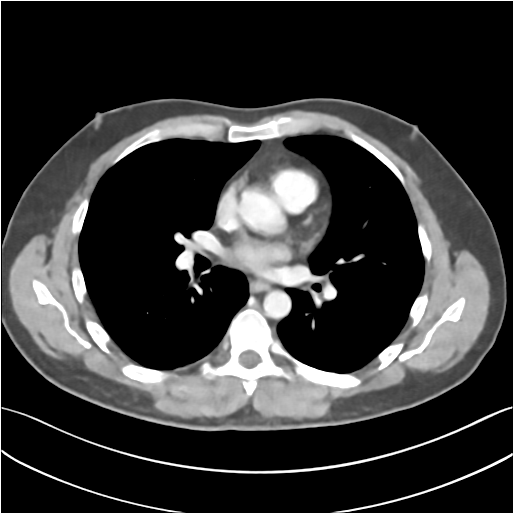}};
 		  			\spy on (-0.4, -0.9) in node [left] at (-0.9,-1.3);
 		  			\spy on (0.25, 0.25) in node [left] at (1.7,-1.3);
 		  			\node [white] at (0.0,1.4) {\small $\text{RMSE} = 22.9$ HU};
			   \end{scope}
		 	   \draw[red, line width=0.2mm] (-1.1,-1.38) circle (1mm);
		       \draw[red, line width=0.2mm] (1.51,-1.40) circle (0.9mm);		   
 		  \end{tikzpicture}} &
 		  
 		  \raisebox{-.5\height}{
 		  	\begin{tikzpicture}
 		  		\begin{scope}[spy using outlines={rectangle,yellow,magnification=1.7,size=8mm, connect spies}]
 		  			\node {\includegraphics[width=34mm]{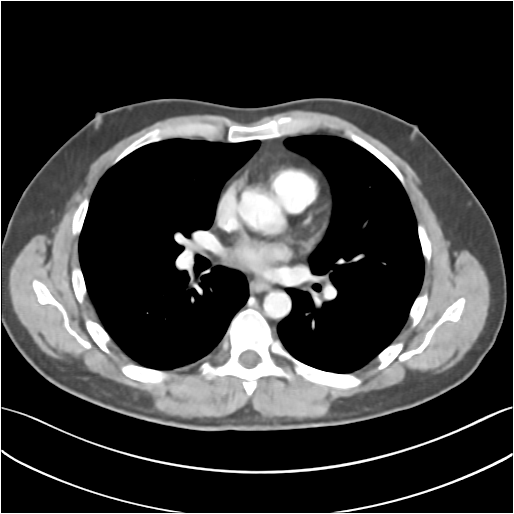}};
 		  			\spy on (-0.4, -0.9) in node [left] at (-0.9,-1.3);
 		  			\spy on (0.25, 0.25) in node [left] at (1.7,-1.3);
 		  			\node [white] at (0.0,1.4)  {\small  \color{yellow}{$\text{RMSE} = 21.9$ HU}};
 		  		\end{scope}
 		        \draw[red, line width=0.2mm] (-1.1,-1.38) circle (1mm);
		        \draw[red, line width=0.2mm] (1.51,-1.40) circle (0.9mm);		
 		  \end{tikzpicture}}

 	 	\vspace{0.2pc}
 		\\
         		
 		\small {Noise2True} &	
 		\boxitfig{6.83cm}{0.65cm}
 		\parbox{3.75cm}{\centering
 		\small ~Noise2Same} &
 		\specialcell[c]{\small {\bfseries Proposed SSRL} in \\ \small Noise2Same setup}
 		  & 
 		\small{Noisy-As-Clean} &
 		\small{Self2Self}

 		\\
 		\raisebox{-.5\height}{
 			\begin{tikzpicture}
 			\begin{scope}[spy using outlines={rectangle,yellow,magnification=1.7,size=8mm, connect spies}]
 			\node {\includegraphics[width=34mm]{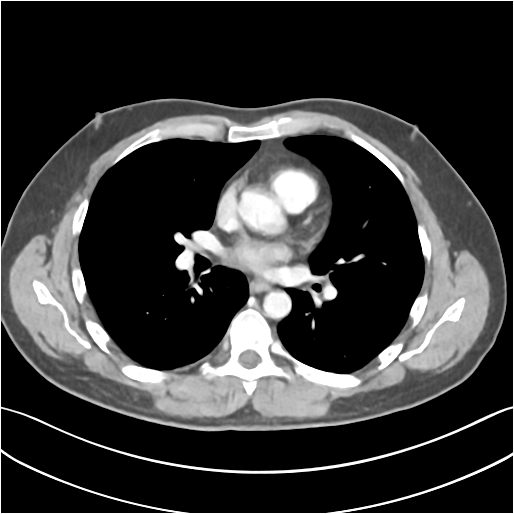}};
 			\spy on (-0.4, -0.9) in node [left] at (-0.9,-1.3);
 			\spy on (0.25, 0.25) in node [left] at (1.7,-1.3);
 			\node [white] at (0.0,1.4) {\small $\text{RMSE} = 16.3$ HU};
 			\end{scope}
 			\draw[red, line width=0.2mm] (-1.1,-1.38) circle (1mm);
		    \draw[red, line width=0.2mm] (1.51,-1.40) circle (0.9mm);
 			\end{tikzpicture}} &  
 		\raisebox{-.5\height}{
 			\begin{tikzpicture}
 			\begin{scope}[spy using outlines={rectangle,yellow,magnification=1.7,size=8mm, connect spies}]
 			\node {\includegraphics[width=34mm]{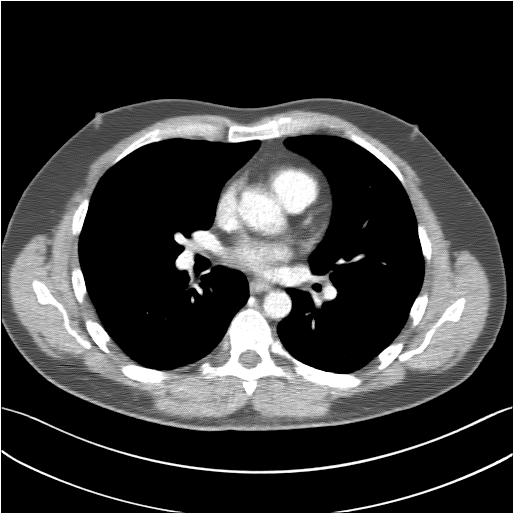}};
 			\spy on (-0.4, -0.9) in node [left] at (-0.9,-1.3);
 			\spy on (0.25, 0.25) in node [left] at (1.7,-1.3);
 			\node [white] at (0.0,1.4) {\small $\text{RMSE} = 28.4$ HU};
 			\end{scope}
 		    \draw[red, line width=0.2mm] (-1.1,-1.38) circle (1mm);
		    \draw[red, line width=0.2mm] (1.51,-1.40) circle (0.9mm); 			
 			\end{tikzpicture}} &
 			
 		\raisebox{-.5\height}{
 			\begin{tikzpicture}
 			\begin{scope}[spy using outlines={rectangle,yellow,magnification=1.7,size=8mm, connect spies}]
 			 \node {\includegraphics[width=34mm]{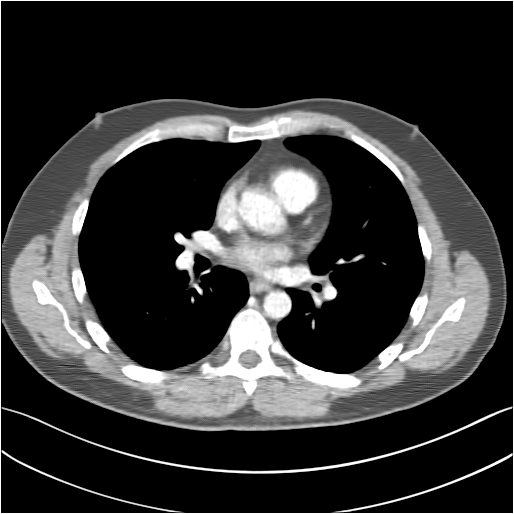}};
 			\spy on (-0.4, -0.9) in node [left] at (-0.9,-1.3);
 			\spy on (0.25, 0.25) in node [left] at (1.7,-1.3);
 			\node [white] at (0.0,1.4)  {\small  \color{yellow}{$\text{RMSE} = 26.0$ HU}};
 			\end{scope}
 			\draw[red, line width=0.2mm] (-1.1,-1.38) circle (1mm);
		    \draw[red, line width=0.2mm] (1.51,-1.40) circle (0.9mm);
 			\end{tikzpicture}} 	&
 			
		\raisebox{-.5\height}{
 			\begin{tikzpicture}
 			\begin{scope}[spy using outlines={rectangle,yellow,magnification=1.7,size=8mm, connect spies}]
 			\node {\includegraphics[width=34mm]{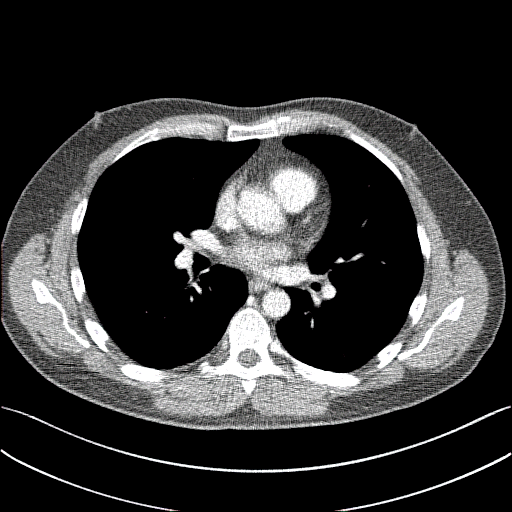}};
 			\spy on (-0.4, -0.9) in node [left] at (-0.9,-1.3);
 			\spy on (0.25, 0.25) in node [left] at (1.7,-1.3);
 			\node [white] at (0.0,1.4) {\small $\text{RMSE} = 47.8$ HU};
 			\end{scope}
 			\draw[red, line width=0.2mm] (-1.1,-1.38) circle (1mm);
		    \draw[red, line width=0.2mm] (1.51,-1.40) circle (0.9mm);
 			\end{tikzpicture}} 	&
		\raisebox{-.5\height}{
 			\begin{tikzpicture}
 			\begin{scope}[spy using outlines={rectangle,yellow,magnification=1.7,size=8mm, connect spies}]
 			\node {\includegraphics[width=34mm]{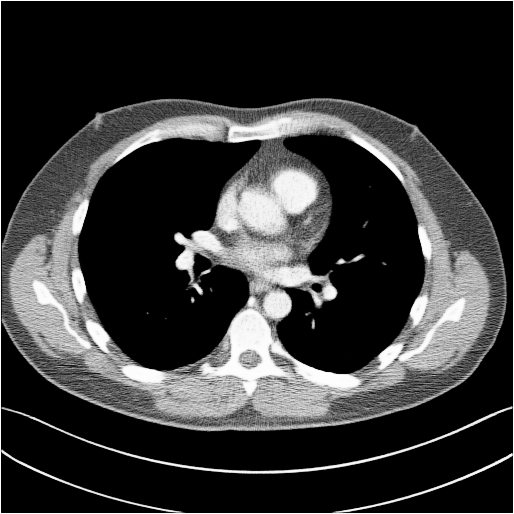}};
 			\spy on (-0.4, -0.9) in node [left] at (-0.9,-1.3);
 			\spy on (0.25, 0.25) in node [left] at (1.7,-1.3);
 			\node [white] at (0.0,1.4) {\small $\text{RMSE} = 31.6$ HU};
 			\end{scope}
 			\draw[red, line width=0.2mm] (-1.1,-1.38) circle (1mm);
		    \draw[red, line width=0.2mm] (1.51,-1.40) circle (0.9mm);
 			\end{tikzpicture}} 	
 		
 	 \vspace{-0.25pc}	
 	\end{tabular}
 	\caption{Comparisons of denoised images from different learning/fitting methods in low-dose CT (display window is $[800, 1200]$ HU).
    We used DnCNN for all self-supervised denoising in red boxes and Noise2True.
 	 RMSE values were averaged across all test samples.
  	\vspace{-0.5pc}
 	 }
 	\label{fig:ct}
 \end{figure*}
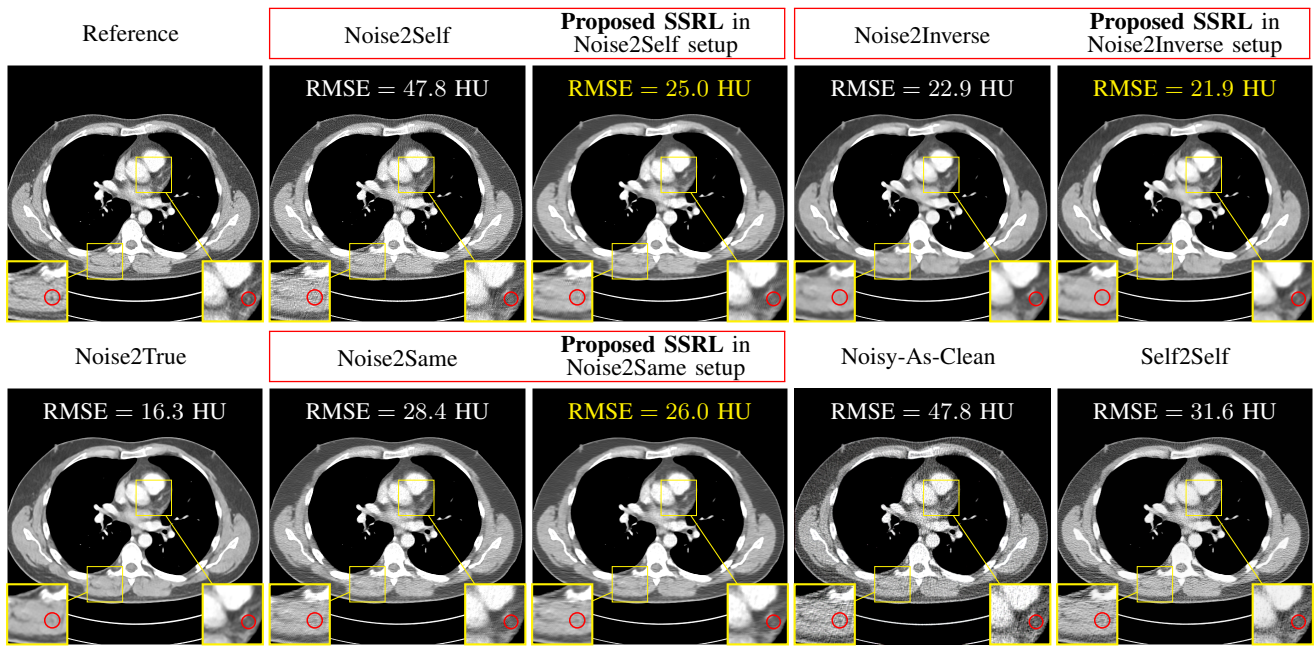
 
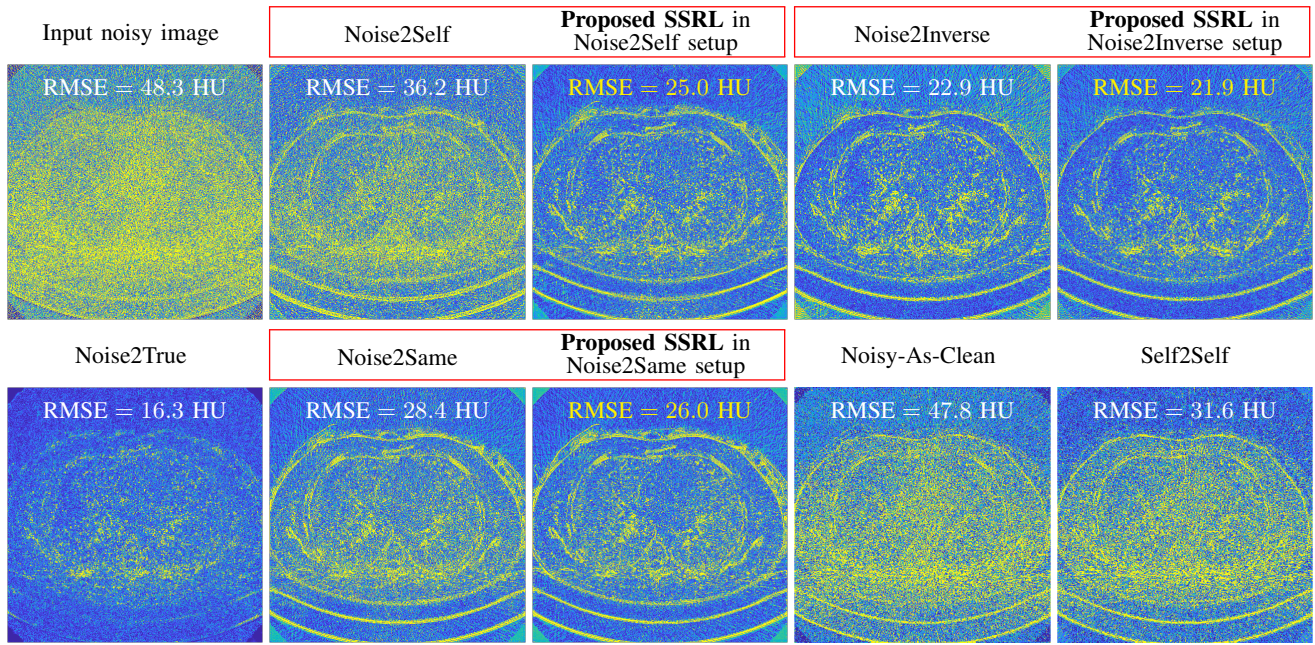
\begin{figure*}[!t]
 	\centering
 	\small\addtolength{\tabcolsep}{-9.5pt}
 	\renewcommand{\arraystretch}{0.8}
 	\begin{tabular}{ccccc}
 		
 		 \small{Input noisy image} & 
 		 \boxitfig{6.83cm}{0.65cm}
 		 \parbox{3.75cm}{\centering
 		 \small ~Noise2Self}  &  \specialcell[c]{\small {\bfseries Proposed SSRL} in \\ \small Noise2Self setup} &
   		  \boxitfig{6.83cm}{0.65cm}
 		  \parbox{3.75cm}{\centering
 		  	\small ~Noise2Inverse} &
 		  \specialcell[c]{\small {\bfseries Proposed SSRL} in \\ \small Noise2Inverse setup}
 		 \\ 
 		
 		\raisebox{-.5\height}{
 			\begin{tikzpicture}
 			\begin{scope}[spy using outlines={rectangle,yellow,magnification=1.55,size=15mm, connect spies}]
 			\node {\includegraphics[width=34mm]{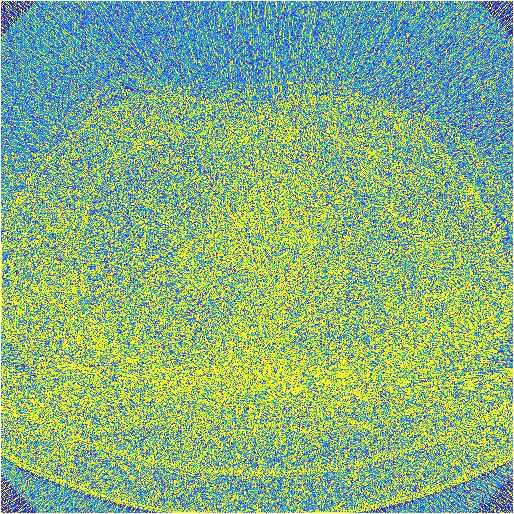}};
 			\node [white] at (0.0,1.4) {\small $\text{RMSE} = 48.3$ HU};
 			\end{scope}
 			\end{tikzpicture}} 
 			&
 		\raisebox{-.5\height}{
 			\begin{tikzpicture}
 				\begin{scope}[spy using outlines={rectangle,yellow,magnification=1.55,size=15mm, connect spies}]
 					\node {\includegraphics[width=34mm]{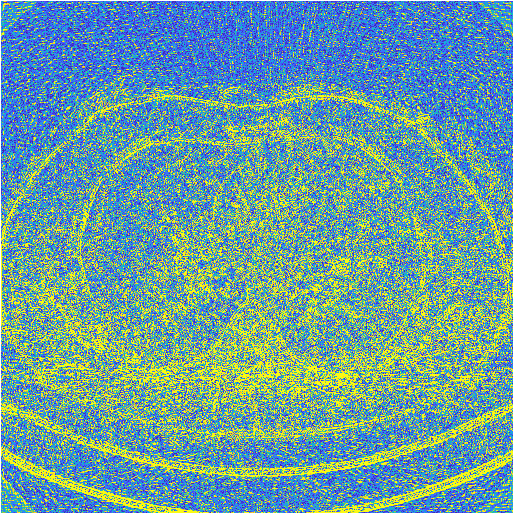}};
 					\node [white] at (0.0,1.4) {\small $\text{RMSE} = 36.2$ HU};
 				\end{scope}
 		\end{tikzpicture}} &
 		
 		\raisebox{-.5\height}{
 			\begin{tikzpicture}
 				\begin{scope}[spy using outlines={rectangle,yellow,magnification=1.55,size=15mm, connect spies}]
 					\node {\includegraphics[width=34mm]{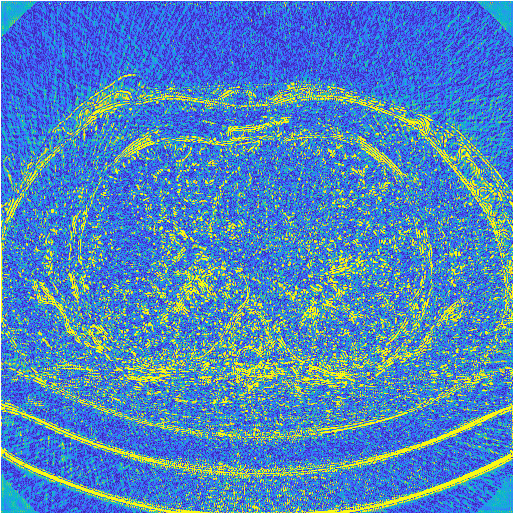}};
 					\node [white] at (0.0,1.4)  {\small  \color{yellow}{$\text{RMSE} = 25.0$ HU}};
 				\end{scope}
 		\end{tikzpicture}}  &  		 
 		
 		   \raisebox{-.5\height}{
 		  	\begin{tikzpicture}
 		  		\begin{scope}[spy using outlines={rectangle,yellow,magnification=1.55,size=15mm, connect spies}]
 		  			\node {\includegraphics[width=34mm]{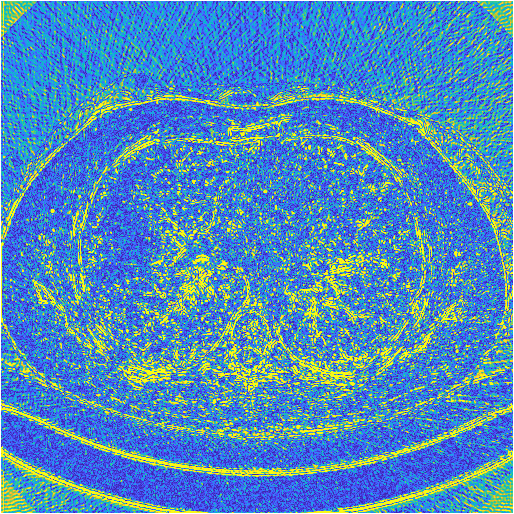}};
 		  			\node [white] at (0.0,1.4) {\small $\text{RMSE} = 22.9$ HU};
 		  		\end{scope}
 		  \end{tikzpicture}} &
 		  
 		  \raisebox{-.5\height}{
 		  	\begin{tikzpicture}
 		  		\begin{scope}[spy using outlines={rectangle,yellow,magnification=1.55,size=15mm, connect spies}]
 		  			\node {\includegraphics[width=34mm]{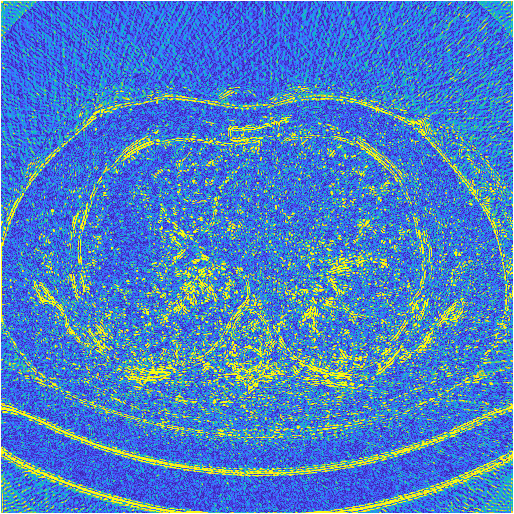}};
 		  			\node [white] at (0.0,1.4)  {\small  \color{yellow}{$\text{RMSE} = 21.9$ HU}};
 		  		\end{scope}
 		  \end{tikzpicture}}
 			\vspace{0.2pc}
 			\\
 			
          \small{Noise2True} &
 		  \boxitfig{6.83cm}{0.65cm}
 		  \parbox{3.75cm}{\centering
 		  	\small ~Noise2Same} & 
 		  \specialcell[c]{\small {\bfseries Proposed SSRL} in \\ \small Noise2Same setup}
 		  & 
          \small{Noisy-As-Clean} &
          \small{Self2Self}

 		  \\
 		\raisebox{-.5\height}{
 			\begin{tikzpicture}
 			\begin{scope}[spy using outlines={rectangle,yellow,magnification=1.55,size=15mm, connect spies}]
 			\node {\includegraphics[width=34mm]{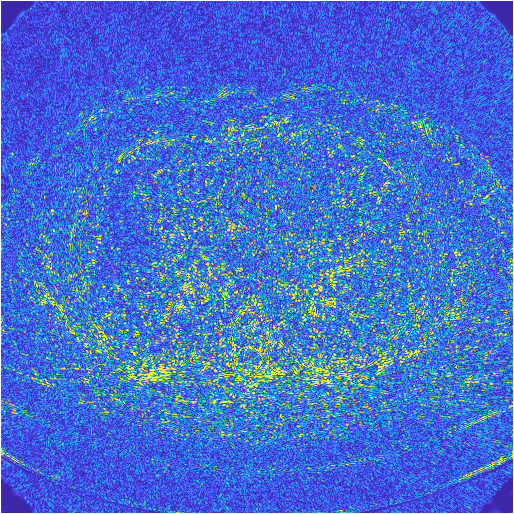}};
 			\node [white] at (0.0,1.4) {\small $\text{RMSE} = 16.3$ HU};
 			\end{scope}
 			\end{tikzpicture}} &  
 			
 		\raisebox{-.5\height}{
 			\begin{tikzpicture}
 			\begin{scope}[spy using outlines={rectangle,yellow,magnification=1.55,size=15mm, connect spies}]
 			\node {\includegraphics[width=34mm]{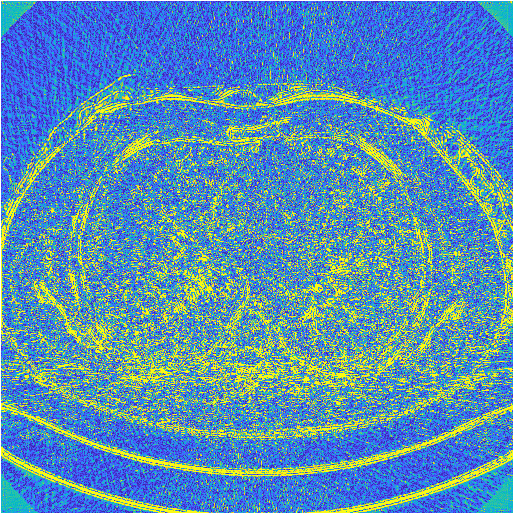}};
 			\node [white] at (0.0,1.4) {\small $\text{RMSE} = 28.4$ HU};
 			\end{scope}
 			\end{tikzpicture}} &
 			
 		\raisebox{-.5\height}{
 			\begin{tikzpicture}
 			\begin{scope}[spy using outlines={rectangle,yellow,magnification=1.55,size=15mm, connect spies}]
 			 \node {\includegraphics[width=34mm]{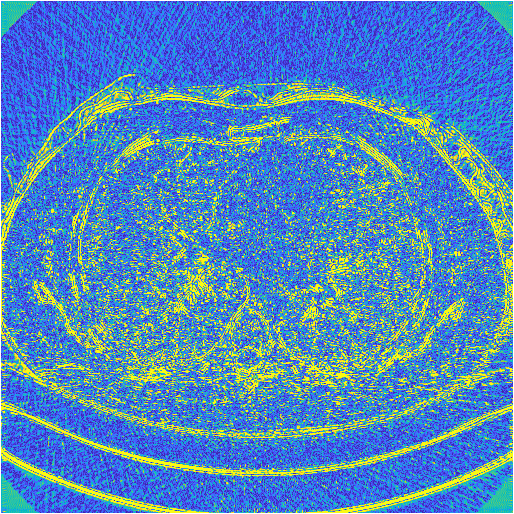}};
 			\node [white] at (0.0,1.4)  {\small  \color{yellow}{$\text{RMSE} = 26.0$ HU}};
 			\end{scope}
 			\end{tikzpicture}}  &
 		
 		\raisebox{-.5\height}{
 			\begin{tikzpicture}
 			\begin{scope}[spy using outlines={rectangle,yellow,magnification=1.55,size=15mm, connect spies}]
 			\node {\includegraphics[width=34mm]{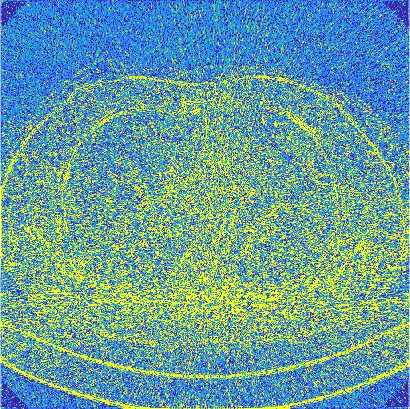}};
 			\node [white] at (0.0,1.4) {\small $\text{RMSE} = 47.8$ HU};
 			\end{scope}
 			\end{tikzpicture}} &
 			
 		\raisebox{-.5\height}{
 			\begin{tikzpicture}
 			\begin{scope}[spy using outlines={rectangle,yellow,magnification=1.55,size=15mm, connect spies}]
 			 \node {\includegraphics[width=34mm]{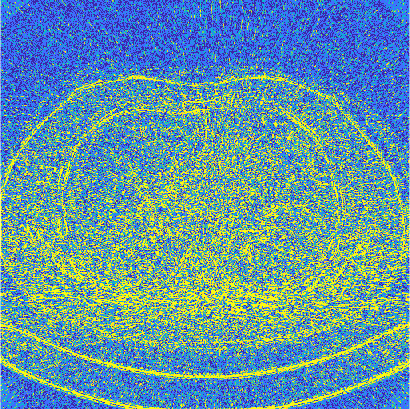}};
 			\node [white] at (0.0,1.4) {\small $\text{RMSE} = 31.6$ HU};
 			\end{scope}
 			\end{tikzpicture}} 

    \vspace{-0.25pc}
 	\end{tabular}
 	\caption{Error map comparisons of denoised images from different learning/fitting methods in low-dose CT (blue and yellow denote $0$ and $50$ absolute errors in HU, respectively). The results correspond to those in Fig.~\ref{fig:ct}.
 	}
 	\label{fig:ct:err}
    \vspace{-0.5pc}
 \end{figure*}

 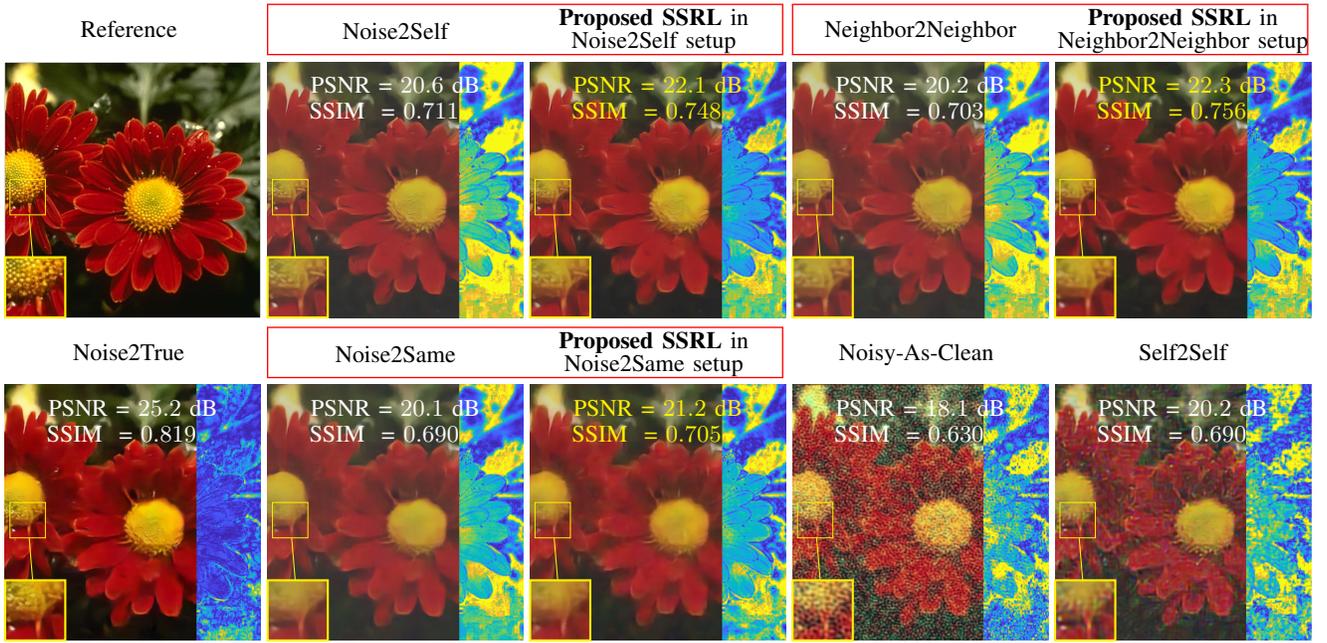
\begin{figure*}[!t]
 	\centering
 	\small\addtolength{\tabcolsep}{-9.5pt}
 	\renewcommand{\arraystretch}{0.8}
 	\begin{tabular}{cccccc}
 		 \small{Reference} &  
 		 \boxitfig{6.83cm}{0.65cm}
 		  \parbox{3.75cm}{\centering
 		  \small ~Noise2Self} &
 		 \specialcell[c]{\small {\bfseries Proposed SSRL} in \\ \small Noise2Self setup} &  
 		 \boxitfig{6.83cm}{0.65cm}
 		  \parbox{3.75cm}{\centering
 		  \small ~Neighbor2Neighbor} &
 		 \specialcell[c]{\small {\bfseries Proposed SSRL} in \\ \small Neighbor2Neighbor setup}
 		 \\ 
 		
 		\raisebox{-.5\height}{
 			\begin{tikzpicture}
 			\begin{scope}[spy using outlines={rectangle,yellow,magnification=1.7,size=8mm, connect spies}]
 			\node {\includegraphics[width=34mm,height=34mm]{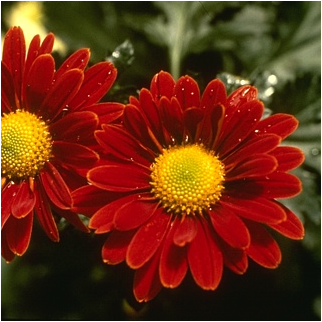}};
 			\spy on (-1.4, -0.1) in node  [left] at (-0.9,-1.3);
 			\end{scope}
 			\end{tikzpicture}} &
 			
 		\raisebox{-.5\height}{
 			\begin{tikzpicture}
 			\begin{scope}[spy using outlines={rectangle,yellow,magnification=1.7,size=8mm, connect spies}]
 			\node {\includegraphics[width=34mm,height=34mm]{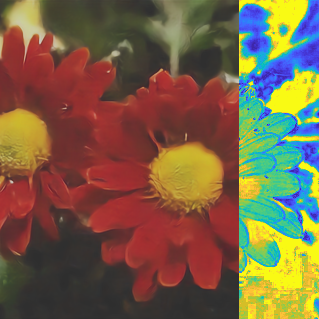}};
 			\spy on (-1.4, -0.1) in node  [left] at (-0.9,-1.3);
 			\node [white] at (0.0,1.4) {\small PSNR = $20.6$ dB};
 			\node [white] at (-0.15,1.05) {\small SSIM \ = $0.711$};
 			\end{scope}
 			\end{tikzpicture}} &

 		\raisebox{-.5\height}{
 			\begin{tikzpicture}
 			\begin{scope}[spy using outlines={rectangle,yellow,magnification=1.7,size=8mm, connect spies}]
 			\node {\includegraphics[width=34mm,height=34mm]{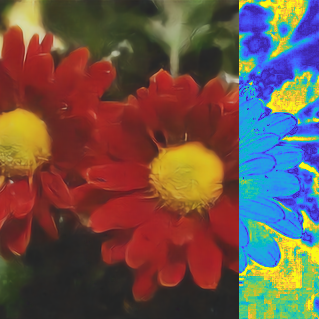}};
 			\spy on (-1.4, -0.1) in node  [left] at (-0.9,-1.3);
 			\node [white] at (0.0,1.4)  {\small  \color{yellow}{PSNR = $22.1$ dB}};
 			\node [white] at (-0.15,1.05)  {\small  \color{yellow}{SSIM \ = $0.748$}};
 			\end{scope}
 			\end{tikzpicture}} &
 		   \raisebox{-.5\height}{
 		  	\begin{tikzpicture}
 			\begin{scope}[spy using outlines={rectangle,yellow,magnification=1.7,size=8mm, connect spies}]
 		  	\node {\includegraphics[width=34mm,height=34mm]{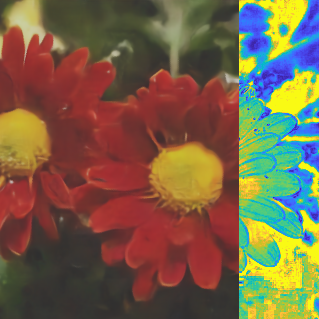}};
 			\spy on (-1.4, -0.1) in node  [left] at (-0.9,-1.3);
 			\node [white] at (0.0,1.4) {\small PSNR = $20.2$ dB};
 			\node [white] at (-0.15,1.05) {\small SSIM \ = $0.703$};
 		  	\end{scope}
 		  	\end{tikzpicture}} &
 		
 		\raisebox{-.5\height}{
 			\begin{tikzpicture}
 			\begin{scope}[spy using outlines={rectangle,yellow,magnification=1.7,size=8mm, connect spies}]
 		  	\node {\includegraphics[width=34mm,height=34mm]{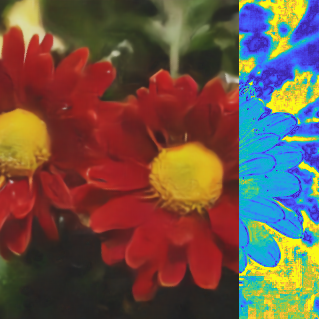}};
 			\spy on (-1.4, -0.1) in node  [left] at (-0.9,-1.3);
 			\node [white] at (0.0,1.4)  {\small  \color{yellow}{PSNR = $22.3$ dB }};
 			\node [white] at (-0.15,1.05)  {\small  \color{yellow}{SSIM \ = $0.756$}};
 			\end{scope}
 			\end{tikzpicture}}
 	 	\vspace{0.2pc}
 			\\

 		  \small {Noise2True} &
          \boxitfig{6.83cm}{0.65cm} 
 		  \parbox{3.75cm}{\centering
 		  \small ~Noise2Same}  & 
 		  \specialcell[c]{\small {\bfseries Proposed SSRL} in \\ \small Noise2Same setup}
 		  & 
 		  \small {Noisy-As-Clean} &
 		  \small {Self2Self}
 		  \\
 		\raisebox{-.5\height}{
 			\begin{tikzpicture}
 			\begin{scope}[spy using outlines={rectangle,yellow,magnification=1.7,size=8mm, connect spies}]
 			\node {\includegraphics[width=34mm,height=34mm]{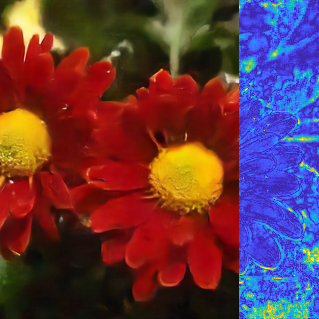}};
 			\spy on (-1.4, -0.1) in node  [left] at (-0.9,-1.3);
 			\node [white] at (0.0,1.4) {\small PSNR  = $25.2$ dB};
 			\node [white] at (-0.15,1.05) {\small SSIM \ = $0.819$};
 			\end{scope}
 			\end{tikzpicture}} &  
 		  	
 		\raisebox{-.5\height}{
 			\begin{tikzpicture}
 			\begin{scope}[spy using outlines={rectangle,yellow,magnification=1.7,size=8mm, connect spies}]
 		  	\node {\includegraphics[width=34mm,height=34mm]{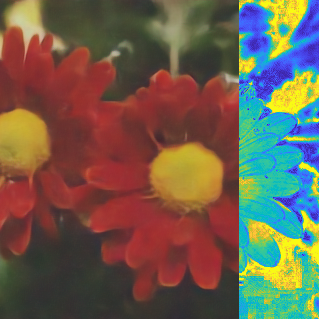}};
 			\spy on (-1.4, -0.1) in node  [left] at (-0.9,-1.3);
 			\node [white] at (0.0,1.4) {\small PSNR = $20.1$ dB };
 			\node [white] at (-0.15,1.05)  {\small SSIM \ = $0.690$};
 			\end{scope}
 			\end{tikzpicture}} &
 			
 		\raisebox{-.5\height}{
 			\begin{tikzpicture}
 			\begin{scope}[spy using outlines={rectangle,yellow,magnification=1.7,size=8mm, connect spies}]
            \node {\includegraphics[width=34mm,height=34mm]{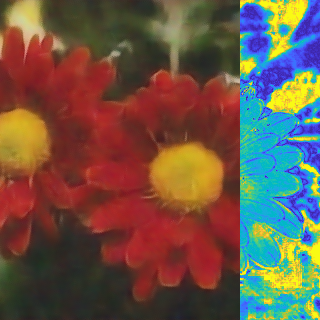}}; 			 
 			\spy on (-1.4, -0.1) in node  [left] at (-0.9,-1.3);
 			\node [white] at (0.0,1.4)  {\small  \color{yellow}{PSNR = $21.2$ dB}};
 			\node [white] at (-0.15,1.05)  {\small \color{yellow}{SSIM \ = $0.705$}};
 			\end{scope}
 			\end{tikzpicture}} &
 		
 		\raisebox{-.5\height}{
 			\begin{tikzpicture}
 			\begin{scope}[spy using outlines={rectangle,yellow,magnification=1.7,size=8mm, connect spies}]
 		  	\node {\includegraphics[width=34mm,height=34mm]{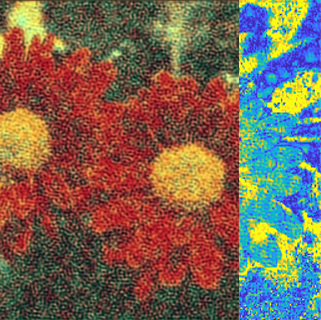}};
 			\spy on (-1.4, -0.1) in node  [left] at (-0.9,-1.3);
 			\node [white] at (0.0,1.4) {\small PSNR = $18.1$ dB };
 			\node [white] at (-0.15,1.05)  {\small SSIM \ = $0.630$};
 			\end{scope}
 			\end{tikzpicture}} &
 			
 		\raisebox{-.5\height}{
 			\begin{tikzpicture}
 			\begin{scope}[spy using outlines={rectangle,yellow,magnification=1.7,size=8mm, connect spies}]
            \node {\includegraphics[width=34mm,height=34mm]{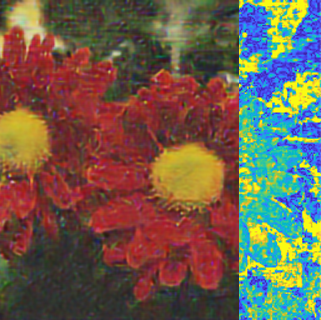}}; 			 
 			\spy on (-1.4, -0.1) in node  [left] at (-0.9,-1.3);
 			\node [white] at (0.0,1.4)  {\small  PSNR = $20.2$ dB};
 			\node [white] at (-0.15,1.05)  {\small SSIM \ = $0.690$};
 			\end{scope}
 			\end{tikzpicture}} 
 			
    \vspace{-0.25pc}	
 	\end{tabular}
 	\caption{Comparisons of denoised images {\bfseries (left)} from different learning/fitting methods and their saturation error maps {\bfseries (right)} in camera imaging (blue and yellow denote $0$ and $0.5$ absolute error, respectively). We used DnCNN for all self-supervised denoising methods in red boxes and Noise2True.
 	 PSNR \& SSIM values were averaged across all test samples.} 
    \vspace{-0.5pc}
 	\label{fig:nat}
 \end{figure*}
  \begin{figure*}[!tbh]

 	\centering
 	\small\addtolength{\tabcolsep}{-9.5pt}
 	\renewcommand{\arraystretch}{0.8}
 	
 	\begin{tabular}{ccccc}
 		
 		 \small{Input noisy image} & 
 		 \boxitfig{6.83cm}{0.65cm}
 		  \parbox{3.75cm}{\centering
 		  \small ~Noise2Self} &
 		 \specialcell[c]{\small {\bfseries Proposed SSRL} in \\ \small Noise2Self setup} &
        \boxitfig{6.83cm}{0.65cm} 
 		  \parbox{3.75cm}{\centering
 		  \small ~Neighbor2Neighbor}  & 
 		  \specialcell[c]{\small {\bfseries Proposed SSRL} in \\ \small Neighbor2Neighbor setup} 		 
 		 \\ 
 		
 		\raisebox{-.5\height}{
 			\begin{tikzpicture}
 			\begin{scope}[spy using outlines={rectangle,yellow,magnification=1.7,size=8mm, connect spies}]
 			\node {\includegraphics[width=34mm,height=34mm]{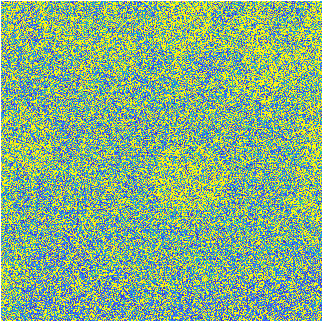}};
 			\node [white,fill=black] at (0.00,1.4) {\small PSNR = $9.8$ dB\,};
 			\node [white,fill=black] at (0.00,1.04)  {\small SSIM \ = $0.134$\,\,};
 			\end{scope}
 			\end{tikzpicture}} &
 		
 		\raisebox{-.5\height}{
 			\begin{tikzpicture}
 			\begin{scope}[spy using outlines={rectangle,yellow,magnification=1.7,size=8mm, connect spies}]
 			\node {\includegraphics[width=34mm,height=34mm]{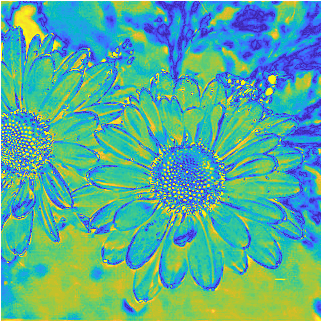}};
 			\node [white,fill=black] at (0.0,1.4) {\small PSNR = $20.6$ dB};
 			\node [white,fill=black] at (0.0,1.04) {\small SSIM \ = $0.711$ \;\;};
 			\end{scope}
 			\end{tikzpicture}} &

 		\raisebox{-.5\height}{
 			\begin{tikzpicture}
 			\begin{scope}[spy using outlines={rectangle,yellow,magnification=1.7,size=8mm, connect spies}]
 			\node {\includegraphics[width=34mm,height=34mm]{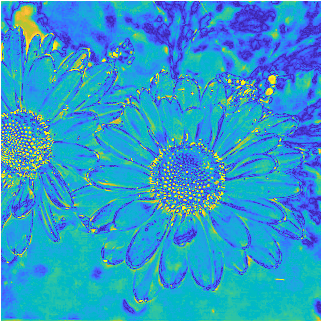}};
 			\node [yellow,fill=black] at (0.0,1.4)  {\small  \color{yellow}{PSNR = $22.1$ dB}};
 			\node [yellow,fill=black] at (0.0,1.04)  {\small  \color{yellow}{SSIM \ = $0.748$ \;\;}}; 			\end{scope}
 			\end{tikzpicture}}  &  		
 		   \raisebox{-.5\height}{
 		  	\begin{tikzpicture}
 			\begin{scope}[spy using outlines={rectangle,yellow,magnification=1.7,size=8mm, connect spies}]
 		  	\node {\includegraphics[width=34mm,height=34mm]{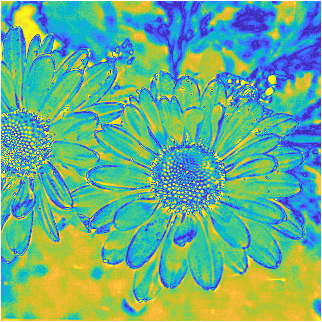}};
 			\node [white,fill=black] at  (0.0,1.4)  {\small PSNR = $20.2$ dB};
 			\node [white,fill=black] at (0.0,1.04) {\small SSIM \ = $0.703$ \;\;};
 			\end{scope}
 		  	\end{tikzpicture}} &
 		
 		\raisebox{-.5\height}{
 			\begin{tikzpicture}
 			\begin{scope}[spy using outlines={rectangle,yellow,magnification=1.7,size=8mm, connect spies}]
 		  	\node {\includegraphics[width=34mm,height=34mm]{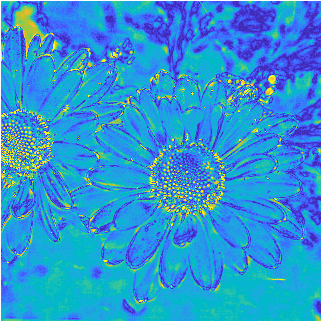}};
 			\node [yellow,fill=black] at (0.0,1.4)  {\small  \color{yellow}{PSNR = $22.3$ dB }};
 			\node [yellow,fill=black] at (0.0,1.04)  {\small  \color{yellow}{SSIM \ = $0.756$ \;\;}}; 			\end{scope}
 			\end{tikzpicture}}
 			\vspace{0.2pc}
 			
 			\\
          \small{Noise2True} &
          \boxitfig{6.83cm}{0.65cm}
 		  \parbox{3.75cm}{\centering
 		  \small ~Noise2Same}  & 
 		  \specialcell[c]{\small {\bfseries Proposed SSRL} in \\ \small Noise2Same setup}
 		  & 
 		  \small{Noisy-As-Clean} & 
 		  \small{Self2Self}
 		  \\
 		\raisebox{-.5\height}{
 			\begin{tikzpicture}
 			\begin{scope}[spy using outlines={rectangle,yellow,magnification=1.7,size=8mm, connect spies}]
 			\node {\includegraphics[width=34mm,height=34mm]{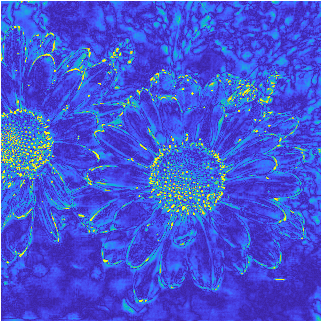}};
 			\node [white,fill=black] at (0.0,1.4) {\small PSNR  = $25.2$ dB};
 			\node [white,fill=black] at (0.0,1.04) {\small SSIM \ = $0.819$ \;\;};
 			\end{scope}
 			\end{tikzpicture}} &  	
 		\raisebox{-.5\height}{
 			\begin{tikzpicture}
 			\begin{scope}[spy using outlines={rectangle,yellow,magnification=1.7,size=8mm, connect spies}]
 		  	\node {\includegraphics[width=34mm,height=34mm]{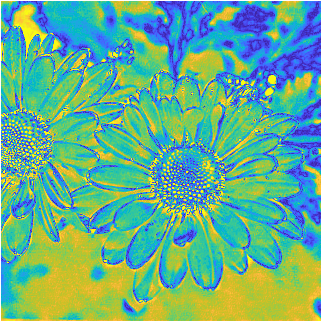}};
 			\node [white,fill=black] at  (0.0,1.4) {\small PSNR = $20.1$ dB };
 			\node [white,fill=black] at (0.0,1.04)  {\small SSIM \ = $0.690$ \;\;}; 			
 			\end{scope}
 			\end{tikzpicture}} &
 			
 		\raisebox{-.5\height}{
 			\begin{tikzpicture}
 			\begin{scope}[spy using outlines={rectangle,yellow,magnification=2.5,size=8mm, connect spies}]
 			 \node {\includegraphics[width=34mm,height=34mm]{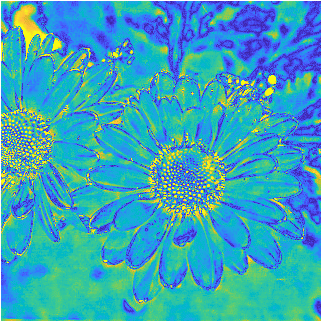}};
 			\node [yellow,fill=black] at  (0.0,1.4)  {\small  \color{yellow}{PSNR = $21.2$ dB}};
 			\node [yellow,fill=black] at (0.0,1.04)  {\small \color{yellow}{SSIM \ = $0.705$ \;\;}}; 		

 			\end{scope}
 			\end{tikzpicture}} &
 		
 		\raisebox{-.5\height}{
 			\begin{tikzpicture}
 			\begin{scope}[spy using outlines={rectangle,yellow,magnification=1.7,size=8mm, connect spies}]
 		  	\node {\includegraphics[width=34mm,height=34mm]{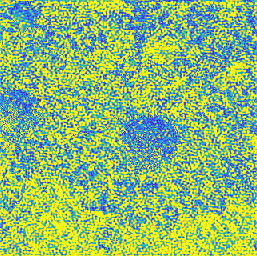}};
 			\node [white,fill=black] at  (0.0,1.4) {\small PSNR = $18.1$ dB };
 			\node [white,fill=black] at (0.0,1.04)  {\small SSIM \ = $0.630$ \;\;}; 			
 			\end{scope}
 			\end{tikzpicture}} &
 			
 		\raisebox{-.5\height}{
 			\begin{tikzpicture}
 			\begin{scope}[spy using outlines={rectangle,yellow,magnification=2.5,size=8mm, connect spies}]
 			 \node {\includegraphics[width=34mm,height=34mm]{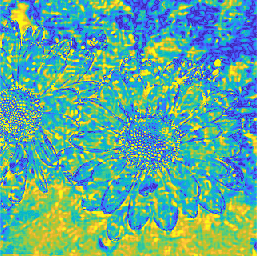}};
 			\node [white,fill=black] at  (0.0,1.4) {\small PSNR = $20.2$ dB };
 			\node [white,fill=black] at (0.0,1.04)  {\small SSIM \ = $0.690$ \;\;}; 			

 			\end{scope}
 			\end{tikzpicture}} 
 			
    \vspace{-0.25pc}
 	\end{tabular}
 	
 	\caption{Error map comparisons of denoised images from different learning/fitting methods in camera imaging (blue and yellow denote $0$ and $50$ absolute errors, respectively). The results correspond to those in Fig.~\ref{fig:nat}.
 	} 	 
 	\label{fig:nat:err}
    \vspace{-0.5pc}
 \end{figure*}

\subsection{Comparisons between different self-supervised denoising methods with simulated imaging datasets}
\label{sec:comp}

Compare each existing self-supervised denoising method using a training dataset to its corresponding SSRL setup in Figs.~\ref{fig:ct}--\ref{fig:nat:err} and Tables~\ref{tab:ct}--\ref{tab:nat:BSD}; see three comparison sets, each grouped by red box. 
For both applications, proposed SSRL achieves significantly better image denoising quality, i.e., closer to the Noise2True quality, compared to the existing self-supervised learning methods,
Noise2Self, Noise2Inverse or Neighbor2Neighbor, and Noise2Same, regardless of the regression NN architecture. (Fig.~\ref{fig:err_bar} shows the prediction uncertainty of all the self-supervised learning methods.)
In addition, Figs.~\ref{fig:ct}--\ref{fig:nat:err} show that for both applications, proposed SSRL regardless of its setup achieves significantly better image denoising performance, compared to existing single-image self-supervised denoising method, Noisy-As-Clean and Self2Self.

\subsubsection{Low-dose CT denoising}
\label{sec:comp:ct:syn}

In all the three comparison sets, SSRL better recovers low-contrast regions (e.g., soft tissues) and small details, and 
significantly reduces noise and artifacts throughout the image, over existing methods, Noise2Self, Noise2Inverse, and Noise2Same. 
See zoom-ins and circled small details in Fig.~\ref{fig:ct}, and error images in Fig.~\ref{fig:ct:err}, particularly in \quotes{Noise2Self vs.~Propose SSRL in Noise2Self setup} and \quotes{Noise2Same vs.~Proposed SSRL in Noise2Same setup.}
Proposed SSRL in the Noise2Inverse setup can provide images with image quality that is comparable to conventional FBP at $10$ times higher dose (when $\rho_0 = 5 \times 10^5$, $\text{RMSE} = 20.5$ HU on average; see DnCNN results in Fig.~\ref{fig:ct}).

Next, comparing Noise2Self and Noise2Same result sets to that of Noise2Inverse in Figs.~\ref{fig:ct}--\ref{fig:ct:err}, and Table~\ref{tab:ct} shows that Noise2Inverse setup significantly improves the denoising quality, compared to Noise2Self and Noise2Same setups. 
In this application,
Assumption~2 (see Section~\ref{sec:prelim}) holds in the Noise2Inverse setup, but unlikely to be satisfied in the Noise2Self and Noise2Same setups.
We conjecture that violation of Assumption~2 particularly degraded the performance in the Noise2Self and Noise2Same setups.
We make a similar conjecture about the worse performance of Self2Self than that of Noise2Inverse in Figs.~\ref{fig:ct}--\ref{fig:ct:err}.

Figs.~\ref{fig:ct}--\ref{fig:ct:err} show that Noisy-As-Clean gives worse denoising performance compared to the other existing self-supervised denoising methods.
The potential reason is that its \dquotes{weak} noise assumption is not satisfied in low-dose CT; this well corresponds to the claim in \cite[\S 5.D]{Xu&etal:20TIP}.

\subsubsection{Camera image denoising}
\label{sec:comp:camera:syn}

Figs.~\ref{fig:nat}--\ref{fig:nat:err} demonstrate that
in all the three comparison sets, SSRL
gives closer image quality, particularly color saturation, to Noise2True than existing methods, Noise2Self, Neighbor2Neighbor, and Noise2Same.
Setting $g$ as weighted median filtering avoids bias in SSRL loss caused by salt-and-pepper noise.
Yet, compared to Noise2True, denoised images obtained by proposed SSRL lack saturation and detail preservation. 
For saturation and detail preservation comparisons, see Figs.~\ref{fig:nat}--\ref{fig:nat:err}.

Comparing the three comparison sets in Figs.~\ref{fig:nat}--\ref{fig:nat:err} and Tables~\ref{tab:nat:BSD} shows that 
all the Noise2Self, Neighbor2Neighbor, and Noise2Same setups have comparable results in terms of PSNR values.
The potential reason is that in this application, all the three setups satisfy Assumptions~1--2 (see Section~\ref{sec:prelim});
in particular, Assumption~2 is well satisfied by pixel-wise i.i.d.~noise.
Potentially with a similar reason, Self2Self gives comparable results to Noise2Self, Neighbor2Neighbor, and Noise2Same in Figs.~\ref{fig:nat}--\ref{fig:nat:err}.

Figs.~\ref{fig:nat}--\ref{fig:nat:err} show that 
Noisy-As-Clean gives worse image denoising performance compared to the other aforementioned existing methods.
We conjecture similarly as in  Section~\ref{sec:comp:ct:syn} 
that the mixed noise setup in Section~\ref{sec:setup:sim:nat} has too strong noise for Noisy-AS-Clean to reduce.

In this application, the zero-mean noise assumption of the existing self-supervised denoising methods is violated, whereas its counterpart in proposed SSRL, Assumption~3 (see Section~\ref{sec:prelim}), is \dquotes{approximately} satisfied by $g$ in Section~\ref{sec:know:nat}.
In addition, we expect that this particular $g$ design potentially reduces the prediction error in Proposition~\ref{thm:soln-all}.


 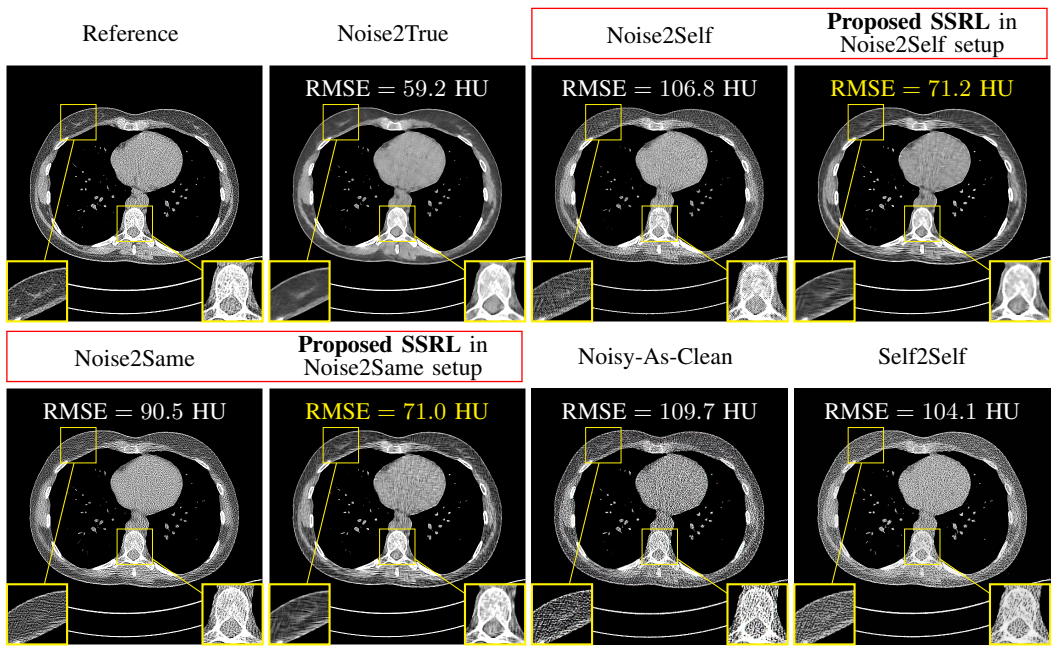
\begin{figure*}[!th]
 	\centering
 	\small\addtolength{\tabcolsep}{-9.5pt}
 	\renewcommand{\arraystretch}{0.8}
 	
 	\begin{tabular}{cccc}
 		
 		 \small{Reference}  & 
 		 \small{Noise2True} & 
 		 \boxitfig{6.83cm}{0.65cm}
 		 \parbox{3.75cm}{\centering
 		 \small ~Noise2Self}  & 
 		 \specialcell[c]{\small {\bfseries Proposed SSRL} in \\ \small Noise2Self setup} 
 		 \\ 
 		
 		\raisebox{-.5\height}{
 			\begin{tikzpicture}
 			\begin{scope}[spy using outlines={rectangle,yellow,magnification=1.7,size=8mm, connect spies}]
 			\node {\includegraphics[width=34mm]{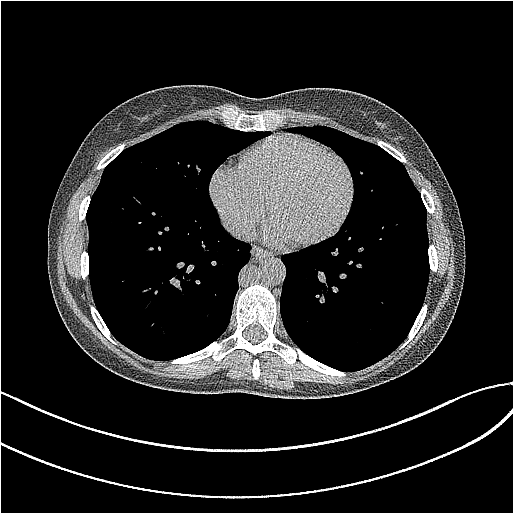}};
 			\spy on (-0.75, 0.95) in node [left] at (-0.9,-1.3);
 			\spy on (0.0, -0.4) in node [left] at (1.7,-1.3);
 			\end{scope}
 			\end{tikzpicture}} &

 		\raisebox{-.5\height}{
 			\begin{tikzpicture}
 			\begin{scope}[spy using outlines={rectangle,yellow,magnification=1.7,size=8mm, connect spies}]
 			\node {\includegraphics[width=34mm]{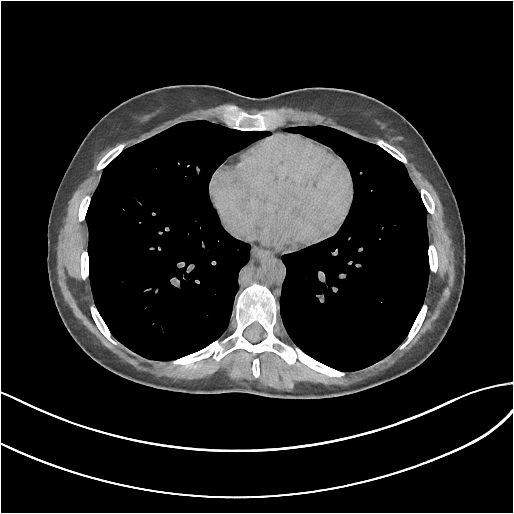}};
 			\spy on (-0.75, 0.95) in node [left] at (-0.9,-1.3);
 			\spy on (0.0, -0.4) in node [left] at (1.7,-1.3);
 			\node [white] at (0.0,1.4) {\small $\text{RMSE} = 59.2$ HU};
 			\end{scope}
 			\end{tikzpicture}} &
 		
 		 \raisebox{-.5\height}{
 			\begin{tikzpicture}
 				\begin{scope}[spy using outlines={rectangle,yellow,magnification=1.7,size=8mm, connect spies}]
 					\node {\includegraphics[width=34mm]{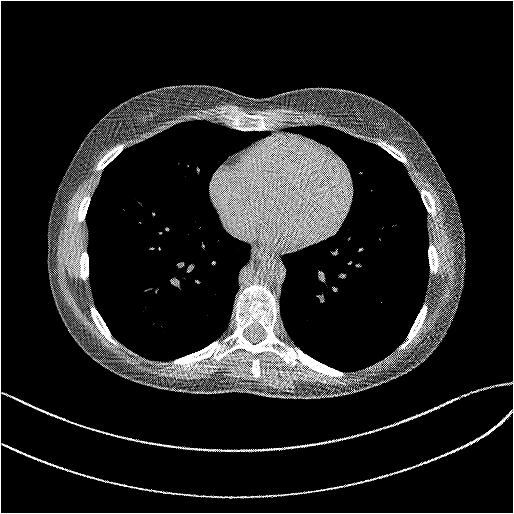}};
 					\spy on (-0.75, 0.95) in node [left] at (-0.9,-1.3);
 					\spy on (0.0, -0.4) in node [left] at (1.7,-1.3);
 					\node [white] at (0.0,1.4) {\small $\text{RMSE} = 106.8$ HU};
 				\end{scope}
 		\end{tikzpicture}} &
 		
 		\raisebox{-.5\height}{
 			\begin{tikzpicture}
 				\begin{scope}[spy using outlines={rectangle,yellow,magnification=1.7,size=8mm, connect spies}]
 					\node {\includegraphics[width=34mm]{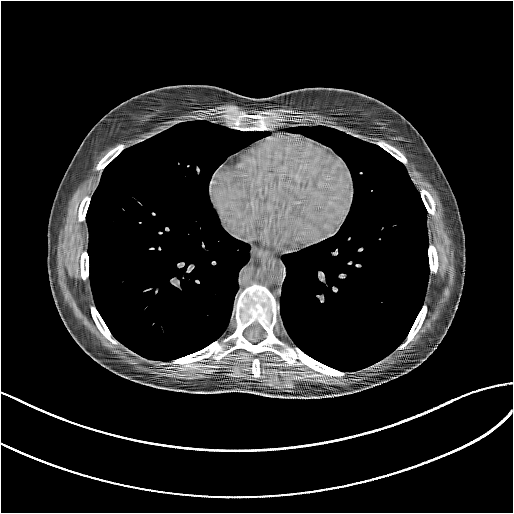}};
 					\spy on (-0.75, 0.95) in node [left] at (-0.9,-1.3);
 					\spy on (0.0, -0.4) in node [left] at (1.7,-1.3);
 					\node [white] at (0.0,1.4)  {\small  \color{yellow}{$\text{RMSE} = 71.2$ HU}};
 				\end{scope}
 		\end{tikzpicture}}  		
 	 	\vspace{0.2pc}
 		\\

 		\boxitfig{6.83cm}{0.65cm}
 		\parbox{3.75cm}{\centering
 		\small ~Noise2Same} & 
 		\specialcell[c]{\small {\bfseries Proposed SSRL} in \\ \small Noise2Same setup} &
         \small {Noisy-As-Clean} &
        \small {Self2Self} 
 		\\
 		\raisebox{-.5\height}{
 			\begin{tikzpicture}
 			\begin{scope}[spy using outlines={rectangle,yellow,magnification=1.7,size=8mm, connect spies}]
 			\node {\includegraphics[width=34mm]{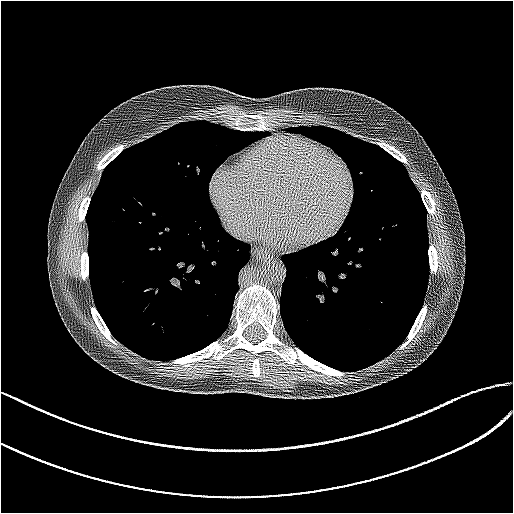}};
 			\spy on (-0.75, 0.95) in node [left] at (-0.9,-1.3);
 			\spy on (0.0, -0.4) in node [left] at (1.7,-1.3);
 			\node [white] at (0.0,1.4) {\small $\text{RMSE} = 90.5$ HU};
 			\end{scope}
 			\end{tikzpicture}} &
 			
 		\raisebox{-.5\height}{
 			\begin{tikzpicture}
 			\begin{scope}[spy using outlines={rectangle,yellow,magnification=1.7,size=8mm, connect spies}]
 			 \node {\includegraphics[width=34mm]{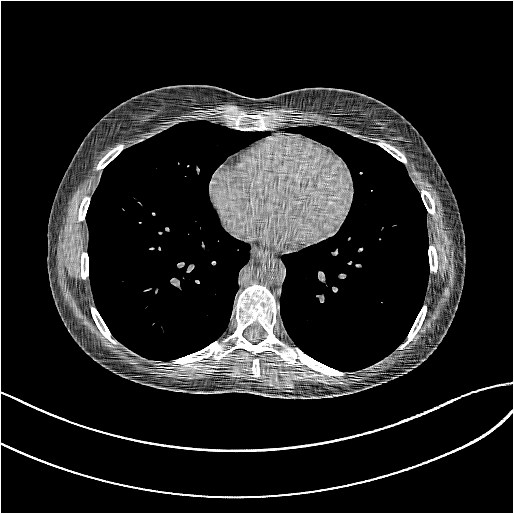}};
 			\spy on (-0.75, 0.95) in node [left] at (-0.9,-1.3);
 			\spy on (0.0, -0.4) in node [left] at (1.7,-1.3);
 			\node [white] at (0.0,1.4)  {\small  \color{yellow}{$\text{RMSE} = 71.0$ HU}};
 			\end{scope}
 			\end{tikzpicture}}& 		
 		\raisebox{-.5\height}{
 			\begin{tikzpicture}
 				\begin{scope}[spy using outlines={rectangle,yellow,magnification=1.7,size=8mm, connect spies}]
 					\node {\includegraphics[width=34mm]{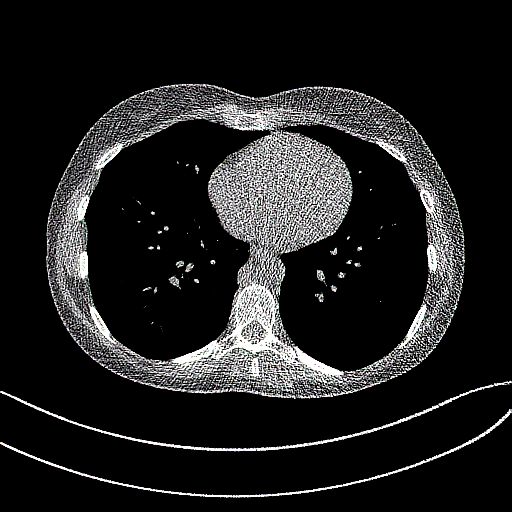}};
 					\spy on (-0.75, 0.95) in node [left] at (-0.9,-1.3);
 					\spy on (0.0, -0.4) in node [left] at (1.7,-1.3);
         			\node [white] at (0.0,1.4) {\small $\text{RMSE} = 109.7$ HU};
 				\end{scope}
 		\end{tikzpicture}}  	
 		&
 		\raisebox{-.5\height}{
 			\begin{tikzpicture}
 			\begin{scope}[spy using outlines={rectangle,yellow,magnification=1.7,size=8mm, connect spies}]
 			\node {\includegraphics[width=34mm]{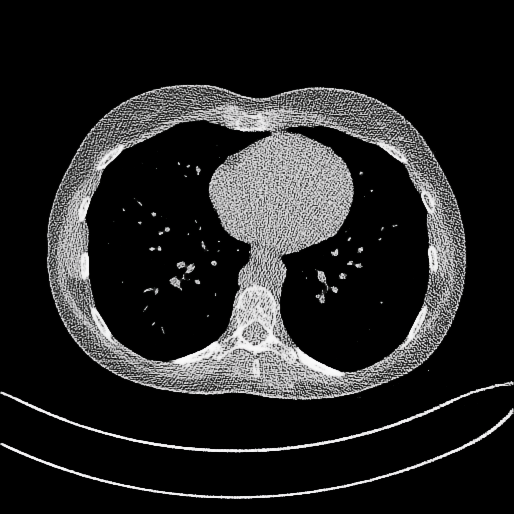}};
 			\spy on (-0.75, 0.95) in node [left] at (-0.9,-1.3);
 			\spy on (0.0, -0.4) in node [left] at (1.7,-1.3);
 			\node [white] at (0.0,1.4) {\small $\text{RMSE} = 104.1$ HU};
 			\end{scope}
 			\end{tikzpicture}} 
 		
 	 \vspace{-0.25pc}	
 	\end{tabular}
 	\caption{Comparisons of denoised images  from different learning methods in low-dose CT (display window is $[800, 1200]$ HU). We used DnCNN for all self-supervised denoising methods in red boxes and Noise2True.
 	 RMSE values were averaged across all test samples.
 	 }
 	\label{fig:ct:real}
 	\vspace{-0.5pc}
 \end{figure*}
 \begin{figure*}[!th]
 	\centering
 	\small\addtolength{\tabcolsep}{-9.5pt}
 	\renewcommand{\arraystretch}{0.8}

 	\begin{tabular}{cccccc}

 		 \small{Reference} &  \quad \quad \small{Noise2True}  	\quad 
 		 \boxitfig{5.7cm}{0.65cm}
 		 \parbox{3.75cm}	  
 		 &

 		  \small{Neighbor2Neighbor} &
 		 \specialcell[c]{\small {\bfseries Proposed SSRL} \\ \small Neighbor2Neighbor} &\small {Noisy-As-Clean} & \small{Self2Self}
 		 
 		 \\ 
 		\raisebox{-.5\height}{
 			\begin{tikzpicture}
 			\begin{scope}[spy using outlines={rectangle,yellow,magnification=1.7,size=8mm, connect spies}]
 			\node {\includegraphics[width=28mm,height=28mm]{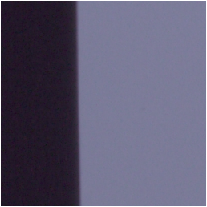}};
 			\end{scope}
 			\end{tikzpicture}} &
 		\raisebox{-.5\height}{
 			\begin{tikzpicture}
 			\begin{scope}[spy using outlines={rectangle,yellow,magnification=1.7,size=8mm, connect spies}]
 			\node {\includegraphics[width=28mm,height=28mm]{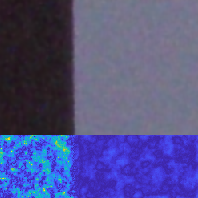}};
 			\node [white,fill=black] at  (0.0,1.15)  {\small  PSNR = $31.1$ dB};
 			\end{scope}
 			\end{tikzpicture}} 
 		&
 		\raisebox{-.5\height}{
 			\begin{tikzpicture}
 			\begin{scope}[spy using outlines={rectangle,yellow,magnification=1.7,size=8mm, connect spies}]
 			\node {\includegraphics[width=28mm,height=28mm]{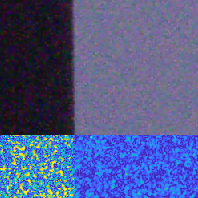}};
 			\node [white,fill=black] at  (0.0,1.15)  {\small  PSNR = $27.2$ dB};

 			\end{scope}
 			\end{tikzpicture}} &

 		\raisebox{-.5\height}{
 			\begin{tikzpicture}
 			\begin{scope}[spy using outlines={rectangle,yellow,magnification=1.7,size=8mm, connect spies}]
 			\node {\includegraphics[width=28mm,height=28mm]{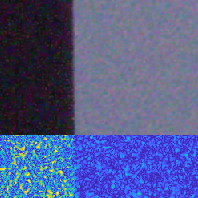}};
 			\node [yellow,fill=black] at  (0.0,1.15)  {\small  \color{yellow}{PSNR = $28.2$ dB}};
 			\end{scope}
 			\end{tikzpicture}} 	&
 		 \raisebox{-.5\height}{
 			\begin{tikzpicture}
 				\begin{scope}[spy using outlines={rectangle,yellow,magnification=1.7,size=8mm, connect spies}]
 			\node {\includegraphics[width=28mm]{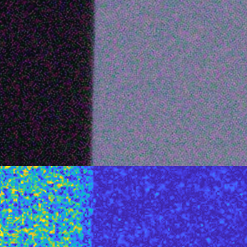}};
 					\node [white,fill=black] at (0.0,1.15) {\small $\text{PSNR} = 26.8$ dB};

 				\end{scope}
 		\end{tikzpicture}} 	
 			& 	
 		 \raisebox{-.5\height}{
 			\begin{tikzpicture}
 				\begin{scope}[spy using outlines={rectangle,yellow,magnification=1.7,size=8mm, connect spies}]
 			\node {\includegraphics[width=28mm]{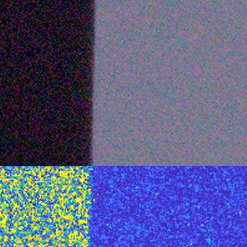}};
 					\node [white,fill=black] at (0.0,1.15) {\small $\text{PSNR} = 26.8$ dB};

 				\end{scope}
 		\end{tikzpicture}} 	 			

 	 \vspace{-0.25pc}	
 	\end{tabular}
 	\caption{Comparisons of denoised images {\bfseries (top)} from different learning methods and their saturation error maps {\bfseries (bottom)} in camera image denoising (blue and yellow denote $0$ and $0.5$ absolute errors, respectively).
 	We used DnCNN for all self-supervised denoising methods in a red box and Noise2True.
 	 PSNR values were averaged across all test samples.} 	
  	 \vspace{-0.5pc}
 	\label{fig:nat:real}
 \end{figure*}
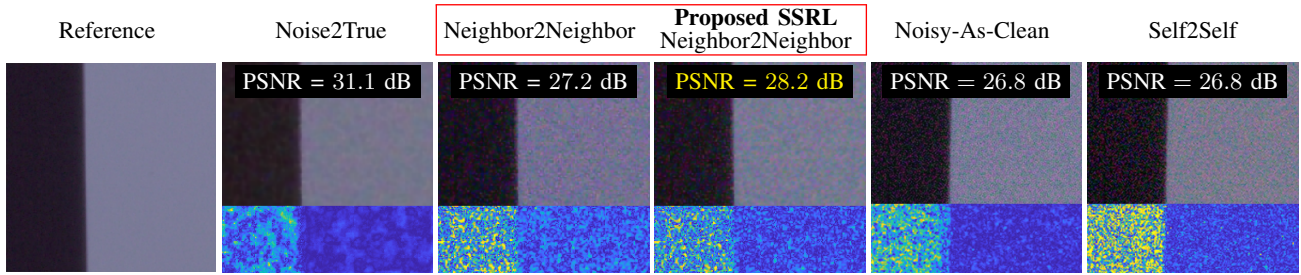

\subsection{Comparisons between different self-supervised denoising methods with intrinsically noisy imaging datasets}
\label{sec:comp:real}

Figs.~\ref{fig:ct:real}--\ref{fig:nat:real} from two different imaging applications with intrinsically noisy datasets \cite{Abdelhamed&etal:18CVPR, Moen&etal:21MP} demonstrate that SSRL significantly improves existing self-supervised denoising methods, particularly Noise2Self, Noise2Same, Neighbor2Neighbor, Noisy-As-Clean, and Self2Self, \emph{without} having their exact noise properties. 
The results correspond to our understanding in Section~\ref{sec:add:guidelines}.

\subsubsection{Low-dose CT denoising}

Fig.~\ref{fig:ct:real} demonstrates that SSRL better recovers low-contrast regions and small details, and significantly reduces noise and artifacts throughout the image, over existing methods, Noise2Self, Noise2Same, Noisy-As-Clean, and Self2Self. 
These results combined with those from Fig.~\ref{fig:ct} might imply that simply setting $g$ as pre-trained NN by existing self-supervised denoising methods works like a charm in SSRL.

In practice, 
using the Noise2Inverse learning setup may be limited in low-dose CT,
if a vendor does not provide software to partition the sinogram into two subsets, e.g., even and odd angles, or reconstruct images from partitioned sinogram data.

\subsubsection{Camera image denoising}

Fig.~\ref{fig:nat:real} \quotes{Neighbor2Neighbor vs.~Propose SSRL in Neighbor2Neighbor setup} shows that 
proposed SSRL gives significantly closer image quality to Noise2True than Neighbor2Neighbor.

The intrinsically noisy dataset unlikely satisfies Assumption~2 (i.e., pixel-wise noise is likely correlated) due to its complicated camera imaging physics~\cite{Zamir&etal:20CVPR}.
We observed that setting $g$ as weighted median filtering is expected to lead its prediction $g(x_J)$ closer to $y$ than $g = \cI$, given $x_{J^c}$. See empirical results in Section~\ref{sec:add:empirical}.

The intrinsically noisy dataset may satisfy the \dquotes{weak} noise assumption of Noisy-As-Clean. We observed that its performance is similar to Self2Self.

\subsection{Summary about designing pseudo-predictors}
\label{sec:summary}

The results from the four imaging experimental sets validate our expectation in Section~\ref{sec:know} that well-designed $g$ by following the suggestions in Sections~\ref{sec:SSRL:ind} improves the denoising performance of learned $f$ via SSRL.
If an application satisfies Assumption~1 (but not Assumption~2), 
we design $g$ to reduce the expected prediction error in Theorem~\ref{thm:soln}.
Sections~\ref{sec:comp:ct:syn} \& \ref{sec:comp:real} show that such $g$ improves the performance of learned $f$ via SSRL compared to $g=\cI$.
If an application satisfies Assumptions~1--2, 
we design $g$ to better satisfy Assumption~3 to approach closer to the optimal performance in Proposition~\ref{thm:soln-all}.
Section~\ref{sec:comp:camera:syn} shows that such $g$ improves the denoising performance of learned $f$ via SSRL compared to $g=\cI$.

\subsection{Limitations}
\label{sec:limit}

The proposed SSRL model \R{sys:SSRL} uses domain knowledge of a specific application in designing $g$. A pseudo-predictor $g$ designed with good domain knowledge -- see, e.g., two examples in Section~\ref{sec:know} -- can improve the prediction quality of learned regression NNs, leading the properties of SSRL \R{sys:SSRL} close to the supervision counterparts (see Sections~\ref{sec:SSRL:ind}--\ref{sec:SSRL}). 
However, $g$ designed with poor domain knowledge could degrade the prediction performance of learned regression NNs via \R{sys:SSRL}.
The limitation is empirically shown in Section~\ref{sec:motivation}.

We conjecture that violation of Assumption~2 (see Section~\ref{sec:prelim}) degrades the prediction accuracy of learned regression NNs.
This is empirically shown with low-dose CT denoising, and is pointed out in Section~\ref{sec:comp}.
Reconstructed images by linear image reconstruction in many ill-conditioned inverse imaging problems, such as light-field photography using focal stack \cite{Chun&etal:20PAMI} and undersampled magnetic resonance imaging \cite{AjaFernandez&VegasSanchez&TristanVega:14MRI, Chun&etal:15TCI}, are likely to violate the assumption.

\section{Conclusion}
\label{sec:conclusion}

It is important to develop SSRL that enables comparable prediction performances to supervised learning, 
because it is extremely challenging to collect many ground-truth target samples in many practical computational imaging and computer vision applications.
The proposed SSRL framework bridges the gap between SSRL and supervised regression learning via domain knowledge of applications.
To achieve closer prediction performance to supervised learning, SSRL uses domain knowledge to design a better pseudo-predictor such that pseudo-prediction becomes closer to ground-truth in the expected sense given some condition.
For image denoising in low-dose CT and camera imaging with both synthetic and intrinsically noisy datasets, SSRL achieves significantly improved prediction compared to the existing self-supervised methods \cite{Batson&Royer:19ICML, Huang&eta:21CVPR, Hendriksen&etal:20TCI, Xie&Wang&Ji:20NIPS, Quan&etal:20CVPR, Xu&etal:20TIP}.

Our future work is extending proposed SSRL to other machine learning problems such as teacher-student models (see Section~\ref{sec:tea-stu}) and meta-learning.
On the application side, our future work is applying SSRL to regression problems beyond image denoising \cite{Chun&etal:22RAL}.


\renewcommand{\theHtable}{A.\thetable}
\renewcommand\thetable{A.\arabic{table}}
\renewcommand{\theHfigure}{A.\thefigure}
\renewcommand\thefigure{A.\arabic{figure}}

\renewcommand{\theHsection}{A.\thesection}
\renewcommand\thesection{A.\arabic{section}}


\setcounter{table}{0}
\setcounter{figure}{0}
\setcounter{section}{0}

\section*{Appendix}
\section{Proofs for Theorem~\ref{thm:soln}}
\label{sec:prf:thm:soln}

We claim that $f^\star (x) = \bbE [g(x_{J}) | x_{J^c}]$.
Observing under the $\cJ$-complement setup that 
\bes{
\bbE_{x} \| f(x_{J^c}) - g(x_J) \|_2^2
= 
\bbE_{x_{J^c}} \bbE_{x_J | x_{J^c}} \| f(x_{J^c}) - g(x_J) \|_2^2,
}
we first investigate $\bbE_{x_J | x_{J^c}} \| f(x_{J^c}) - g(x_J) \|_2^2$ as follows:
\begingroup
\thinmuskip=1.5mu\relax
\medmuskip=2mu plus 1mu minus 2mu\relax
\thickmuskip=2.5mu plus 2.5mu\relax
\eas{
&~\bbE_{x_J | x_{J^c}} \| f(x_{J^c}) - g(x_J) \|_2^2
\\
&=
\bbE_{x_J | x_{J^c}} \! \big[
\| f(x_{J^c}) -  \bbE_{x_{J} | x_{J^c}} [g(x_{J})] \|_2^2 
\\
& \hspace{2.4pc} 
+ \| \bbE_{x_{J} | x_{J^c}} [g(x_{J})] - g(x_J) \|_2^2
\\
& \hspace{2.4pc} 
+ 2 \ip{ f(x_{J^c}) - \bbE_{x_{J} | x_{J^c}} [g(x_{J})] }{ \bbE_{x_{J} | x_{J^c}} [g(x_{J})] - g(x_J) } \big]
\\
&=
\bbE_{x_J | x_{J^c}} \! \big[
\| f(x_{J^c}) - \bbE_{x_{J} | x_{J^c}} [g(x_{J})] \|_2^2 
\\
& \hspace{2.4pc} + \| \bbE_{x_{J} | x_{J^c}} [g(x_{J})] - g(x_J) \|_2^2 \big].
}
\endgroup The second equality holds by the following result under Assumption~1:
\begingroup
\thinmuskip=1.5mu\relax
\medmuskip=2mu plus 1mu minus 2mu\relax
\thickmuskip=2.5mu plus 2.5mu\relax
\eas{
&~\bbE_{x_J | x_{J^c}} \ip{ f(x_{J^c}) -  \bbE_{x_{J} | x_{J^c}} [g(x_{J})] }{ \bbE_{x_{J} | x_{J^c}} [g(x_{J})] - g(x_J) }
\\
&= 
\ip{ f(x_{J^c}) -  \bbE_{x_{J} | x_{J^c}} [g(x_{J})] }{ \bbE_{x_J | x_{J^c}} [ \bbE_{x_{J} | x_{J^c}} [g(x_{J})] - g(x_J) ] }
\\
&=
\ip{ f(x_{J^c}) - \bbE_{x_{J} | x_{J^c}} [g(x_{J})] }{ \bbE_{x_J | x_{J^c}} [g(x_{J})] - \bbE_{x_J | x_{J^c}} [g(x_{J})] }
\\
&= 
\ip{ f(x_{J^c}) - \bbE_{x_{J} | x_{J^c}} [g(x_{J})] }{ 0 }
=
0,
}
\endgroup
noting that $f(x_{J^c}) - \bbE_{x_{J} | x_{J^c}} [g(x_{J})]$ (function of $x_{J^c}$) is considered a constant given $x_{J^c}$.
Thus, we obtain
\eas{
&~ \bbE_{x_J | x_{J^c}} \| f(x_{J^c}) - g(x_J) \|_2^2
\\
& \geq 
\bbE_{x_J | x_{J^c}} \| \bbE_{x_{J} | x_{J^c}} [g(x_{J})] - g(x_J) \|_2^2
}
Taking $\bbE_{x_{J^c}}$ both sides on the above result yields
\begingroup
\thinmuskip=1.5mu\relax
\medmuskip=2mu plus 1mu minus 2mu\relax
\thickmuskip=2.5mu plus 2.5mu\relax
\bes{
\bbE_{x} \| f(x_{J^c}) - g(x_J) \|_2^2
\geq 
\bbE_{x} \| \bbE_{x_{J} | x_{J^c}} [g(x_{J})] - g(x_J) \|_2^2
}
\endgroup
(by the law of iterated expectations, $\bbE_{x_{J}, x_{J^c}} [h(x_J,x_{J^c})] = \bbE_{x_{J^c}} \bbE_{x_{J}} [h(x_J,x_{J^c})| x_{J^c}]$).
This gives the first result \R{eq:thm:soln}.

To obtain the second result \R{eq:thm:soln:err}, we rewrite $\| f^\star(x) - y \|_2^2$ as follows:
\begingroup
\thinmuskip=1.5mu\relax
\medmuskip=2mu plus 1mu minus 2mu\relax
\thickmuskip=2.5mu plus 2.5mu\relax
\eas{
\| f^\star(x) - y \|_2^2
&= \| f^\star(x) - f^\ast(x) \|_2^2 + \| f^\ast(x) - y \|_2^2 
\\
&\hspace{1pc} + 2 \ip{ f^\star(x) - f^\ast(x) }{ f^\ast(x) - y } 
}
\endgroup
Taking conditional expectation of $y$ at $x=x'$ gives the second result as follows: 
\begingroup
\thinmuskip=1.5mu\relax
\medmuskip=2mu plus 1mu minus 2mu\relax
\thickmuskip=2.5mu plus 2.5mu\relax
\eas{
&~ \bbE_{y} [ \| f^\star(x) - y \|_2^2 | x=x' ]
\\
&= \| f^\star(x') - f^\ast(x') \|_2^2 + \mathrm{Var}(y | x=x')
\\
&\hspace{1pc} + 2 \ip{ f^\star(x) - f^\ast(x) }{ \bbE_{y} [ f^\ast(x) - y | x = x' ] }
\\
&= \| f^\star(x') - f^\ast(x') \|_2^2 + \mathrm{Var}(y | x=x')
}
\endgroup
where the first equality holds by the conditional variance definition, and the second equality holds by the fact that the second quantity in the inner product is zero.

\section{Proofs for Proposition~\ref{thm:soln-all}}
\label{sec:prf:thm:soln-all}

Observe first that combining Assumptions~1--2 and the $\cJ$-complement implies that 
$f (x)_m$ and $g (x)_m$ are conditionally independent, given $y$, i.e.,
\bes{
f(x_{J^c})_m | y \indep g (x_J)_m | y, \quad \forall m.
}
Using this result with Assumption~3 and reminding that the $\cJ$-complement implies $\bbE_x \| f(x) - g(x) \|_2^2 
= \bbE_x \| f(x_{J^c}) - g(x_{J}) \|_2^2 
$, we obtain the following result from \R{sys:SSRL,rewrite1}:
\bes{
\bbE_x \| f(x_{J^c}) - g(x_{J}) \|_2^2 
= \bbE_{x,y} \| f(x_{J^c}) - y \|_2^2 + \| g(x_{J}) - y \|_2^2 
}
where the equality uses
\begingroup
\allowdisplaybreaks
\eas{
&~\bbE_{x,y} \ip{ f(x_{J^c}) - y }{ g(x_{J}) - y } 
\\
&= \bbE_y \bbE_{x|y} \ip{ f(x_{J^c}) - y }{ g(x_{J}) - y }
\\ 
&= \sum_{m} \bbE_y ( \bbE_{x|y} [ f(x_{J^c})_m - y_m ] ) ( \bbE_{x|y} [ g(x_{J})_m - y_m ] )
\\
&= 0
}
\endgroup
in which the second equality uses the first result above and the third equality holds by Assumption~3.
(The following equality similarly holds for any $K \in \cK $: $\bbE_{x} \| f(x)_K - g(x)_K \|_2^2 = \bbE_{x,y} \| f(x)_K - y_K  \|_2^2 +  \| g(x)_K - y_K \|_2^2$.)
Finding the optimal solution (noting that the second term in the result above is irrelevant to $f$) completes the proof.

\section{Proofs for Proposition~\ref{thm:loss:b}}
\label{sec:prf:thm:loss:b}

We first obtain the following bound:
\begingroup
\thinmuskip=1.5mu\relax
\medmuskip=2mu plus 1mu minus 2mu\relax
\thickmuskip=2.5mu plus 2.5mu\relax
\allowdisplaybreaks
 \eas{
 &~ \bbE_{x,y} \ip{f(x) - y}{g(x_J) - y} 
 \\
 &= \bbE_y \bbE_{x|y} \sum_m (f(x)_m - y_m) (g(x_J)_m - y_m)
 \\
 &= \sum_m \bbE_y \left[ \bbE_{x|y} (f(x)_m - y_m) (g(x_J)_m - y_m) \right. 
 \\
 &\hspace{2.7pc} \left. - \bbE_{x|y} (f(x)_m - y_m) \bbE_{x|y} ( g(x_J)_m  - y_m )  \right]
 \\
 &= \sum_m \bbE_y \left[ \mathrm{Cov} ( f(x)_m - y_m, g(x_J)_m - y_m | y ) \right]
 \\
 &= \sum_m \bbE_y \left[ \mathrm{Cov} ( f(x)_m, g(x_J)_m | y ) \right]
 \\
 &= \sum_m \bbE_y  \left[ \mathrm{Cov} ( f(x)_m - f(x_{J^c})_m, g(x_J)_m  | y ) \right] 
 \\
 &\leq \sum_m \bbE_y \left[ \left( \mathrm{Var} ( f(x)_m - f(x_{J^c})_m | y ) \cdot \mathrm{Var} ( g(x_J)_m | y  ) \right)^{1/2} \right]
 \\
 &\leq \sum_m \bbE_y \left[ \mathrm{Var} ( f(x)_m - f(x_{J^c})_m | y ) \cdot \mathrm{Var} ( g(x_J)_m | y  )  \right]^{1/2}
 \\
 &\leq \left( M \cdot \sum_{m=1}^M \bbE_y \left[ \mathrm{Var} ( f(x)_m  - f(x_{J^c})_m | y ) \cdot \mathrm{Var} ( g(x_J)_m | y  )  \right] \right)^{\!\!\!\!\!\!1/2}
 \\
 &\leq \left( M \cdot \sum_{m=1}^M \bbE_y \left[ \mathrm{Var} ( f(x)_m - f(x_{J^c})_m | y ) \cdot \sigma^2  \right] \right)^{\!\!\!\!\!\!1/2},
 }
 \endgroup
where 
the second equality holds by Assumption~3,
the third equality uses $\mathrm{Cov} (X,Y | Z) = \bbE [XY|Z] - \bbE[X|Z] \bbE [Y|Z] $ where $X$, $Y$, and $Z$ are random variables or vectors,
the fifth equality holds because $f(x_{J^c})_m$ does not correlate with $g(x_J)_m$, $\forall m$ (due to Assumptions 1--2), so subtracting $f(x_{J^c})_m$ from $f(x)_m$ does not change the covariance.
Now, the first inequality uses the Pearson correlation coefficient bound,
the second inequality uses the Jensen's inequality $\bbE \sqrt{X} \leq \sqrt{\bbE X}$,
the third inequality uses the Jensen's inequality $\sum_m \sqrt{ a_m } \leq \sqrt{ M' \sum_m a_m }$ for any $a \in \bbR^{M'}$,
and the last inequality holds by the conditional variance bound specified in Proposition~\ref{thm:loss:b}.
We bound and rewrite the final result above and this completes the proof:
\begingroup
\thinmuskip=1.5mu\relax
\medmuskip=2mu plus 1mu minus 2mu\relax
\thickmuskip=2.5mu plus 2.5mu\relax
\allowdisplaybreaks
\eas{
&~\bbE_{x,y} \ip{f(x) - y}{g(x_J) - y} 
\\
&\leq \sigma \sqrt{M} \cdot \left( \sum_{m=1}^M \bbE_y \left[ \mathrm{Var} ( f(x)_m - f(x_{J^c})_m | y )  \right] \right)^{\!\!\!\!\!\!1/2}
\\
&\leq \sigma \sqrt{M} \cdot \left( \sum_{m=1}^M \bbE_y \left[ \bbE_{x|y} \left[ f(x)_m - f(x_{J^c})_m \right]^2 \right] \right)^{\!\!\!\!\!\!1/2}
\\
&= \sigma \sqrt{M} \cdot \left( \sum_{m=1}^M \bbE_x \left[ f(x)_m - f(x_{J^c})_m \right]^2 \right)^{\!\!\!\!\!\!1/2},
}
\endgroup
where the equality uses the filtration property of conditional expectation.

\section{Examples that support behavior of \R{sys:SSRL:b} depending on $\sigma$}
\label{sec:SSRL:eg}

The first example in general regression models the pseudo-target as follows: $g(x_J) = y+e_1$, where $e_1 \in \bbR^M$ is some arbitrarily additive noise independent of $y$.
This gives $\mathrm{Var} (g(x_J)_m | y) = \mathrm{Var} ( y_m + (e_1)_m | y ) = \mathrm{Var} (e_1)_m \leq \sigma^2$, $m = 1,\ldots,M$.
Under this model, how much $g(x_J)$ moves around $y$ is captured by $\sigma$.

The second example in image denoising assumes that $x$ is corrupted by AWGN $e_2 \in \bbR^N$ that is independent of $y$.
Setting $g$ as a linear mapping $G \in \bbR^{J \times N}$ gives $\mathrm{Var} (g(x_J)_n | y) = \mathrm{Var} ( ( G y_J )_n + ( G ( e_2 )_J )_n | y ) = \mathrm{Var} ( ( G ( e_2 )_J )_n ) \leq \sigma^2$, $\forall n = 1,\ldots,N$.
Under this model, how much $g(x_J)$ moves around $y$ is captured by $\sigma$.

\section{Preliminary results with the teacher-student learning perspective}
\label{sec:tea-stu}

The proposed SSRL framework is applicable to teacher-student learning~\cite{Wang&Yoon:21TPAMI, Hinton&etal:15NIPSW, Bucilua&etal:06SIGKDD} that aims to learn a smaller student network from bigger teacher network(s).
We ran preliminary SSRL experiments for low-dose CT denoising (with simulated data).
The teacher model $g$ is the pre-trained $8$-layer DnCNN by Noise2Self (with checkerboard masking), and we set the student model $f$ as \{$8,7,6,5,4,3$\}-layer DnCNN. Applying SSRL-Noise2Self, we obtained the following numerical results: the RMSE (HU) values of student models with \{$8,7,6,5,4,3$\}-layer DnCNNs are \{$25.0,25.3,25.5,25.4,25.2,27.5$\}, respectively. The student DnCNNs that have the equal or lower complexity compared to its teacher network, significantly improves its teacher model of which RMSE value is 30.9 (in HU). The results might imply that student models learned from SSRL can outperform their teacher model, if they retain sufficiently high network complexity as compared to their teacher's (e.g., $3$-layer DnCNN). In addition, we have additional SSRL experiment in low-dose CT denoising with the \dquotes{iterative} teacher-student perspective. The teacher model $g$ is pre-trained $5$-layer DnCNN from the non-iterative teacher-student SSRL method above, and we set the student model $f$ as $3$-layer DnCNN. We obtained comparable performance ($27.3$ RMSE (in HU)) to that of the $3$-layer DnCNN obtained by the non-iterative teacher-student SSRL method. We conjecture that iterative teacher-student SSRL needs more sophisticated $g$-setups, such as an ensemble of teacher models~\cite{Hinton&etal:15NIPSW}.

\ifCLASSOPTIONcaptionsoff
  \newpage
\fi

\bibliographystyle{IEEEtran}
\bibliography{referenceBibs_Bobby}


\renewcommand{\theHtable}{S.\thetable}
\renewcommand\thetable{S.\arabic{table}}
\renewcommand{\theHfigure}{S.\thefigure}
\renewcommand\thefigure{S.\arabic{figure}}
\renewcommand{\theHsection}{S.\thesection}
\renewcommand\thesection{S.\arabic{section}}

\setcounter{section}{0}
\setcounter{figure}{0}
\setcounter{table}{0}



\begin{strip}
\vspace{-2.0pc}
\begin{center}
\fontsize{23.4}{23.4}\selectfont 
Self-supervised regression learning using domain knowledge: Applications to improving\\ self-supervised denoising in imaging\\ (Supplement)
\end{center}
\vspace{-3.0pc}
\end{strip}

\begin{strip}
\begin{center}
\fontsize{10}{10}\selectfont Il Yong Chun$^{\dagger,\ast}$,~\IEEEmembership{Member,~IEEE,} Dongwon Park$^\dagger$,~\IEEEmembership{Student Member,~IEEE,} Xuehang Zheng$^\dagger$, \\ Se Young Chun$^\ast$,~\IEEEmembership{Member,~IEEE,} and Yong Long$^\ast$,~\IEEEmembership{Member,~IEEE}
\end{center}
\vspace{-0.5pc}
\end{strip}

\blfootnote{$\dagger$\textit{These authors contributed equally to this work.}}
\blfootnote{$\ast$\textit{Corresponding authors.}}

\section{Empirical results to support some claims in main paper}
\label{sec:add:empirical}

\begin{figure}[!t]
\centering
\small\addtolength{\tabcolsep}{-9.5pt}

\begin{tabular}{c}
\includegraphics[width=45mm]{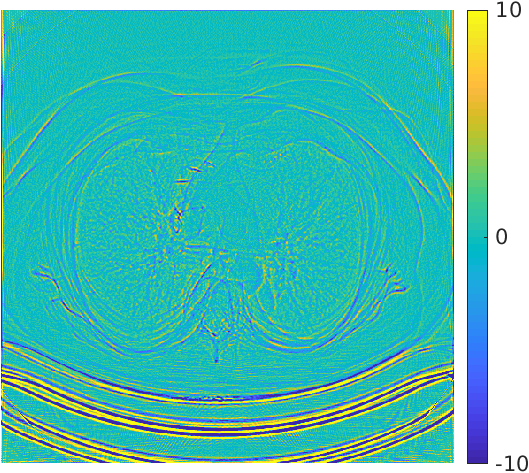}
\end{tabular}

\vspace{-0.5pc}
\caption{Sample mean of noise in low-dose FBP images --  $e$ in \R{eq:noise:ct}.
We calculated the sample mean with $200$ samples.}
\label{fig:exp_noise_ct}
\vspace{-3.5pc}
\end{figure}

Figure~\ref{fig:exp_noise_ct} empirically supports our claim in Section~\ref{sec:know:ct} that noise of FBP-reconstructed images in low-dose CT, i.e., $e$ in \R{eq:noise:ct}, has approximately zero-mean.
The position of the patient table base is similar across FBP images, so it gave higher errors in the calculated sample mean; see the bottom of the image in Figure~\ref{fig:exp_noise_ct}.

\begin{figure}[!t]
\centering
\small\addtolength{\tabcolsep}{-9.5pt}

\begin{tabular}{c}
        \small{(a) Synthetic low-dose CT dataset ($30$ samples)}\vspace{-0.2pc}
        \\
        
 		\raisebox{-.5\height}{
 			\begin{tikzpicture}
 			\begin{scope}[spy using outlines={rectangle,yellow,magnification=1.7,size=8mm, connect spies}]
			\node {\includegraphics[width=90mm]{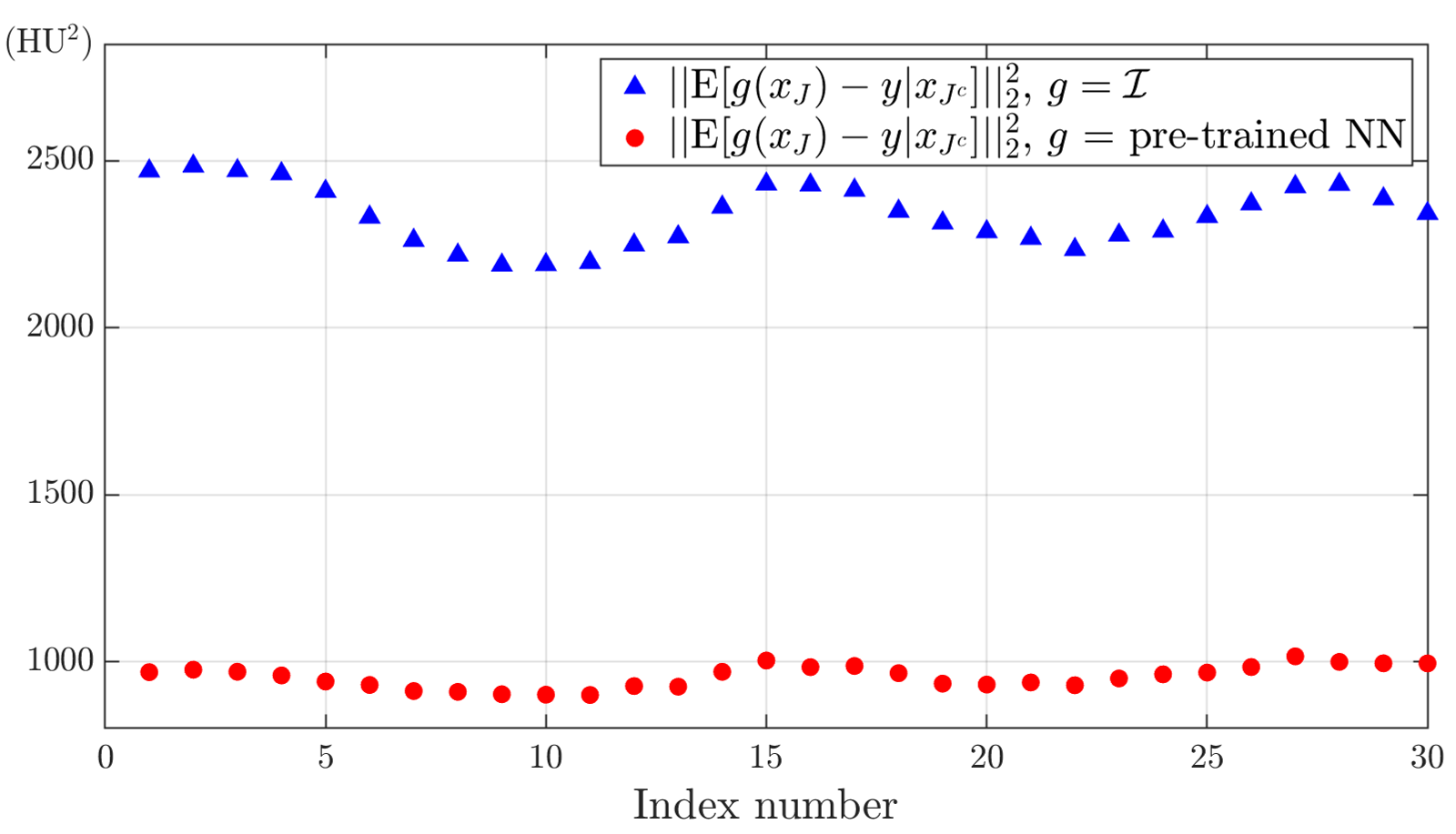}};  			
 			\end{scope}
 			\end{tikzpicture}}
 			\\
 			
        \small{(b) Real-world camera imaging dataset ($30$ samples)}\vspace{-0.2pc}
        \\
        
 		\raisebox{-.5\height}{
 			\begin{tikzpicture}
 			\begin{scope}[spy using outlines={rectangle,yellow,magnification=1.55,size=15mm, connect spies}]
			\node {\includegraphics[width=90mm]{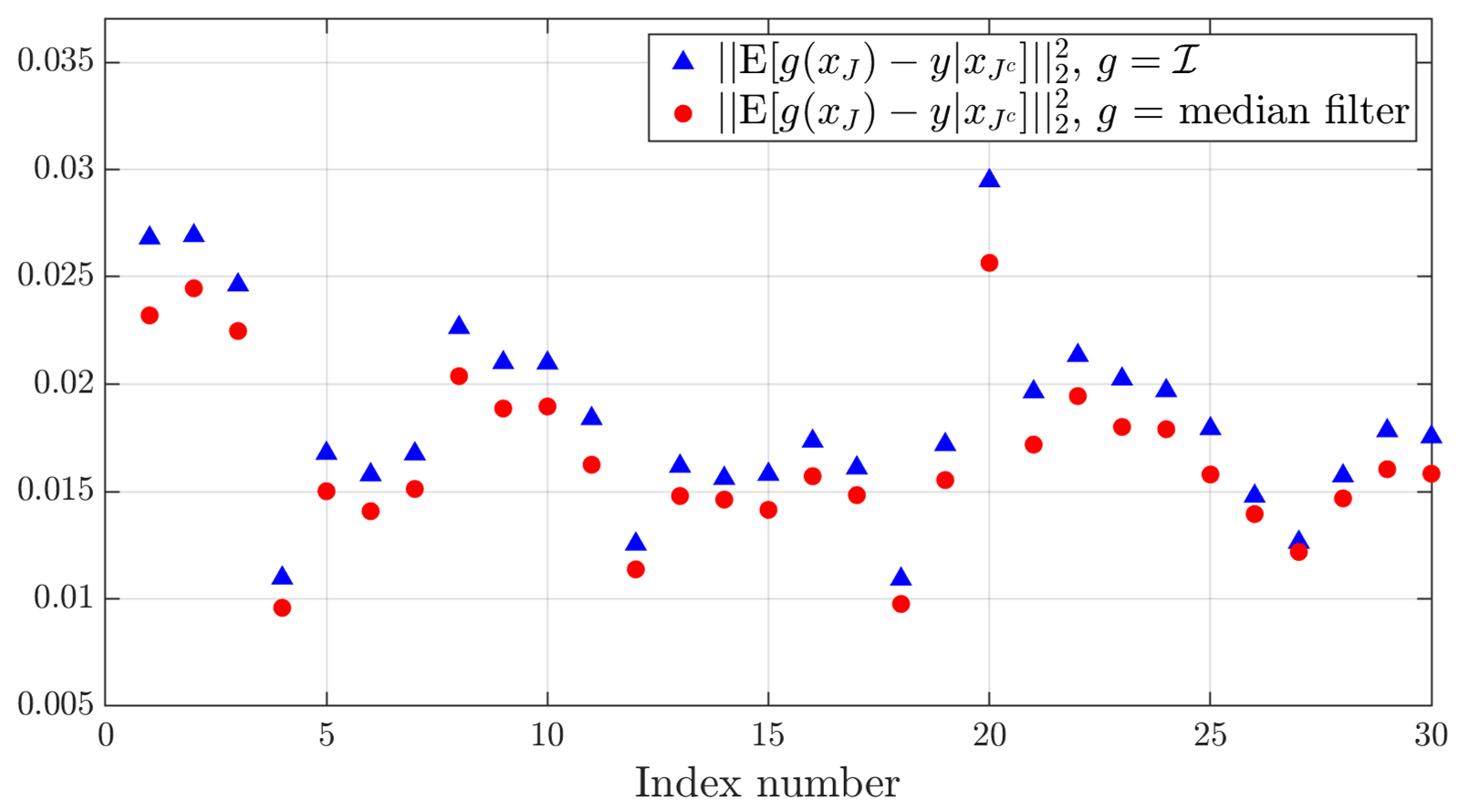}};
 			\end{scope}
 			\end{tikzpicture}} 

\end{tabular}

\vspace{-0.5pc}
\caption{Empirical observation of $|| \bbE[g(x_J) - y|x_{J^c}] ||_2^2$ with different $g$ functions for different applications. 
We estimated the conditional expected values from samples.}
\label{fig:cond_exp}
\vspace{-0.5pc}
\end{figure}

Figure~\ref{fig:cond_exp} supports our claim in Section~\ref{sec:know:ct} and Section~\ref{sec:comp:real} that given $x_J$, designed $g$ leads $g(x_J)$ closer to $y$ than $g=\cI$.
In the simulated low-dose CT experiment,
we calculated empirical $||\bbE[g(x_J)-y|x_{J^c}]||_2^2$ with $50$ noise realizations for each test sample for  pre-trained NN $g$ and $g=\cI$.
See Figure~\ref{fig:cond_exp}(a).
In the real-world camera image denoising experiment,
we calculated empirical $||\bbE[g(x_J)-y|x_{J^c}]||_2^2$ for each randomly selected test sample for weighted median filtering $g$ and $g=\cI$. (The dataset includes only a single noise realization for each sample.)
See Figure~\ref{fig:cond_exp}(b).

\begin{figure}[!t]
 	\centering
 	\addtolength{\tabcolsep}{-9.5pt}
 	\renewcommand{\arraystretch}{0.8}
 	
 	\begin{tabular}{cc}

 		 \specialcell[c]{\small 
 		 Simulated camera imaging dataset (BSD 300 test samples)} 		 
 		 \vspace{-0.25pc}
 		 \\

 		\raisebox{-.5\height}{
 			\begin{tikzpicture}
 			\begin{scope}[spy using outlines={rectangle,yellow,magnification=1.7,size=8mm, connect spies}]
			\node {\includegraphics[width=34mm,height=34mm]{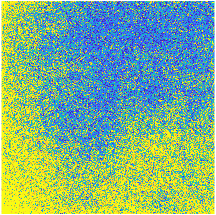}};  			
 			\end{scope}
 			\end{tikzpicture}}     \hspace{-1pc} 
 		\raisebox{-.5\height}{
 			\begin{tikzpicture}
 			\begin{scope}[spy using outlines={rectangle,yellow,magnification=1.55,size=15mm, connect spies}]
			\node {\includegraphics[width=34mm,height=34mm]{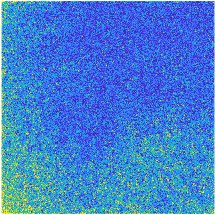}};
 			\end{scope}
 			\end{tikzpicture}} 

    \vspace{-0.4pc}
 	\end{tabular}
 	
	\caption{Empirical observations of $|\bbE[x-y|y]|$ {\bfseries (left)} and $|\bbE[g(x)-y|y]|$ {\bfseries (right)} in simulated camera image denoising where $g$ is median filtering (blue and yellow denote $0$ and $0.1$ absolute errors, respectively). We used the model $y = g(x) + e'$, where $e'$ is independent of $y$.}
    \label{fig:nat:sample_mean}
    \vspace{-1.0 pc}
 \end{figure}

Figure~\ref{fig:nat:sample_mean} empirically supports our claim in Section~\ref{sec:know:nat} that designed $g$ \dquotes{approximately} satisfies $\bbE[g(x) | y] = y$ in camera image denoising using simulated datasets. 
We modeled that $y = g(x) + e'$, where $e'$ is an error term independent of $y$. 
Using this model, we rewrote Assumption~3 by $\bbE[g(x) | y] = \bbE [e'] = 0$.
We calculated the sample mean to estimate $\bbE [e']$ using BSD 300 test samples (with noise model \R{eq:noise:nat}) for median filtering $g$ and $g = \cI$. 
The empirical measures for \{$\mathrm{avg} ( | \bbE[x-y|y] | )$, $\mathrm{avg} ( | \bbE[g(x)-y|y] | )$\} are \{$0.0201$, $0.0098$\}, where $\mathrm{avg}$ denotes averaging across pixels and $g$ is set as median filtering. These support our claim that $\bbE[g(x)|y] = y$ is approximately satisfied with median filtering $g$.

\section{Implementation details for simulated data experiments}

\subsection{Common implementation details in both applications}
\label{sec:com_details}

For all the existing self-supervised denoising methods using training datasets specified in Section~\ref{sec:result} and Noise2True,
we finely tuned their hyperparameters, 
including
the initial learning rate, learning rate decay parameters, minibatch size, number of DnCNN layers, and balancing parameter $\sigma$ (see, e.g., \R{sys:SSRL:b}), to achieve the best numerical results.
We simply applied the chosen hyperparameter sets to corresponding SSRL setups.
We applied the chosen learning rate decay parameters, minibatch size, and number of DnCNN layers in Noise2True experiments to all the aforementioned self-supervised learning methods.

For the existing self-supervised denoising methods, Noise2Self \citesupp{Batson&Royer:19ICML:supp} and Noise2Same \citesupp{Xie&Wang&Ji:20NIPS:supp}, we used their default masking setups.
The Noise2Self default setup uses the deterministic masking scheme for each $J$ that equi-spacedly samples $6.25\%$ of the number of pixels in each training image (i.e., a single pixel is selected in each $4 \times 4$ window). 
The Noise2Same default setup uses the saturated sampling scheme \citesupp{Xie&Wang&Ji:20NIPS:supp, Krull&etal:19CVPR:supp} for each $J$ that randomly samples $\approx\! 0.5\%$ of the number of pixels in each training image (i.e., a single pixel is sampled in each $14 \times 14$ window). 
In training denoising NNs, both methods interpolate missing pixels in $x_{J^c}$ by applying weighted average to their $8$ neighboring pixels, and use interpolated $x_{J^c}$ as input to denoisers.

For the existing Noise2Noise-motivated methods that emulate pairs of two independent noisy images, 
Noise2Inverse \citesupp{Hendriksen&etal:20TCI:supp} and Neighbor2Neighbor \citesupp{Huang&eta:21CVPR:supp},
we calculated their loss with non-masked images as proposed.

We tested all trained regression NNs to non-masked images -- rather than masked images with $J^c$ -- as this setup gave higher denoising accuracy than prediction with masked images \citesupp{Batson&Royer:19ICML:supp, Xie&Wang&Ji:20NIPS:supp}. 

For the existing state-of-the-art single-image self-supervised denoising methods, Noisy-As-Clean \citesupp{Xu&etal:20TIP:supp} and Self2Self \citesupp{Quan&etal:20CVPR:supp}, we used their default implementations.
In the Noisy-As-Clean experiments, we selected \quotes{Blind DnCNN+NAC} for simulated noisy datasets (this gave better results than \quotes{Blind Resnet+NAC}), and used \quotes{Blind Resnet+NAC} for intrinsically noisy datasets following their setup.
In the Self2Self experiments, we used \quotes{Self2Self-NN} for all the experiments.

\subsection{Implementation details for self-supervised learning experiments with simulated low-dose CT denoising data}

We used the Michigan image reconstruction toolbox (MIRT) \citesupp{fessler:16:irt:supp} in simulating low-dose FBP images.
We used fan-beam geometry for sinogram simulation,
where width of each detector column is $1.2858$ mm, source to detector distance is $1085.6$ mm, and source to rotation center distance is $595$ mm. 
For the FBP method, we used a ramp filter 
because in general, it better preserves the sharpness of edges on reconstructed images than Hanning filter (but overall noise increases).

The common hyperparameters for all self-supervised learning methods and Noise2True were defined as follows.
We used $8$-layer DnCNN \citesupp{Zhang&etal:17TIP:supp} with its default setup and the modified U-Net used in Noise2Self \citesupp{Batson&Royer:19ICML:supp}, and trained all DnCNNs and U-Nets with the mini-batch version of Adam \citesupp{Kingma&Ba:15ICLR:supp}.
We selected the batch size and the number of epochs as $2$ and $1,\!000$, respectively,
and decayed the learning rates by a factor of $0.95$ every $10$ epochs.
In training DnCNNs and U-Nets, we selected the initial learning rates as $0.1$ and $5 \times 10^{-5}$, respectively, unless stated otherwise.

For proposed SSRL-Noise2Self and SSRL-Noise2Same, we used complementary checkboard masks $J$ and $J^c$ in Figure~\ref{fig:diagram} (top). 
We observed in this application that if $g$ is set to use small amount of information, i.e., $| J | \ll | J^c |$, then pre-trained $g$ makes poor prediction.
For SSRL-Noise2Self,
we set $g$ as pre-trained NN by Noise2Self with complementary checkerboard masks (see its inference results with $8$-layer DnCNN in Figure~\ref{fig:ct:checker}(a)).
In training DnCNNs, we used the same initial learning rate as that used in pre-training $g$, $0.01$.
We computed $\cL_{\text{ind}}$ as given in \R{sys:SSRL:Jind} (see Figure~\ref{fig:diagram}(top)). 

For SSRL in the Noise2Inverse setup,
we set $f$ and $g$ as $f/2$ and $\cI - g/2$, respectively, where $g$ is pre-trained NN by Noise2Inverse.
In training DnCNNs, we used the same initial learning rate as that used in pre-training $g$, $0.001$.
In inference, we averaged the predictions from $f$ and $g$, as this corresponds to training setup above.
In all Noise2Inverse inferences (including SSRL-Noise2Inverse), 
we input full-view FBP images
since this is consistent with other experiments and gave better denoising performance than denoising half-view FBP images.
For SSRL-Noise2Same, 
we set $g$ as pre-trained NN by Noise2Same with complementary checkerboard masks (see its test results with DnCNN in Figure~\ref{fig:ct:checker}(b)), and computed $\cL$ (both terms) in \R{sys:SSRL:b} only on $J^c$.
We chose the balancing parameter $\sigma$ as $15$, setting the ratio of the first term to the squared second term $2\sigma \sqrt{M} \| f(x)_{J^c} - f(x_{J^c})_{J^c} \|_2^2$ in \R{sys:SSRL:b} as $10$.
For Noise2Same with either the default setup and complementary checkerboard masks, we chose the balancing parameter $\sigma$ as $500$.
For self-supervised denoising methods with complementary checkerboard masks,
we interpolated missing pixels in both $x_{J^c}$ and $x_J$, by averaging their $4$ neighboring pixels.

The DnCNN and U-Net training time for existing self-supervised denoising methods and proposed SSRL methods was less than $10$ hours and $12$ hours, respectively, with an NVIDIA TITAN Xp GPU.
It took total less than $22$ hours to train both $f$ and $g$.

\begin{table*}[!b]	
\vspace{-0.5pc}
	\centering
	\small\addtolength{\tabcolsep}{-3pt}
    \caption{Averaged test RMSE (HU) comparisons from different learning methods with U-Net in low-dose CT denoising (simulated low-dose CT dataset).}
    \label{tab:ct}
    
    \vspace{-0.4pc}
	\begin{tabular}{ccccccc}
	\toprule
    Noise2True   &
    Noise2Self  &
    \specialcell[c]{{\bfseries Proposed SSRL~~~} \\  in Noise2Self \\ setup}  &
    \specialcell[c]{Noise2-\\Inverse}  &
    \specialcell[c]{{\bfseries Proposed SSRL} \\ in Noise2- \\ Inverse setup} &
    ~Noise2Same  & 
    \specialcell[c]{{\bfseries Proposed SSRL} \\ in Noise2Same \\ setup}      
    \\
    \midrule
    \midrule
    18.5   &
    \boxittabbb{3.95cm}{1.7cm}
    32.3  &
    {\bfseries 26.7} &
    \boxittabbb{3.7cm}{1.7cm}
    24.0   &  
    {\bfseries 23.5}  &
    \boxittabb{4.1cm}{1.7cm}
    31.1   &  
    {\bfseries 28.2} \\    
    \bottomrule    
	\end{tabular}
  \vspace{-0.5pc}
\end{table*}
\begin{table*}[!bh]	
	\centering
	\small\addtolength{\tabcolsep}{-3pt}
	\caption{Averaged test PSNR (dB) ({\bfseries first} row) and SSIM ({\bfseries second} row) comparisons with from different learning methods with U-Net in camera image denoising (simulated noisy \emph{BSD 300} dataset).}
    \label{tab:nat:BSD}
    
    \vspace{-0.4pc}
	\begin{tabular}{ccccccc}
	\toprule
    Noise2True~~   & Noise2Self
    &
    \specialcell[c]{{\bfseries Proposed SSRL~} \\  in Noise2Self \\ setup}  &
    \specialcell[c]{
    Neighbor2- \\
    Neighbor}
    &
    \specialcell[c]{{\bfseries Proposed SSRL~} \\ in Neighbor2- \\ Neighbor setup} &
    Noise2Same  & 
    \specialcell[c]{{\bfseries Proposed SSRL~} \\ in Noise2Same \\ setup}      
    \\
    \midrule
    
    26.0    & 
    20.9    &  
    {\bfseries 22.5 }  &
    20.8   &
    {\bfseries 22.4 } & 
    19.5    &  
    {\bfseries 20.7 } \\    
    0.849    &\;
    \boxittabbb{3.94cm}{1.95cm}
    0.730    &  
    {\bfseries 0.765 }  &
    \boxittabbb{4.0cm}{1.95cm}
    0.728   &
    {\bfseries 0.767 }  &
    \boxittabb{4.15cm}{1.95cm}
    0.612  &  
    {\bfseries 0.660} \\    
    \bottomrule    
	\end{tabular}
\end{table*}


 \begin{figure*}[!b]
 	\centering
 	\addtolength{\tabcolsep}{-9.5pt}
 	\renewcommand{\arraystretch}{0.8}
 	
 	\begin{tabular}{cc}
 		
 		 \specialcell[c]{\small (a) Noise2Self with complementary \\ \small checkerboard masks} &  
 		 \specialcell[c]{\small (b) Noise2Same with complementary \\ \small checkerboard masks} 
 		 \vspace{-0.2pc}
 		 \\

 		\raisebox{-.5\height}{
 			\begin{tikzpicture}
 			\begin{scope}[spy using outlines={rectangle,yellow,magnification=1.7,size=8mm, connect spies}]
 			\node {\includegraphics[width=34mm]{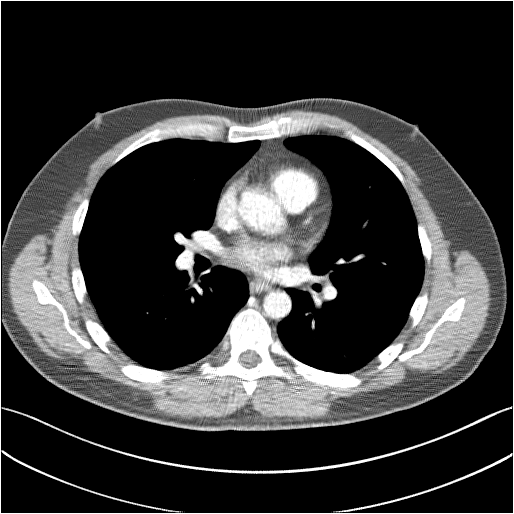}};
 			\spy on (-0.4, -0.9) in node [left] at (-0.9,-1.3);
 			\spy on (0.25, 0.25) in node [left] at (1.7,-1.3);
			\node [white] at (0.0,1.4) {\small $\text{RMSE} = 30.9$ HU};
 			\end{scope}
 			\draw[red, line width=0.2mm] (-1.1,-1.38) circle (1mm);
		    \draw[red, line width=0.2mm] (1.51,-1.40) circle (0.9mm); 			
 			\end{tikzpicture}}     \hspace{-1pc} 
 		\raisebox{-.5\height}{
 			\begin{tikzpicture}
 			\begin{scope}[spy using outlines={rectangle,yellow,magnification=1.55,size=15mm, connect spies}]
 			\node {\includegraphics[width=34mm]{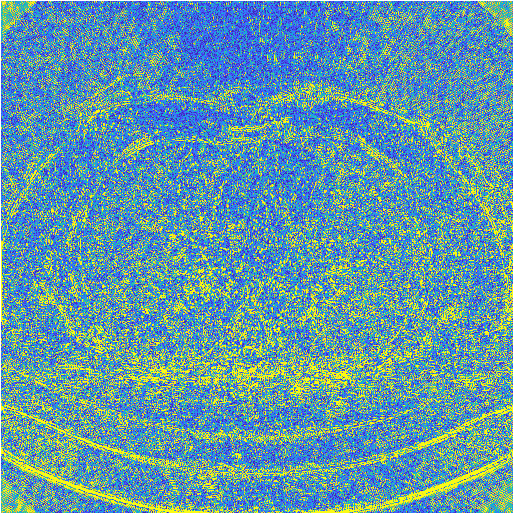}};
 			\end{scope}
 			\end{tikzpicture}} &

        \hspace{0.25pc}

 		\raisebox{-.5\height}{
 			\begin{tikzpicture}
 			\begin{scope}[spy using outlines={rectangle,yellow,magnification=1.7,size=8mm, connect spies}]
 			\node {\includegraphics[width=34mm]{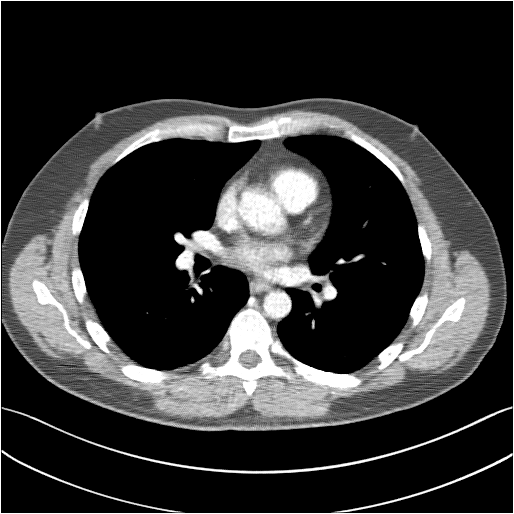}};
 			\spy on (-0.4, -0.9) in node [left] at (-0.9,-1.3);
 			\spy on (0.25, 0.25) in node [left] at (1.7,-1.3);
			\node [white] at (0.0,1.4) {\small $\text{RMSE} = 28.7$ HU};
 			\end{scope}
 			\draw[red, line width=0.2mm] (-1.1,-1.38) circle (1mm);
		    \draw[red, line width=0.2mm] (1.51,-1.40) circle (0.9mm);
 			\end{tikzpicture}}  \hspace{-1pc}    
 		\raisebox{-.5\height}{
 			\begin{tikzpicture}
 			\begin{scope}[spy using outlines={rectangle,yellow,magnification=1.55,size=15mm, connect spies}]
 			\node {\includegraphics[width=34mm]{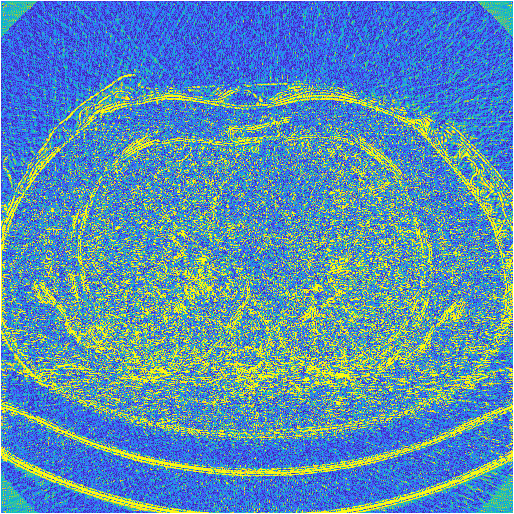}};
 			\end{scope}
 			\end{tikzpicture}}

    \vspace{-0.4pc}
 	\end{tabular}
 	
	\caption{Denoised images via DnCNNs and their corresponding error maps from Noise2Self and Noise2Same with complementary checkerboard masks in low-dose CT. 
	(The display window of denoised images is $[800, 1200]$ HU; in error maps, 
    blue and yellow denote $0$ and $50$ absolute errors in HU, respectively.)
	RMSE values were averaged across all test samples.}
	\label{fig:ct:checker}
 \end{figure*}

\subsection{Implementation details for self-supervised learning experiments with simulated camera image denoising data}
\label{sec:impl:sim:camera}

The common hyperparameters for all self-supervised learning methods and Noise2True were defined as follows. (In this application, these gave good image denoising performance across all existing self-supervised learning methods and Noise2True, since we rescaled or normalized training images; see details below.)
We used the default $17$-layer DnCNN \citesupp{Zhang&etal:17TIP:supp} and the modified U-Net used in Noise2Self \citesupp{Batson&Royer:19ICML:supp},
and trained all DnCNNs and U-Nets with the mini-batch version of Adam \citesupp{Kingma&Ba:15ICLR:supp}.
We selected the initial learning rate, the batch size, and the number of epochs as $8 \times 10^{-4}$, $8$, and $190$, respectively,
and decayed the learning rates by a factor of $0.5$ every $50,\!000$ iterations.
(For Neighbor2Neighbor \citesupp{Huang&eta:21CVPR:supp}, we set the batch size as $32$, as it reduces training image size with $2\times 2$ sub-sampling window.)
We used the data augmentations in Noise2Same~\citesupp{Xie&Wang&Ji:20NIPS:supp}, i.e., random crops with size $256 \times 256$, rotation and flipping. 
Except for Noise2Same, we rescaled all images to $[0,1]$, following \citesupp{Batson&Royer:19ICML:supp, Huang&eta:21CVPR:supp}.
In Noise2Same experiments (including SSRL-Noise2Same), we normalized each image by subtracting its mean and dividing by its standard deviation at each channel,
following \citesupp{Xie&Wang&Ji:20NIPS:supp}.

For proposed SSRL in the Noise2Self and Noise2Same setups (referred to as SSRL-Noise2Self and SSRL-Noise2Same, respectively), we used the deterministic masking scheme in Figure~\ref{fig:diagram}(bottom) for each $J$ with $\approx\! 11.1\%$ and $\approx\! 1.2\%$ sampling ratio, respectively
-- i.e., a single pixel is selected in each $3 \times 3$ and $9 \times 9$ window, respectively. These setups gave more appealing results than the default masking parameters (i.e., $4 \times 4$ window in Noise2Self and $14 \times 14$ window in Noise2Same); compare Figure~\ref{fig:natural:default} to corresponding results in Figure~\ref{fig:nat}.
We observed in this application that using sufficient amount of information for a linear interpolation in $f$ is useful for giving good prediction.
For weighted median filtering \citesupp{brownrigg1984weighted:84ACM:supp} $g$ in all SSRL setups, we used the following weights: $[1,2,1; 2,\boxed{9},2; 1,2,1]$, where a box denotes the central weight. 
The dilation rates of weighted median filtering for SSRL in the Noise2Self, Noise2Same, and Neighbor2Neighbor setups are $3$, $9$, and $1$, respectively, 
corresponding to the distances between pixels in $J$.
For SSRL-Noise2Self, 
we computed $\cL_{\text{ind}}$ in \R{sys:SSRL:Jind} only on $J$ (see Figure~\ref{fig:diagram} (bottom)).
For SSRL-Noise2Same, we used the same balancing parameter $\sigma$ as Noise2Same used, i.e., $\sigma = 1$, and computed $\cL$ (both terms) in \R{sys:SSRL:b} only on $J$.

The DnCNN and U-Net training time for each experiment was less than $72$ hours with an NVIDIA TITAN V GPU.

\section{Supplementary experimental results for Section~\ref{sec:result}}
\label{sec:add:results}

 \begin{figure*}[!t]
 	\centering
 	\addtolength{\tabcolsep}{-9.5pt}
 	\renewcommand{\arraystretch}{0.8}
 	
 	\begin{tabular}{cc}
 		
 		 \specialcell[c]{\small (a) SSRL-Noise2Self with the default setup \\ 
 		 \small in Noise2Self (i.e., $4 \times 4$ window)} &  
 		 \specialcell[c]{\small (b) SSRL-Noise2Same with the default setup \\ 
 		 \small in Noise2Same (i.e., $14 \times 14$ window)}
 		 \vspace{-0.2pc}
 		 \\

 		\raisebox{-.5\height}{
 			\begin{tikzpicture}
 			\begin{scope}[spy using outlines={rectangle,yellow,magnification=1.7,size=8mm, connect spies}]
 			\node {\includegraphics[width=34mm]{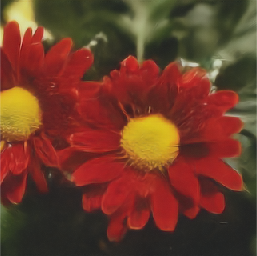}};
  			\spy on (-1.4, -0.1) in node  [left] at (-0.9,-1.3);

			\node [white] at (0.0,1.4) {\small $\text{PSNR} = 22.0$ dB};
 			\end{scope}
 			\end{tikzpicture}}     \hspace{-1pc} 
 		\raisebox{-.5\height}{
 			\begin{tikzpicture}
 			\begin{scope}[spy using outlines={rectangle,yellow,magnification=1.55,size=15mm, connect spies}]
 			\node {\includegraphics[width=34mm]{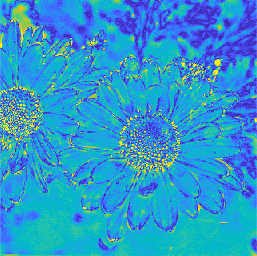}};
 			\end{scope}
 			\end{tikzpicture}} &

        \hspace{0.25pc}

 		\raisebox{-.5\height}{
 			\begin{tikzpicture}
 			\begin{scope}[spy using outlines={rectangle,yellow,magnification=1.7,size=8mm, connect spies}]
 			\node {\includegraphics[width=34mm]{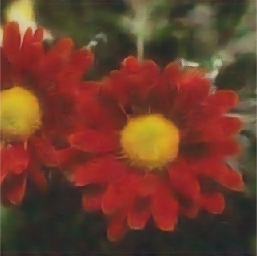}};
 			\spy on (-1.4, -0.1) in node  [left] at (-0.9,-1.3);

			\node [white] at (0.0,1.4) {\small $\text{PSNR} = 21.1$ dB};
 			\end{scope}
 			\end{tikzpicture}}  \hspace{-1pc}    
 		\raisebox{-.5\height}{
 			\begin{tikzpicture}
 			\begin{scope}[spy using outlines={rectangle,yellow,magnification=1.55,size=15mm, connect spies}]
 			\node {\includegraphics[width=34mm]{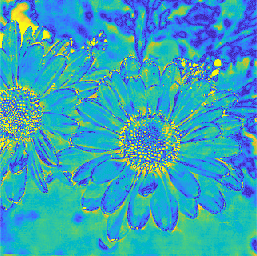}};
 			\end{scope}
 			\end{tikzpicture}}

    \vspace{-0.4pc}
 	\end{tabular}
 	
	\caption{Denoised images via DnCNNs and their corresponding error maps from SSRL-Noise2Self and SSRL-Noise2Same with default masks setup in camera image denoising. (In error maps, blue and yellow denote $0$ and $50$ absolute errors, respectively.) 
	PSNR values were averaged across all BSD 300 test samples.}
	\label{fig:natural:default}
    \vspace{-0.2pc}
 \end{figure*}

Tables~\ref{tab:ct}--\ref{tab:nat:BSD} report quantitative U-Net image denoising results with chest slices of The 2016 Low Dose CT Grand Challenge data in simulated low-dose CT, and with BSD 300 data in simulated camera imaging.
Red boxes in Tables~\ref{tab:ct}--\ref{tab:nat:BSD} compare an existing self-supervised denoising method to proposed SSRL in the corresponding setup.

Figure~\ref{fig:ct:checker} shows denoised images from Noise2Self and Noise2Same with DnCNN and complementary checkerboard masks $J$ and $J^c$, and reports the corresponding quantitative test results, in low-dose CT denoising. 
Comparing the results in Figure~\ref{fig:ct:checker} to 
those of Noise2Self and Noise2Same using the default masking setups in Figures~\ref{fig:ct} and \ref{fig:ct:err}
shows that checkerboard masking improves denoising quality over default masking in Noise2Self, and achieves comparable denoising performance to default masking in Noise2Same.
(See their default masking setups in Section~\ref{sec:com_details}.)
SSRL-Noise2Self with checkerboard masking achieves significantly better denoising quality compared to Noise2Self with checkerboard masking.
We used pre-trained DnCNN from these two setups as $g$ in the corresponding SSRL setup.

\begin{figure*}[!thb] 
	\centering
	\includegraphics[width=0.42\textwidth]{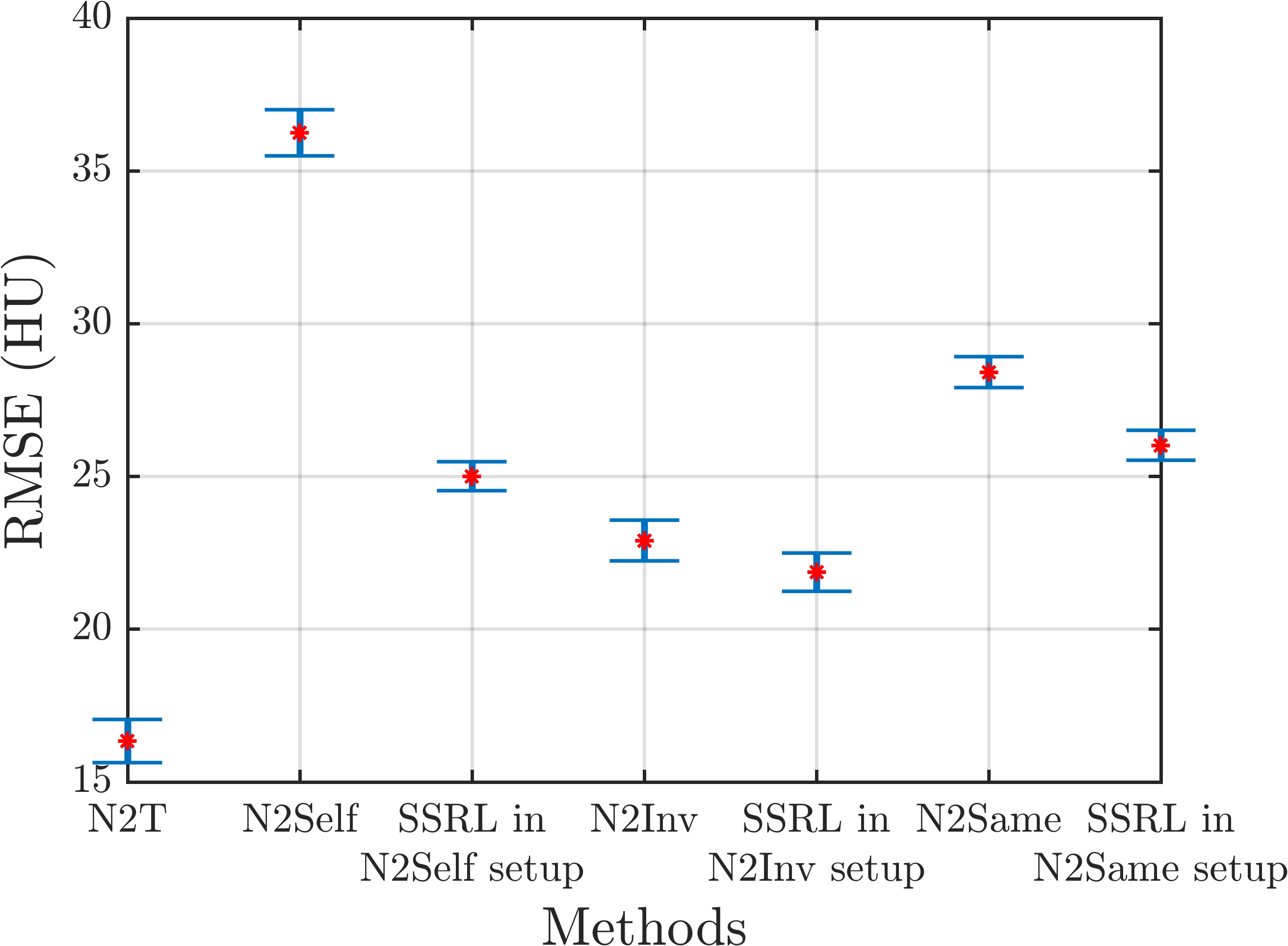}
	\includegraphics[width=0.42\textwidth]{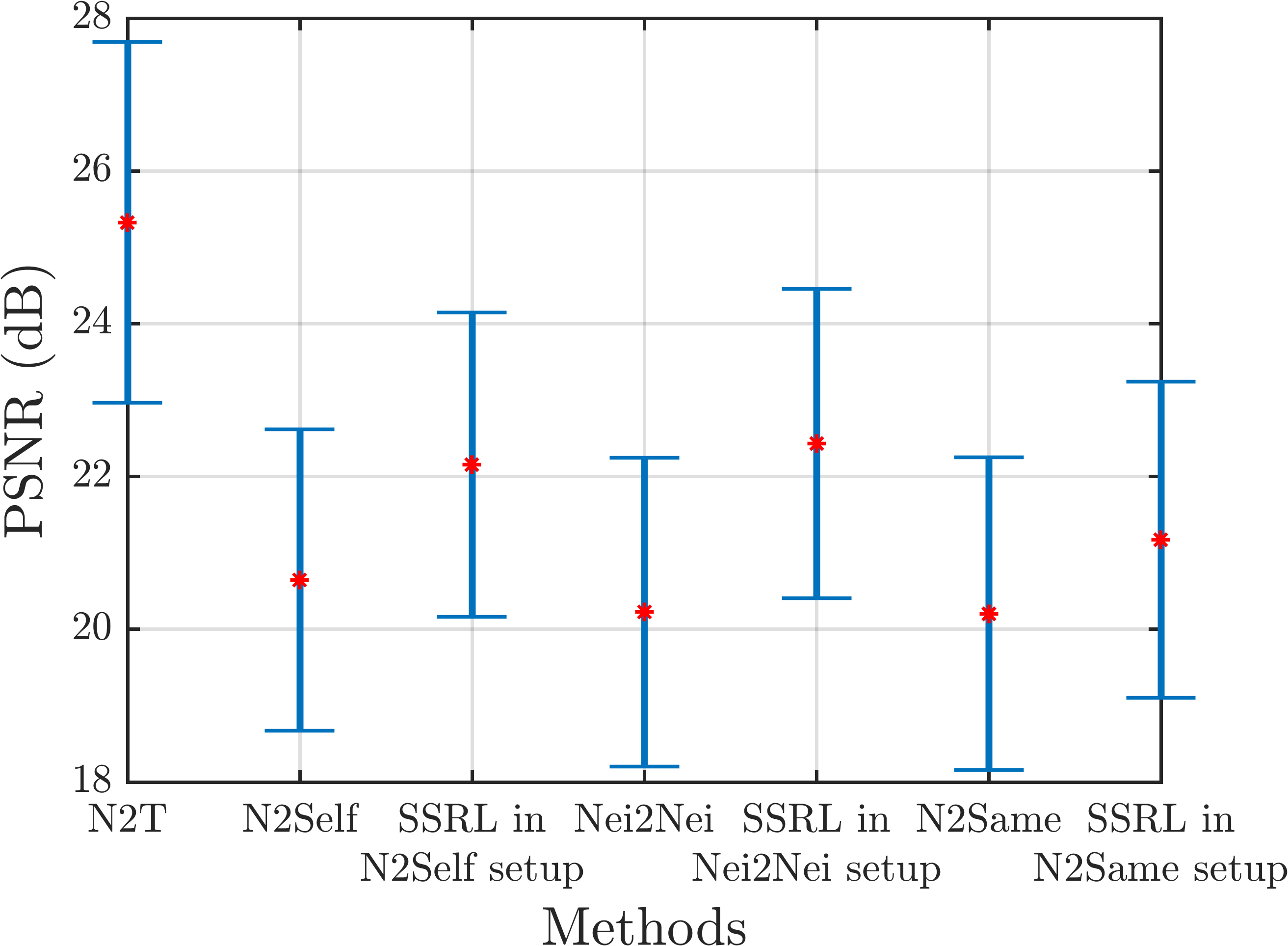} 

    \vspace{-0.4pc}
	\caption{Denoising performance error bars for different learning methods with DnCNN in simulated low-dose CT {\bfseries (left\textnormal{, $30$ test images})} and simulated camera image denoising  (with BSD 300) {\bfseries (right\textnormal{, $300$ test images})}. 
	Red asterisks denote the averaged test RMSE or PSNR values. Error bar represents one standard deviation of test RMSE or PSNR values.}
    \label{fig:err_bar}
\vspace{-0.5pc}
\end{figure*}

Figure~\ref{fig:natural:default} shows denoised camera images from SSRL-Noise2Self and SSLR-Noise2Same using the default masking parameters in Noise2Self and Noise2Same (see Section~\ref{sec:impl:sim:camera}). 
Compare the results in Figure~\ref{fig:natural:default} with the corresponding ones in Figure~\ref{fig:nat} using the designed setups (see Section~\ref{sec:impl:sim:camera}).
The comparisons demonstrate that the default and designed setups give very similar PSNR results, i.e., $\leq 0.1$ dB, but the designed setups gives slightly more visually appealing results than the default ones in Noise2Self and Noise2Same.

Figure~\ref{fig:err_bar} compares prediction uncertainty of trained DnCNN denoisers via different learning methods in both applications.
The error bar graphs in Figure~\ref{fig:err_bar} show that for both applications, in all the three comparison sets, proposed SSRL gives similar or lower prediction uncertainty over the existing self-supervised learning methods, Noise2Self, Noise2Inverse or Neighbor2Neighbor, and Noise2Same.

\ifCLASSOPTIONcaptionsoff
  \newpage
\fi

\bibliographystylesupp{IEEEtran}
\bibliographysupp{reference_Supp}

\end{document}